\documentclass[prd,superscriptaddress,altaffilletter,showpacs,nofootinbib]{revtex4}
\usepackage[dvips]{graphicx}
\usepackage{amsmath}
\newcommand{\be}{\begin{equation}}
\newcommand{\ee}{\end{equation}}
\newcommand{\bea}{\begin{eqnarray}}
\newcommand{\eea}{\end{eqnarray}}
\newcommand{\der}{\partial}
\newcommand{\vphi}{\varphi}

\begin{document}

\title{On the phantom barrier crossing and the bounds on the speed of sound in non-minimal derivative coupling theories}

\author{Israel Quiros}\email{iquiros@fisica.ugto.mx}\affiliation{Dpto. Ingenier\'ia Civil, Divisi\'on de Ingenier\'ia, Universidad de Guanajuato, Gto., M\'exico.}

\author{Tame Gonzalez}\email{tamegc72@gmail.com}\affiliation{Dpto. Ingenier\'ia Civil, Divisi\'on de Ingenier\'ia, Universidad de Guanajuato, Gto., M\'exico.}

\author{Ulises Nucamendi}\email{ulises@ifm.umich.mx}\affiliation{Instituto de F\'isica y Matem\'aticas, Universidad Michoacana de San Nicol\'as de Hidalgo, Edificio C-3, Ciudad Universitaria, CP. 58040 Morelia, Michoac\'an, M\'exico.}

\author{Ricardo Garc\'{\i}a-Salcedo}\email{rigarcias@ipn.mx}\affiliation{CICATA - Legaria del Instituto Polit\'ecnico Nacional, 11500, M\'exico, D.F., M\'exico.}

\author{Francisco Antonio Horta-Rangel}\email{anthort@hotmail.com}\affiliation{Dpto. Ingenier\'ia Civil, Divisi\'on de Ingenier\'ia, Universidad de Guanajuato, Gto., M\'exico.}

\author{Joel Saavedra}\email{joel.saavedra@ucv.cl}\affiliation{Instituto de F\'isica, Pontificia Universidad Cat\'olica de Valpara\'iso,
Casilla 4950, Valpara\'iso, Chile.}

\date{\today}

\begin{abstract} In this paper we investigate the so called ``phantom barrier crossing'' issue in a cosmological model based in the scalar-tensor theory with non-minimal derivative coupling to the Einstein's tensor. Special attention will be paid to the physical bounds on the squared sound speed. The numeric results are geometrically illustrated by means of a qualitative procedure of analysis that is based on the mapping of the orbits in the phase plane onto the surfaces that represent physical quantities in the extended phase space, that is: the phase plane complemented with an additional dimension relative to the given physical parameter. We find that the cosmological model based in the non-minimal derivative coupling theory -- this includes both the quintessence and the pure derivative coupling cases -- has serious causality problems related with superluminal propagation of the scalar and tensor perturbations. Even more disturbing is the finding that, despite that the underlying theory is free of the Ostrogradsky instability, the corresponding cosmological model is plagued by the Laplacian (classical) instability related with negative squared sound speed. This instability leads to an uncontrollable growth of the energy density of the perturbations that is inversely proportional to their wavelength. We show that independent of the self-interaction potential, for the positive coupling the tensor perturbations propagate superluminally, while for the negative coupling a Laplacian instability arises. This latter instability invalidates the possibility for the model to describe the primordial inflation.\end{abstract}


\maketitle

%%%%%%%%%%%%%%%%%%%%%%%%%%%%%%%%%%%%%%%

\section{Introduction}\label{sec-intro}

Scalar-tensor theories \cite{maeda-book, faraoni-book}, among which the Brans-Dicke (BD) theory \cite{bd-theory} is the prototype, have a long and hesitating history \cite{brans}. Despite that until very recently no fundamental scalar particle was found in nature, these theories have found a variety of applications both in gravitational and in cosmological contexts. In the list of famous scalar fields (this includes the prototype BD scalar field) we encounter the Higgs particle of the standard model of particles \cite{smp-book}, the dilaton -- and other moduli fields -- of the effective (low-energy) string theory \cite{wands-rev}, the inflaton that accounts for the early inflationary stage of the cosmic evolution \cite{linde, inflaton} and the quintessence field that embodies the so called dark energy that inflates the Universe at late times \cite{quintessence}, among others. Starting in 2013 year things changed and it seems that the first fundamental scalar particle has been finally discovered \cite{higgs-discovery}. This entails that scalars and, consequently, scalar-tensor theories have to be taken seriously as feasible scenarios for physical phenomena. 

The BD theory \cite{bd-theory}, as well as the more general scalar-tensor theories \cite{maeda-book, faraoni-book}, are classical theories of the gravitational field and as such these are not intended to describe quantum gravitational phenomena. However, there are indications that including higher order terms into the gravitational action makes the given theory of gravity more compatible with quantum (renormalizable) variants \cite{stelle-prd} whose predictions can be trusted back enough into the past. One example is the addition of four-order terms like $R_{\mu\nu\tau\rho}R^{\mu\nu\tau\rho}$, $R_{\mu\nu}R^{\mu\nu}$ and $R^2$ into the Einstein-Hilbert action that gives a class of multimass models of gravity \cite{stelle-grg} where, in addition to the usual massless excitations of the fields, there are massive scalar and spin-2 excitations with a total of 8 degrees of freedom.\footnote{The unwanted (yet tractable) property of this theory is that the massive spin-2 mode is ghost-like \cite{ovrut-prd}.} In this vein it is interesting to complement the action of standard scalar-tensor theories with higher-order terms in order to have a theory more compatible with a would be quantum version. This modification would include not only terms quadratic in the curvature invariants but, also, higher-derivative terms like: $c_1R\der_\mu\phi\der^\mu\phi,$ $c_2R_{\mu\nu}\der^\mu\phi\der^\nu\phi,$ $c_3R_{\mu\nu}\phi\nabla^\mu\nabla^\nu\phi,$ $c_4\nabla_\mu R\phi\der^\mu\phi,$ $c_5R\phi\nabla^2\phi,$ $c_6\nabla^2R\phi^2,$ where $\nabla^2\equiv\nabla_\mu\nabla^\mu$ and $c_1,...,c_6$ are coupling constants with the dimensions of length-squared. 

The problem with the undiscriminated addition of higher-derivative terms is that the resulting equations of motion contain derivatives higher than second-order and this, in turn, leads to the appearance of awful and catastrophic Ostrogradsky ghosts in the theory, that makes it strongly unstable and untenable as an adequate model of gravitational phenomena. The most general possible scalar-tensor theories that contain higher order derivatives and derivative couplings in the Lagrangian and that, at the same time, lead to second-order motion equations -- so that these are free of the Ostrogradsky instability -- are called as ``Horndeski'' theories \cite{horndeski, nicolis, deffayet, tsujikawa} (see \cite{beyond-horndeski} for a class of theories generalizing the Horndeski ones). These theories have been applied with success to describe the cosmological evolution of our Universe in different contexts \cite{kazuya, japan, chow, also}. An interesting subset of the Horndeski theories is composed of the so called scalar-tensor theories with a non-minimal derivative (kinetic) coupling, in particular those where the kinetic coupling is to the Einstein's tensor \cite{sushkov, saridakis-sushkov, matsumoto, granda, gao, germani, germani-1}: $\propto G^{\mu\nu}\der_\mu\phi\der_\nu\phi$. The latter theory is characterized by its relative mathematical simplicity when compared with other Horndeski theories and also by its ability to account for the early (transient) inflationary stage, since it is able to explain in a unique manner both a quasi-de Sitter phase and an exit from it without any fine-tuned potential \cite{sushkov}. 

The action for the typical theory with non-minimal derivative coupling of the scalar with the Einstein's tensor: $G_{\mu\nu}\equiv R_{\mu\nu}-g_{\mu\nu}R/2$, is given by:

\bea &&S=\int d^4x\frac{\sqrt{|g|}}{2}\left[R-\left(\epsilon g^{\mu\nu}-\alpha G^{\mu\nu}\right)\der_\mu\phi\der_\nu\phi-2V(\phi)\right]+S_m,\label{action}\eea where we set $8\pi G_N=c=h=1$, and the coupling constant $\alpha$ is a real number. The parameter $\epsilon$ can take the following values: $\epsilon=+1$ (quintessence), $\epsilon=-1$ (phantom cosmology), and $\epsilon=0$ (pure derivative coupling).\footnote{In this paper we refer to 'pure derivative coupling' -- independent of the presence or absence of the self-interacting potential -- to the models based in the action principle \eqref{action} without the standard kinetic term ($\epsilon=0$), i. e., there is only kinetic coupling to the Einstein's tensor.} In the above equation $S_m$ is the action of the matter degrees of freedom other than the scalar field. 

Theories of the type \eqref{action} have been studied in different contexts. For instance, in \cite{rinaldi} static, spherically symmetric solutions to the gravitational field equations derived from \eqref{action} were explored and black hole solutions with a single regular horizon were found, and their thermodynamical properties were examined. Related work regarding asymptotically locally AdS and flat black holes can be found in \cite{anabalon}, while in \cite{cisterna} the authors constructed the first neutron stars based in \eqref{action}. The obtained construction may -- in principle -- constrain in a phenomenological way the free parameters of the model. Cosmological scenarios based in theories with kinetic coupling with the Einstein's tensor have been studied in \cite{saridakis-sushkov} in order to examine quintessence (and phantom) models of dark energy with zero and constant self-interaction potentials. It has been shown that, in general, the universe transits from one de Sitter solution to another, depending on the coupling parameter. A variety of behaviors -- including Big Bang and Big Crunch solutions, and also cosmological bounce -- reveals the capabilities of the corresponding cosmological model. A dynamical systems analysis of the derivative coupling model with the Higgs-like potential can be found in \cite{matsumoto}, while a similar study for the exponential potential has been performed in \cite{huang}. It was found that, for the quintessence case, the stable fixed points are the same with and without the non-minimal derivative coupling, while for the pure derivative coupling (no standard canonical kinetic term) only the de Sitter attractor exists and the dark matter solution is unstable. Cosmology based in \eqref{action} has been also investigated in \cite{jinno}. The latter paper points out the existence of the Laplacian instability in the theory with kinetic coupling of the scalar field with the Einstein's tensor in the context of reheating after inflation. Particle production after inflation in the model \eqref{action} tensor has also been studied in the reference \cite{jinno-others} by the same authors.

A very interesting -- and to our opinion, central -- aspect of the theory \eqref{action} was investigated in \cite{gao}. In that reference it was found that in the cosmological model based in \eqref{action} with pure derivative coupling to the Einstein's tensor ($\epsilon=0$) and with vanishing potential $V=0$ -- in the absence of other matter sources ($S_m=0$) -- the scalar behaves as pressureless matter with vanishing sound speed, so that it could be a candidate of cold dark matter. By also considering the scalar potential ($V\neq 0$), it was found that the scalar field may play the role of both dark matter and dark energy. In this case, the effective equation of state (EOS) of the scalar field $\omega_\text{eff}$ can cross the phantom divide \cite{caldwell, crossing-odintsov, crossing-observ, crossing-nesseris, mohseni}: $\omega_\Lambda=-1$ (this is properly the EOS parameter of the cosmological constant), but this can lead to the sound speed becoming superluminal as it crosses the divide, and so is physically forbidden.\footnote{It is well known that Horndeski theories all possess some configurations with a superluminal propagation.} The possibility of the phantom divide crossing in the model is in itself a very interesting finding, however two results we find particularly interesting in this study: i) that the crossing of the phantom divide may be linked with superluminal sound speed, and ii) that the physical limits on the sound speed are used as a basic criterion for rejection of a given cosmological model. The fact that the physical bounds on the speed of propagation of the perturbations of the field is to be taken carefully and seriously when Horndeski-type theories are under investigation, was understood also by the authors of \cite{sup-lum-gal-1}. In that reference it was shown that, when the Dirac-Born-Infeld (DBI) galileon is considered as a local modification to gravity, such as in the Solar system, the existing stable solutions always exhibit superluminality, casting doubt on the existence of a standard Lorentz invariant UV completion of that theory.\footnote{There exist alternative points of view on this issue. For instance, in \cite{vikman, sup-lum-gal-2} it is shown that k-essence and galileon theories, respectively, satisfy an analogue of Hawking's chronology protection conjecture, an argument that can be extended to include Hordenski theories in general. However, there are strong arguments that contradict such kinds of non-orthodox points of view on causality (for more on this issue see Ref. \cite{sup-lum-other}). In this regard we recommend the clear and pedagogical discussion on this issue given in \cite{ellis-roy}.} 

In view of the importance of the above issue, and given that there does not exist in the bibliography a thorough discussion on the implications for cosmology of the physical bounds on the speed of sound in the theory with the kinetic coupling to the Einstein's tensor,\footnote{In \cite{gao} the subject was only partially investigated -- only connection of the phantom barrier crossing with superluminality of the scalar perturbations was established -- besides only the pure derivative coupling case $\epsilon=0$ was considered in that reference. The issue was also stated but not investigated in \cite{dent}. In this latter reference (see last paragraph of page 8) the authors state that the investigation of the instabilities and superluminality in the model with the kinetic coupling to the Einstein's tensor lies beyond the scope of their paper. A similar statement can be found in \cite{matsumoto} (see the top paragraph of page 3).} in the pres paper we shall be concentrating in the ``$\omega_\Lambda=-1$'' barrier crossing issue in the model \eqref{action} by paying special attention to the physical bounds on the speed of sound squared $c_s^2$. These bounds are imposed by stability and causality, two fundamental principles of classical physical theories: The squared sound speed should be non-negative $c^2_s\geq 0$ since otherwise, the cosmological model will be classically unstable against small perturbations of the background energy density, usually called as Laplacian -- also gradient -- instability. Besides, causality arguments impose that the mentioned small perturbations of the background should propagate at most at the local speed of light $c^2_s\leq 1$. 

In order to implement the numeric investigation we shall explore two specific potentials: the frequently encountered in cosmological applications exponential potential \cite{huang, exp-pot-ferreira, exp-pot-wands}: $V=V_0\exp(\lambda\phi)$ and, also, the power-law potential $V=V_0\phi^{2n}$ \cite{pwl-pot-peebles}. The exponential potential

\bea V=V_0\,e^{\lambda\phi}\;\Rightarrow\;V'=\lambda V,\label{exp-pot}\eea where $V_0$ and $\lambda$ are real constants ($V_0\geq 0$), can be found as well in higher-order or higher-dimensional gravity theories \cite{exp-pot-origin-1}, and in string or Kaluza-Klein type models, where the moduli fields may have effective exponential potentials \cite{exp-pot-origin-2}. Exponential potentials can also arise due to nonperturbative effects such as gaugino condensation \cite{exp-pot-origin-3}. In the present model the exponential potential has been investigated in \cite{huang}, where a dynamical systems analysis was performed. The conclusion of the authors was that the derivative coupling to the Einstein's tensor does not modify the phase space dynamics of the quintessence \cite{exp-pot-wands}. The power-law potential

\bea V=V_0\phi^{2n}\;\Rightarrow\;V'=2n\,V_0^{1/2n}V^{1-1/2n},\label{pow-law-pot}\eea where $V_0$ is a non-negative constant and $n$ is a real parameter, is also frequently found in the cosmological applications \cite{pwl-pot-peebles}. In the quintessence case the inverse-power law potential exhibits the tracker behavior, a very desirable property for the quintessence if one wants to avoid the cosmic coincidence problem \cite{pwl-pot-other}. The origin of this potential might be associated with supersymmetry considerations \cite{pwl-pot-origin}.

%-----------------------------------

\begin{figure*}
\includegraphics[width=5cm]{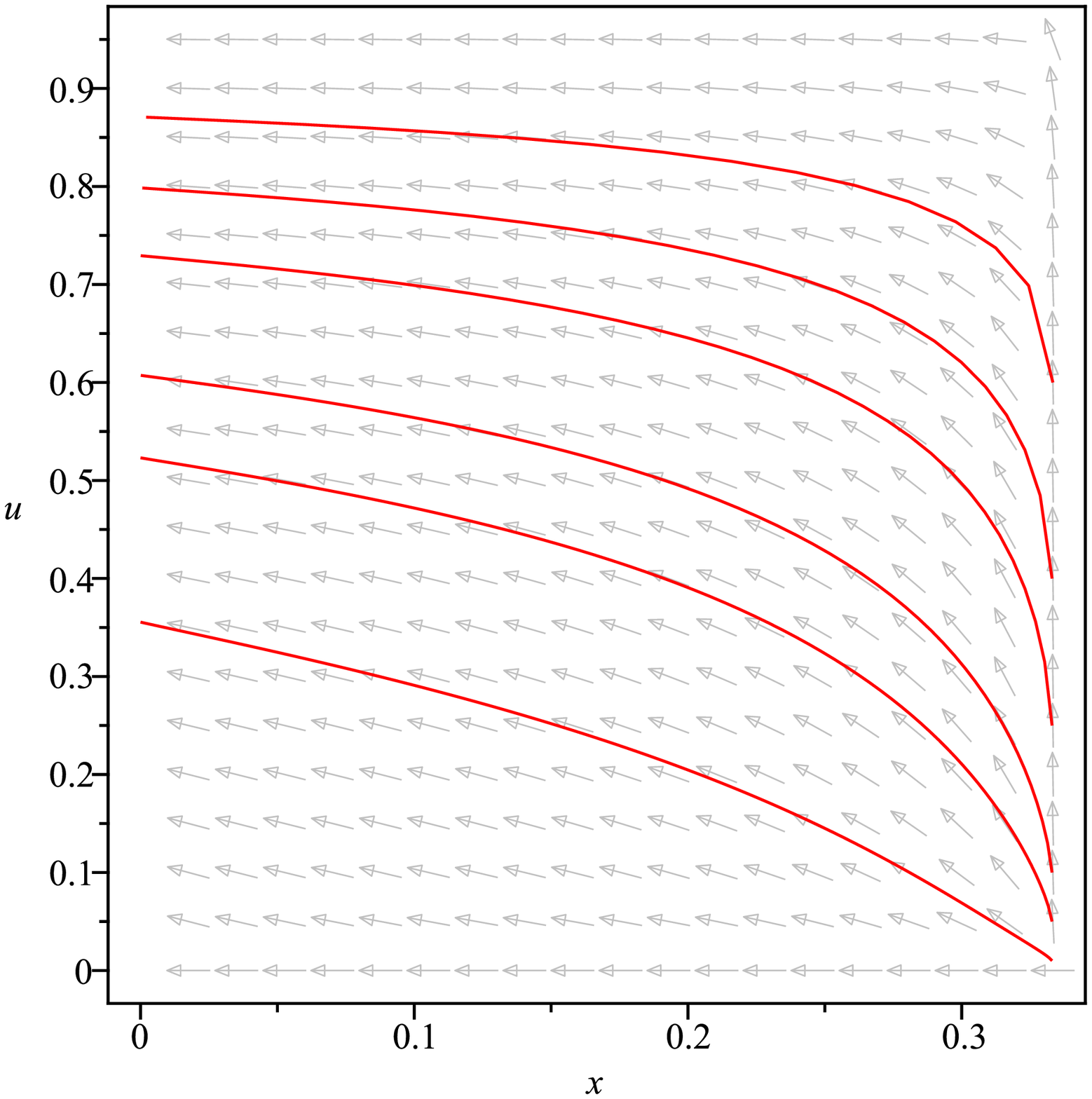}
\includegraphics[width=6cm]{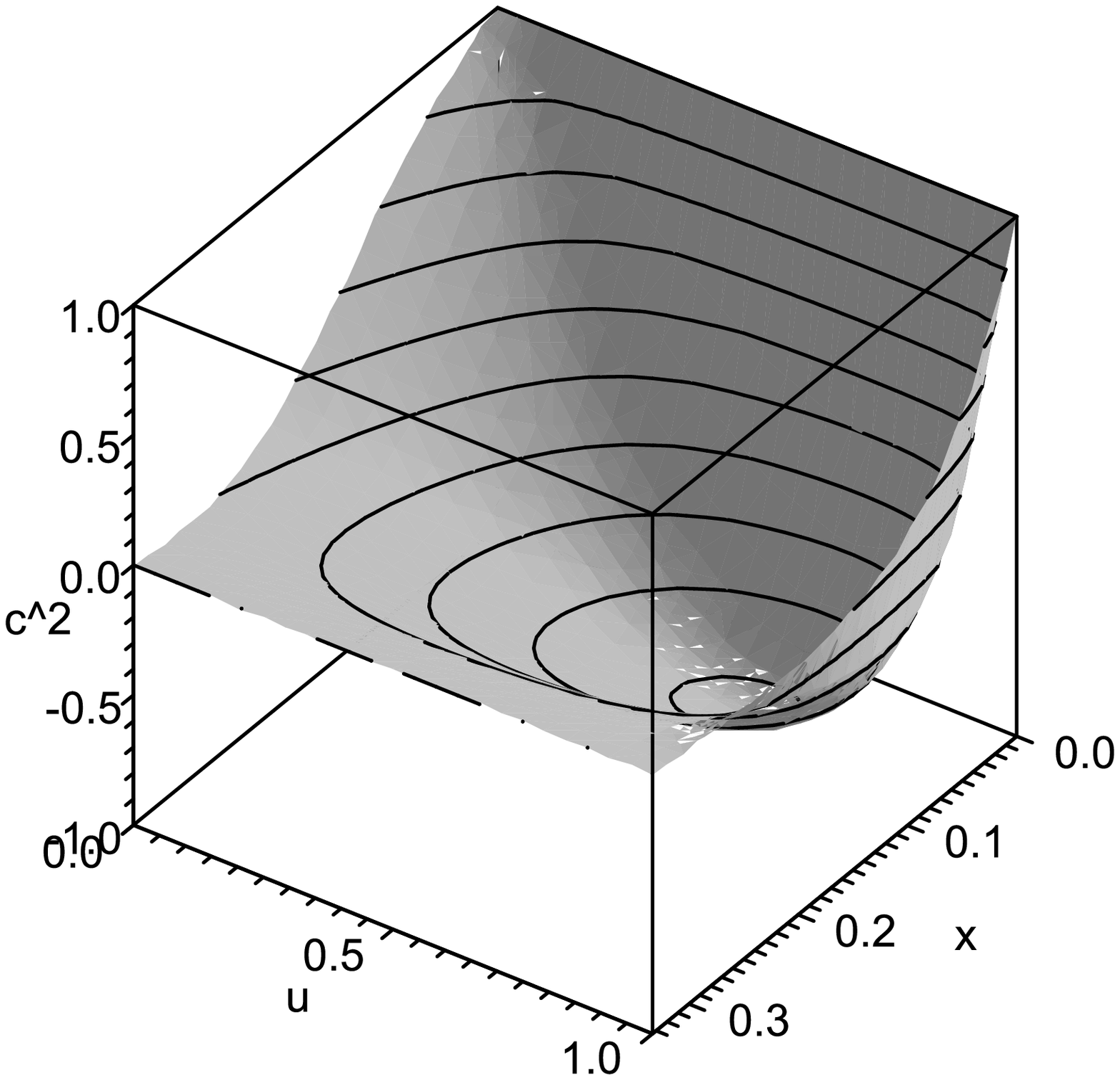}
\includegraphics[width=6cm]{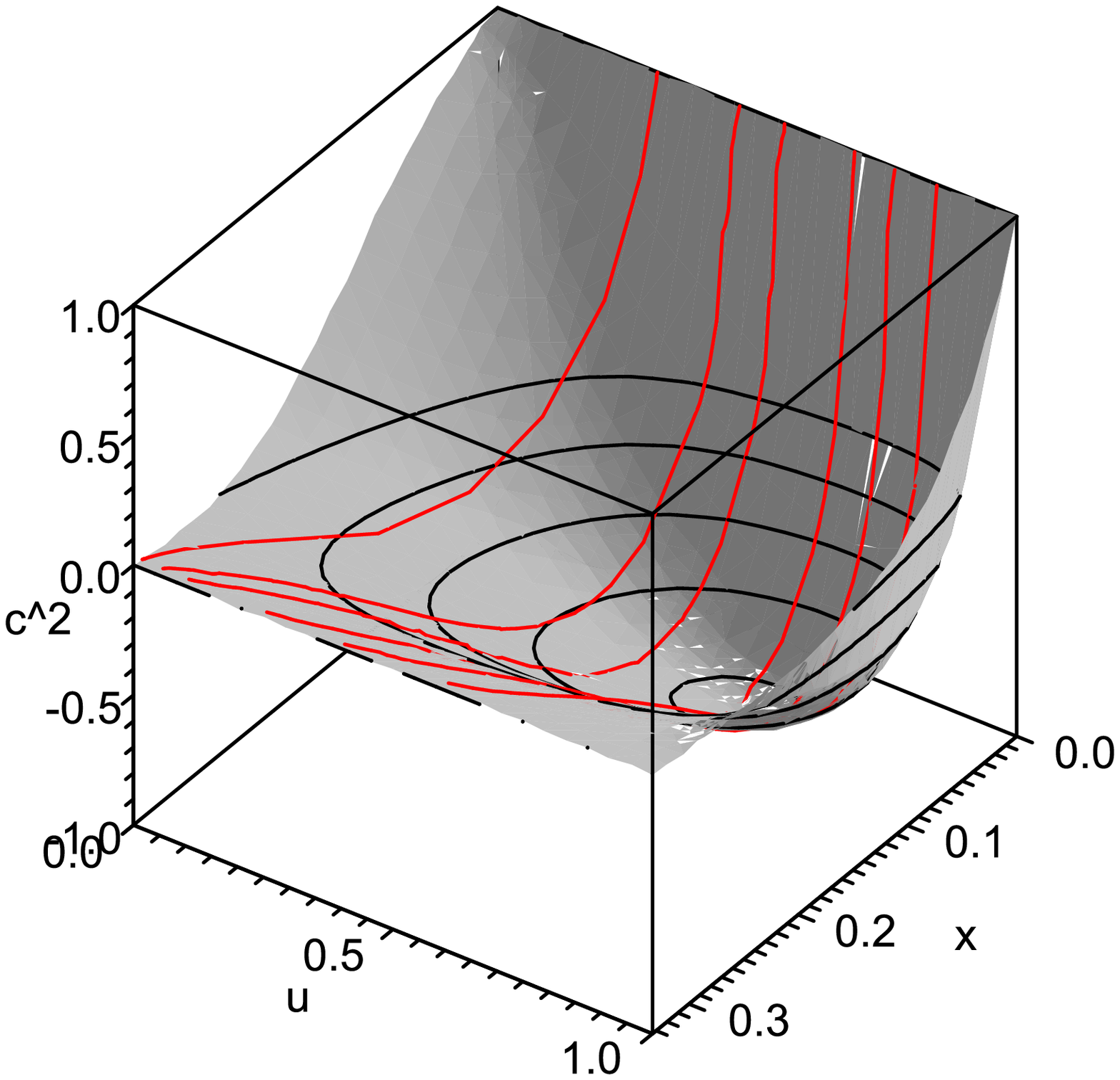}\vspace{0.7cm}
\caption{The $c^2_s$-embedding schematically represented. The phase portrait of the dynamical system \eqref{ode-xu} and the plot of the surface $c^2_s=c^2_s(x,u)$ -- with contours -- in the extended (three-dimensional) phase space that is spanned by the coordinates $x$, $u$ and $c^2_s$, are shown in the left-hand and in the middle figures, respectively. In the right-hand figure the $c^2_s$-embedding diagram is drawn: the orbits (red curves) appearing in the phase portrait (left) have been embedded into the surface $c^2_s=c^2_s(x,u)$. The computations correspond to the cosmological model \eqref{action} with positive coupling ($\alpha>0$) and for the growing exponential potential ($\lambda=5$). The contours drawn in the right-hand figure mark the region where $c^2_s<0$, i. e., where the Laplacian instability develops. The different embedded orbits correspond to whole cosmic evolutionary pathways that are associated with different sets of initial conditions. From the embedding diagram it is seen that independent on the initial conditions chosen the corresponding cosmological histories inevitably go through a stage where $c^2_s<0$, so that the classical gradient instability destroys any chance for the Universe to evolve into its present state.}\label{fig2}\end{figure*}

%-----------------------------------

As a qualitative support to the present discussion, a geometric procedure of analysis based on the properties of the dynamical system is developed. It provides a clear illustration of the failure of causality and/or of the development of Laplacian instability -- as well as of the crossing of the phantom divide -- along given phase space orbits. The mentioned procedure relies on the mapping of phase space orbits into the extended phase space, that is: the phase plane complemented with an additional dimension represented by the physical parameter of interest (the effective EOS or the squared sound speed, for instance). This is why we call the procedure as $P$-embedding, where $P$ refers to the given physical parameter. Although the numeric computations are performed for the exponential and for the power-law potentials only, the constant and vanishing potential cases are implicitly included as their particular cases. The embedding procedure is schematically represented in FIG. \ref{fig2}, where the $c^2_s$-embedding is illustrated for the cosmological model of interest, for the positive coupling case ($\alpha>0$) and for the monotonically growing exponential potential \eqref{exp-pot} with $\lambda=5$.

The numeric investigation is preceded by -- and complimented with -- a throughout analytic study. In this regard we shall go as far as we can before specifying the form of the self-interaction potential, so that our discussion be as independent as it can of the specific choice of the potential. Our results show that the cosmological models based in the scalar-tensor theory with non-minimal derivative coupling to the Einstein's tensor \eqref{action} develop severe causality problems related with superluminal propagation of the perturbations of the scalar field. These problems are critical whenever the crossing of the phantom divide happens, however, these may arise even in the absence of the crossing. More problematic than the violations of causality in the model is the finding that it is plagued by the classical (catastrophic) Laplacian instability, despite that the theory \eqref{action} in which it is based is free of the Ostrogradsky instability. Our results just confirm the inappropriateness of the models based on the kinetic coupling theories of the kind \eqref{action}, as it has been discussed just recently in \cite{new}, on the light of the tight constraint on the difference in speed of photons and gravitons $(c^2_T-c^2)/c^2\leq 6\times 10^{-15}$ ($c_T$ is the speed of the gravitational waves) implied by the announced detection of gravitational waves from the neutron star-neutron star merger GW170817 and the simultaneous measurement of the gamma-ray burst GRB170817A \cite{ligo}.

Before we go further, in order to unify the terminology and to be able to compare our results with other results in the bibliography, we want to make a comment on the sign of the coupling constant $\alpha$. This constant was named as $\kappa$ in \cite{sushkov, saridakis-sushkov}, $\alpha$ in \cite{gao}, $\zeta$ in \cite{dent} and $\omega^2$ in \cite{huang}. If compare the action in \cite{sushkov} (equation (8) of that reference) -- the same action as in \cite{saridakis-sushkov} but in this case the self-interacting potential for the scalar field is considered -- we see that their $\kappa$ corresponds to $-\alpha$ of the present paper, so that, when the authors of \cite{sushkov, saridakis-sushkov} refer to negative coupling $\kappa<0$ this means positive coupling in terms of our $\alpha$ ($\alpha>0$) and vice versa. We recall that in \cite{sushkov, saridakis-sushkov} both cases: $\kappa>0$ ($\alpha<0$) and $\kappa<0$ ($\alpha>0$), were considered. In \cite{gao} it seems that there is a problem with the sign of the Lagrangian density in (1.5) of their work. While a straightforward comparison of the action (2.4) of \cite{gao} -- with the substitution of the Lagrangian density (1.5) -- with our equation \eqref{action} yields the correspondence $\alpha\rightarrow-\alpha$ between the coupling constant in their work and in the present paper, respectively, a comparison of our cosmological field equations \eqref{feqs} with the corresponding equations (2.12) in \cite{gao} yields to a direct correspondence $\alpha\rightarrow\alpha$. Here we give preference to the cosmological field equations so that we shall assume that the sign of the coupling constant in \cite{gao} and in our paper coincides. In a similar way the sign of the coupling constant $\zeta$ in \cite{dent} and $\omega^2$ in \cite{huang} is the same as for our $\alpha$. The only difference is that in \cite{huang} the coupling constant $\omega^2$ is assumed to take positive values exclusively, while in the remaining works (including ours) both signs are considered.

We have organized the paper in the following way. In section \ref{sec-basic} we state the main assumptions on which the present work relies and we write down the basic expressions that will be useful in the subsequent study. Appropriate (dimensionless) variables of the phase space are introduced in order to study in a unified way both the positive and the negative coupling cases. A quite general discussion on the phantom barrier crossing in the model \eqref{action} is given in section \ref{sec-cross}. In section \ref{sec-cs2} we discuss on the behavior of the sound speed squared $c^2_s$ -- the one that accounts for the speed of propagation of the perturbations of the energy density -- in the present model. Especial emphasis is made in the possible violations of the physical bounds $0\leq c^2_s\leq 1$. Section \ref{sec-ds} is dedicated to briefly expose the main properties of the dynamical system corresponding to the present cosmological model in connection with the bounds on the squared sound speed. While in sections \ref{sec-cross} and \ref{sec-cs2} we focus mainly in the quintessence case ($\epsilon=1$), in section \ref{sec-e0} the pure derivative coupling case ($\epsilon=0$) is separately investigated. A thorough discussion of the results obtained in this paper is presented in section \ref{sec-disc}. In particular, the case with the constant potential that can be developed in a fully analytical way, is discussed as a simple illustration of the resultsanal. Finally, brief conclusions are given in section \ref{sec-concl}. For completeness we have included an appendix section \ref{app}. In the appendix an elementary discussion on the so called Laplacian instability is included. 

Throughout the paper we use the units system with $8\pi G_N=c^2=1$, where $G_N$ is the Newton's constant and $c$ is the speed of light in vacuum.

%%%%%%%%%%%%%%%%%%%%%%%%%%%%%%%%%%%%%%%%%%%%%%%%%%%%%

\section{Basic equations and set up}\label{sec-basic}

The main hypothesis of this paper is that the physical bounds on the speed of sound (squared) are viable criteria to reject physical theories like the one being investigated here. Other assumptions considered in this paper are the following: 

\begin{itemize}

\item For simplicity of the discussion we shall focus in the vacuum case, i. e., in \eqref{action} we set $S_m=0$.

\item For definiteness only expanding cosmologies ($H\geq 0$) will be considered and, besides, the scalar field $\phi$ is assumed to be a monotone non-decreasing function of the cosmic time: $\dot\phi\geq 0.$ 

\item We consider non-negative self-interacting potential $V\geq 0$ (non-negative energy density). 

\item Only the cases with $\epsilon=1$ (quintessence) and $\epsilon=0$ (pure derivative coupling) will be of interest.

\end{itemize} 

As a model for the background spacetime here we assume the Friedmann-Robertson-Walker (FRW) metric with flat spatial sections, whose line element is given by: 

\bea ds^2=-dt^2+a^2(t)\delta_{ik}dx^idx^k.\label{frw}\eea The cosmological field equations that can be derived from the action \eqref{action} read:

\bea &&\;\;3H^2=\rho_\text{eff},\;-2\dot H=\rho_\text{eff}+p_\text{eff},\nonumber\\
&&\ddot\phi+3H\dot\phi=\frac{-6\alpha H\dot H\dot\phi-V_\phi}{\epsilon+3\alpha H^2},\label{feqs}\eea where $V_\phi\equiv dV/d\phi$. The effective energy density and pressure of the scalar field are given by

\bea &&\rho_\text{eff}=\frac{\epsilon+9\alpha H^2}{2}\dot\phi^2+V(\phi),\label{rho}\\
&&p_\text{eff}=\frac{\epsilon-3\alpha H^2}{2}\dot\phi^2-V(\phi)-\alpha\dot\phi^2\dot H-2\alpha H\dot\phi\ddot\phi,\label{p}\eea respectively. From the above equations one obtains that:

\bea &&\rho_\text{eff}+p_\text{eff}=\left(\epsilon+9\alpha H^2\right)\dot\phi^2-\alpha\dot\phi^2\dot H-2\alpha H\dot\phi(\ddot\phi+3H\dot\phi).\label{rho+p}\eea  

An interesting property of the effective energy density $\rho_\text{eff}$ in \eqref{rho} and of the effective pressure $p_\text{eff}$ in \eqref{p}, is that these quantities depend not only on the scalar field matter degree of freedom $\phi$ and its derivatives $\dot\phi$ and $\ddot\phi$, but also on the curvature through $H^2$ and $\dot H$. In particular, the effective kinetic energy density of the scalar field in the right-hand-side (RHS) of the Friedmann equation above: $(\epsilon+9\alpha H^2)\dot\phi^2/2$, is contributed not only by $\dot\phi$ but also by the curvature through the squared Hubble rate.\footnote{Notice that when in the above equations the non-minimal derivative coupling vanishes: $\alpha=0$, we recover the standard result of general relativity with minimally coupled scalar field matter.} 

With the help of the first equation in \eqref{feqs} and of \eqref{rho} one can rewrite the Friedmann equation and, correspondingly, the effective energy density, in the following way:

\bea 3H^2=\gamma^2\rho_\phi=\rho_\text{eff},\;\gamma=\frac{1}{\sqrt{1-3\alpha\dot\phi^2/2}},\label{3h2}\eea where $\gamma=\gamma(\dot\phi)$ is the 'boost' function and $$\rho_\phi=\epsilon\frac{\dot\phi^2}{2}+V(\phi),$$ is the standard energy density of the scalar field. Written in the latter form $\rho_\text{eff}$ is a function only of the scalar field degree of freedom $\phi$, and of its derivative $\dot\phi$ since the curvature effects are hidden in the non-canonical form of the effective energy density, i. e., in the boost function.

We point out that for negative coupling ($\alpha<0$), the boost function is bounded from below and also from above: $0<\gamma\leq 1$, while for positive coupling ($\alpha>0$): $1\leq\gamma<\infty$, i. e., it is bounded from below only.

\subsection{Non-negative coupling and upper bound on $|\dot\phi|$}

If we consider non-negative $\alpha\geq 0$, from \eqref{3h2} -- given that we consider non-negative effective energy density exclusively -- it follows that $1-3\alpha\dot\phi^2/2\geq 0$, i. e.

\bea 0\leq\dot\phi^2\leq\frac{2}{3\alpha}\;\Leftrightarrow\;-\frac{1}{3\alpha}\leq X\leq 0,\label{bounds-dphi}\eea where $X=\der_\mu\phi\der^\mu\phi/2=-\dot\phi^2/2$.

We want to point out here the non-conventional nature of the ``effective'' kinetic energy of the scalar field \eqref{rho} under the derivative coupling when $\alpha>0$. Actually, as just seen, the standard kinetic energy $\propto\dot\phi^2$ is bounded from above, a strange feature not arising in standard scalar tensor theories without self-couplings. Notwithstanding, the effective kinetic energy in \eqref{rho}: $\propto(\epsilon+9\alpha H^2)\dot\phi^2,$ is not bounded due the curvature effects encoded in $H^2$.

In reference \cite{sushkov}, since in that presentation the coupling $\kappa$ is of opposite sign as compared with our $\alpha$: $\kappa=-\alpha$, the case where the standard kinetic term is bounded from above corresponds to the condition expressed by Eq. (27) in the mentioned reference (see also equations (19) and (21) of the same reference, recalling that in this paper we have chosen the units where $8\pi G_N=1$, while in \cite{sushkov}: $G_N=1$.)

\subsection{New variables}

In spite of the commonly used variable $X$, in order to study both positive and negative coupling cases in a unified way, in this paper we prefer to use the new variable:

\bea x:=\alpha\dot\phi^2/2,\label{x-var}\eea i. e. the new variable is properly the standard kinetic energy of the scalar field multiplied by the coupling constant. Hence, positive coupling entails that $x\geq 0$, while negative coupling means that $x\leq 0$. Vanishing $x=0$ means that, either the scalar field is a constant $\phi=\phi_0$, or there is not derivative coupling: $\alpha=0$. 

In the same way, in connection with the self-interaction potential term, it will be very useful to introduce the following variable:

\bea y:=\alpha V,\label{y-var}\eea where, for positive $\alpha$ this variable takes non-negative values: $0\leq y<\infty$, while for negative coupling ($\alpha<0$) the variable takes non-positive values instead $-\infty<y\leq 0$.
 
We want to underline that for positive coupling ($\alpha>0$), given that $H^2$: 

\bea 3\alpha H^2=\frac{\epsilon x+y}{1-3x},\label{3h2-xy-eq}\eea should be a non-negative quantity ($H^2\geq 0$), the non-negative variable $x$ should take values in the physically meaningful interval: $0\leq x\leq 1/3$. Meanwhile, for negative coupling ($\alpha<0$) the variable $x$ is non-positive: $-\infty<x\leq 0$.

The above variables will allow us to write the equations in a more compact manner and to make our computations independent of the specific value of the coupling constant.

%-----------------------------------

\begin{figure*}
\includegraphics[width=4.2cm]{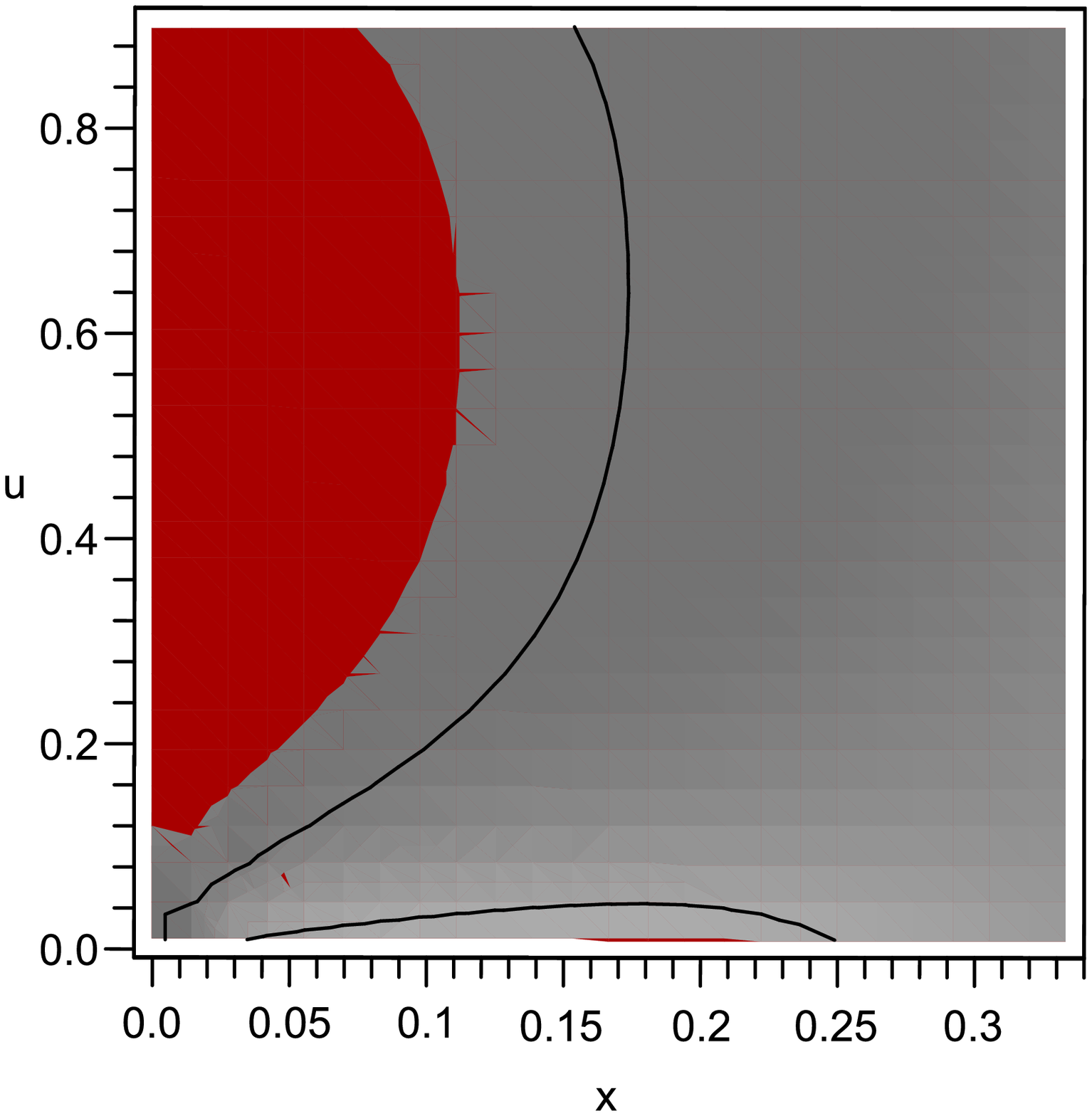}
\includegraphics[width=4.2cm]{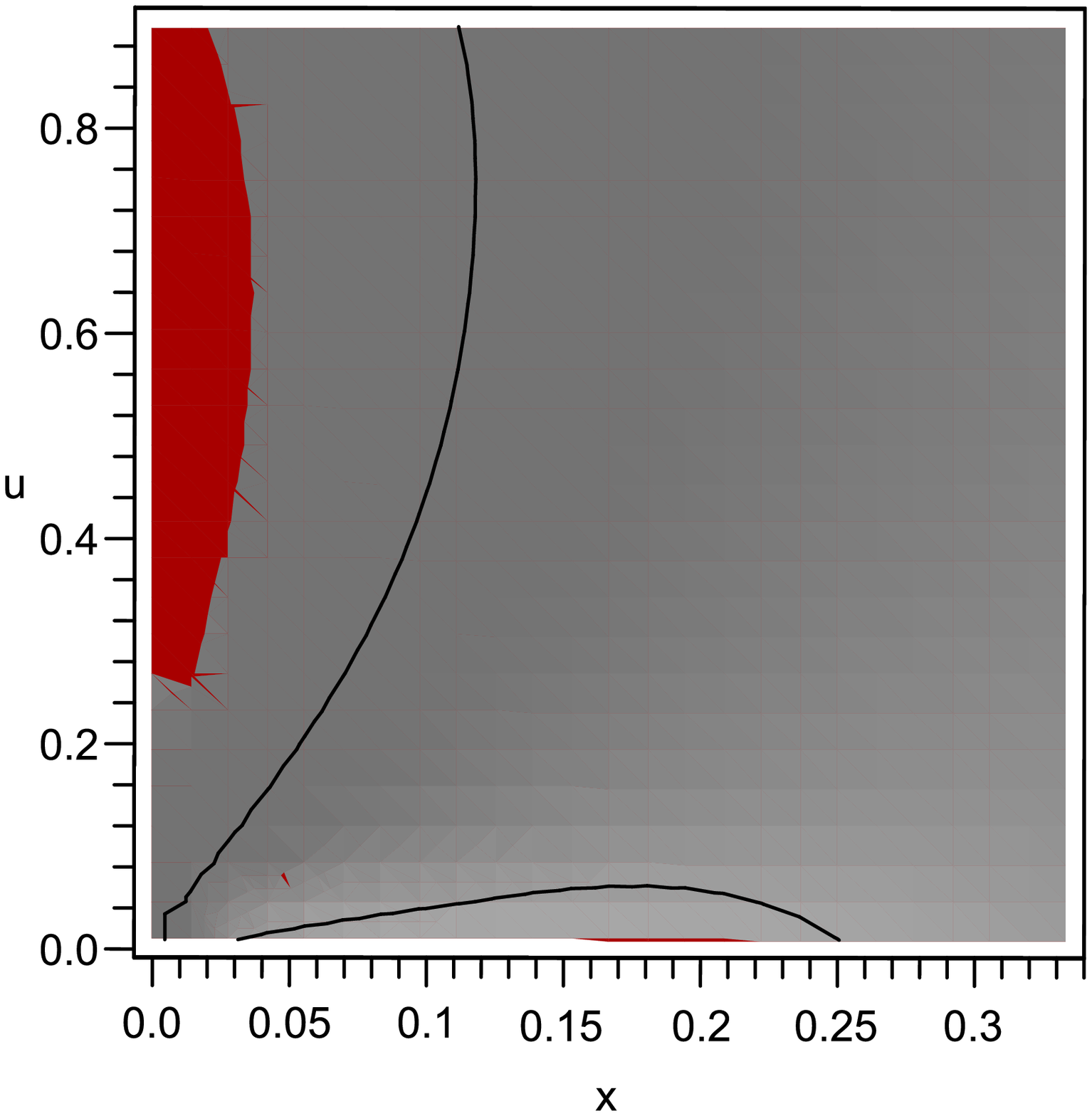}
\includegraphics[width=4.2cm]{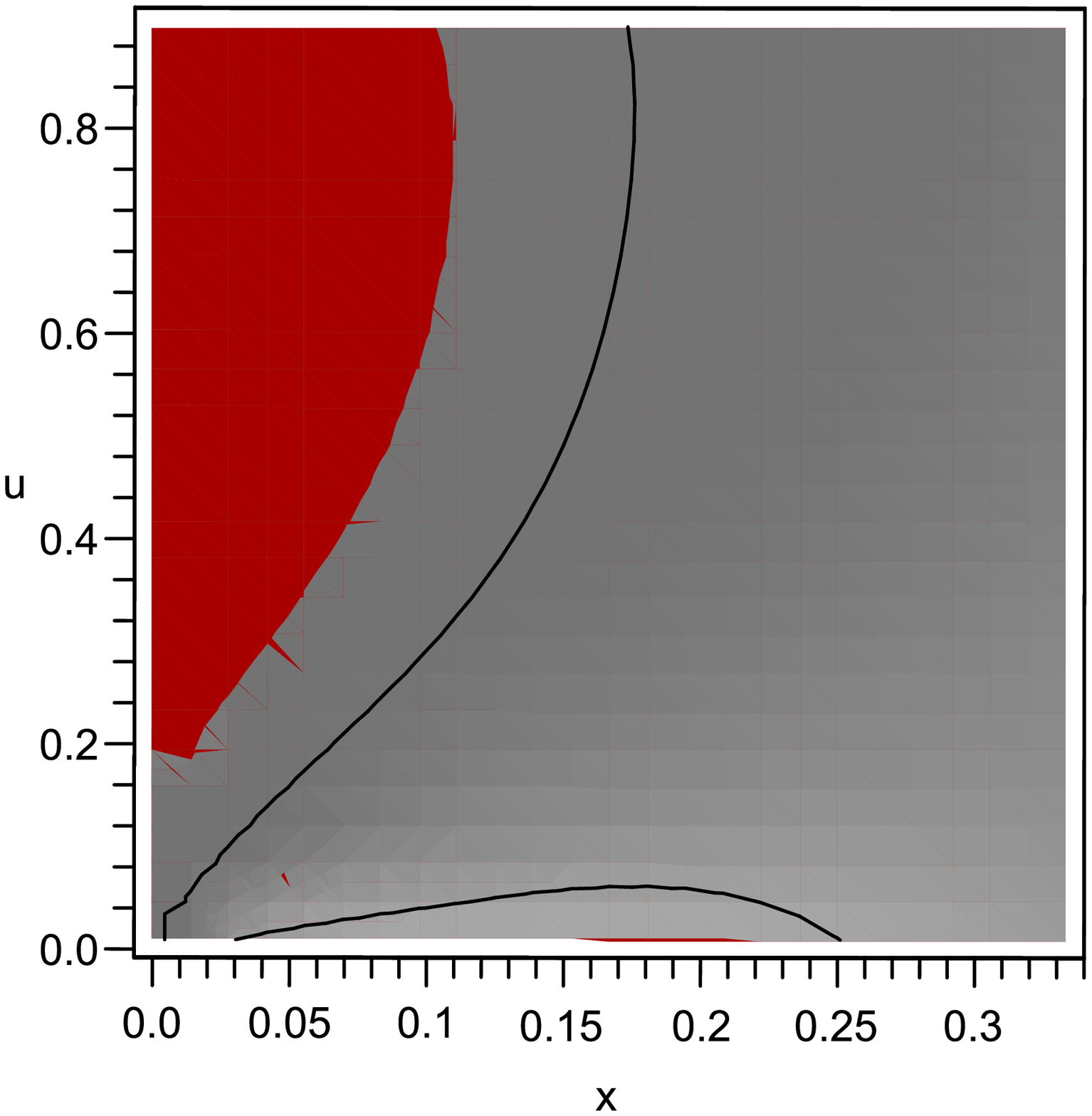}
\includegraphics[width=4.2cm]{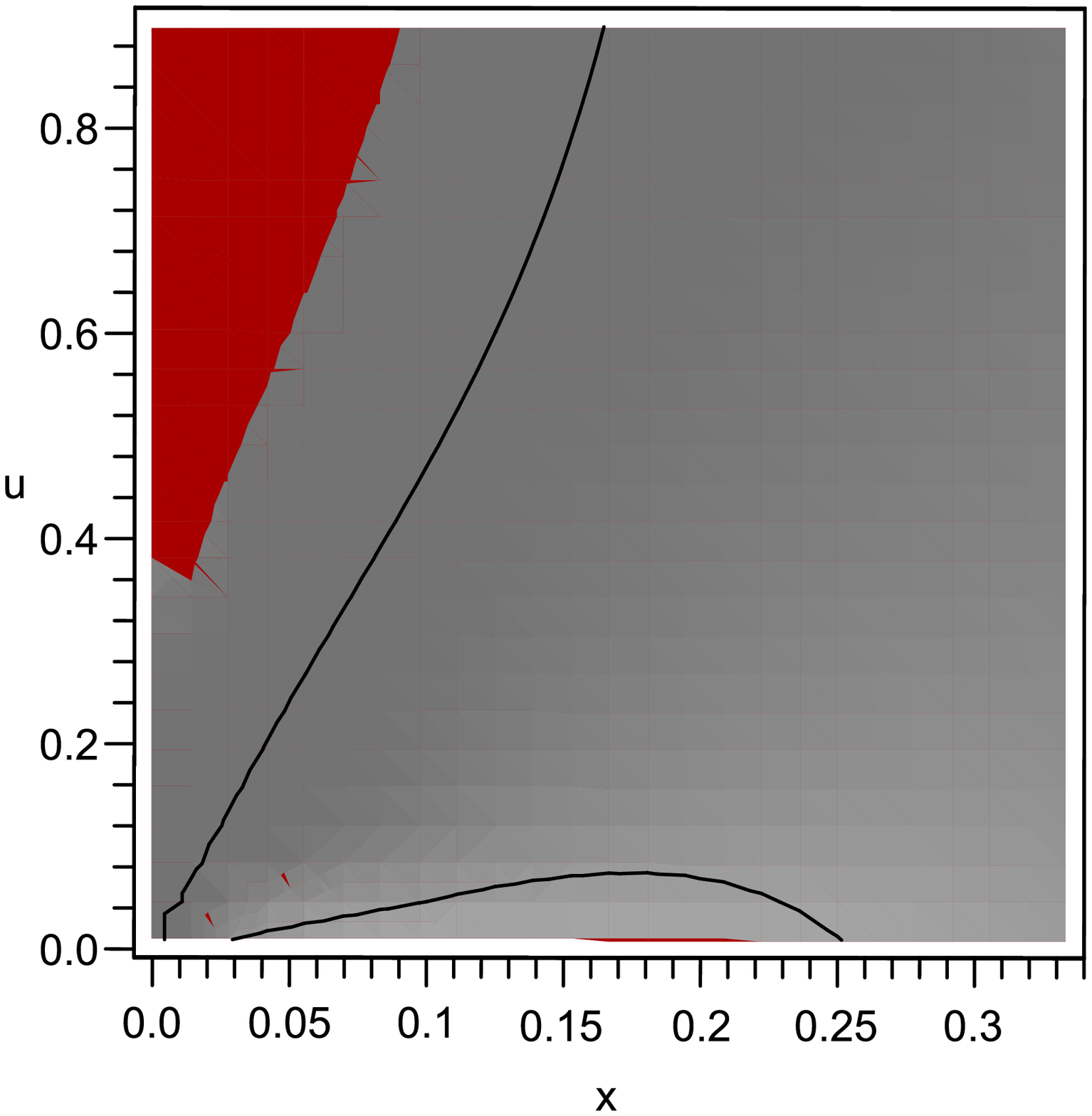}\vspace{0.7cm}
\caption{Geometric representation of the bound $\omega_\text{eff}+1\geq 0$ in the $xu$-plane for positive coupling ($\alpha>0$). For illustrative purposes we have chosen two negative-slope potentials: the decaying exponential potential $V=V_0\exp{(\lambda\phi)}$ with $\lambda<0$ (left hand panels) and the inverse power-law potential $V=V_0\phi^{2n}$ with $n<0$ (right hand panels), for different values of the parameters $\lambda$ and $n$ respectively. In the left hand panels, from left to the right: $\lambda=-5$ and $\lambda=-2$, while in the right-hand panels: $n=-2$ and $n=-1$, respectively. Here we use the bounded variable $u=y/y+1$ ($0\leq u\leq 1$) instead of $y=\alpha V$ (the variable $x=\alpha\dot\phi^2/2$ is already bounded: $0\leq x\leq 1/3$), so that the whole phase plane $xu$ fits into a finite size box. The red-colored regions correspond to the phantom domain where $\omega_\text{eff}+1<0$. For monotonically growing potentials ($\lambda>0|n>0$) the phantom domain is not found so that the crossing is not possible.}\label{fig01}\end{figure*}

%-----------------------------------

%%%%%%%%%%%%%%%%%%%%%%%%%%%%%%%%%%%%%%%%%%%%%%%%%%%%%%%%%%%%%%%%%%%%%%

\section{Phantom barrier crossing: General analysis}\label{sec-cross}

As mentioned in the introduction, one issue of interest when one explores cosmological models of dark energy is the possibility of crossing the so called ``phantom divide'' barrier $\omega_\Lambda=-1$ \cite{caldwell, crossing-odintsov, crossing-observ, crossing-nesseris}. Hence, it will be useful to look for the possibility of the crossing in the theory with non-minimal derivative (kinetic) coupling to the Einstein's tensor \cite{mohseni}. 

If under the assumptions exposed in the section \ref{sec-basic} we combine the second and third equations in \eqref{feqs}, we obtain:

\bea -2\alpha\dot H=R_1+R_2,\label{doth}\eea with (recall that $y_\phi=\alpha V_\phi=\alpha dV/d\phi$):

\bea R_1=\frac{2x\left[\epsilon(1-2x)+y\right](\epsilon+3y)}{(1-3x)F_\epsilon},\;R_2=\frac{2\sqrt{2x(1-3x)(\epsilon x+y)}\,y_\phi}{\sqrt{3}F_\epsilon},\label{r1-r2-def}\eea where, for compactness of writing, we have introduced the following definition:

\bea F_\epsilon\equiv F_\epsilon(x,y):=\epsilon(1-3x+6x^2)+(1+3x)y.\label{f-eps}\eea

The effective EOS parameter of the scalar field is given by:

\bea \omega_\text{eff}=\frac{p_\text{eff}}{\rho_\text{eff}}=-1-\frac{2\dot H}{3H^2}=-1+\frac{R_1+R_2}{3\alpha H^2},\label{weff-def}\eea where $R_1$ and $R_2$ are given by \eqref{r1-r2-def} and, in terms of the variables $x$, $y$, the denominator $3\alpha H^2$ is given by \eqref{3h2-xy-eq}. Hence, for the effective EOS in the general case -- unspecified $\epsilon$ -- we get:

\begin{widetext}\bea \omega_\text{eff}=-1+\frac{2x(\epsilon+3y)\left[\epsilon(1-2x)+y\right]}{(\epsilon x+y)F_\epsilon}+\frac{2}{F_\epsilon}\sqrt\frac{2x(1-3x)^3}{3(\epsilon x+y)}\;y_\phi.\label{eos-master-eq}\eea\end{widetext} 

As it can be seen from \eqref{weff-def}, the crossing of the phantom barrier is achieved only if $-2\dot H$ may change sign during the cosmic evolution. In general $-2\dot H$ is a non-negative quantity. This is specially true for the standard quintessence where in equations \eqref{feqs}, \eqref{rho} and \eqref{p} we set $\alpha=0$ and $\epsilon=1$. In this case $-2\dot H=\dot\phi^2\geq 0$, while the EOS parameter in \eqref{weff-def} can be written as

\bea \omega_\text{eff}=-1+\frac{\dot\phi^2}{3H^2},\label{quint-cross}\eea so that, given that $\dot\phi^2/3H^2$ is always non-negative, then $\omega_\text{eff}\geq -1$. In this case the phantom barrier crossing is not possible unless additional complications are considered such as, for instance: i) non-gravitational interaction of the dark energy and dark matter components \cite{n-m-int}, ii) multiple dark energy fields like in quintom models \cite{quintom-mod, quintom-rev} or iii) extra-dimensional effects \cite{roy-isra}. Here we shall investigate the issue within the frame of the theory \eqref{action} where the derivatives of the scalar field are non-minimally coupled to the Einstein's tensor.

\subsection{Positive coupling ($\alpha>0$)}

For non-negative $x$-s, i. e., for positive coupling ($\alpha>0$), the denominators of $R_1$ and of $R_2$ in \eqref{r1-r2-def} are always positive-valued. So is the numerator of the term $R_1$ which means that this term is always non-negative. Meanwhile, the sign of the numerator of the term $R_2$ is determined by the slope of the self-interaction potential: $y_\phi=\alpha V_\phi=\alpha dV/d\phi$. Consequently, for non-negative $0\leq x\leq 1/3$, the term $R_2$ in \eqref{doth} is the only one that may allow for the crossing of the phantom barrier. 

In this case ($0\leq x\leq 1/3$) two clear conclusions can be done: i) the crossing is due to the derivative coupling with strength $\alpha$, and ii) the crossing is allowed only if $\dot\phi V'=\dot V<0$, i. e., if the self-interaction potential decays with the cosmic expansion. Assuming that this is indeed the case, the competition between the positive term $R_1$ and the negative one $R_2$ during the course of the cosmic evolution is what makes possible the flip of sign of $-2\dot H=R_1+R_2$, and hence the crossing of the phantom barrier. Notice that for the constant potential $V_\phi=0$, as well as for the monotonically growing potentials the crossing is not possible. This is true, in particular, for the growing exponential potential: $V\propto\exp(\lambda\phi)$ with $\lambda>0$ for $\dot\phi>0$ or $\lambda<0$ for $\dot\phi<0$, and for the power-law $V\propto\phi^n$ with $n\geq 0$. 

The above results are illustrated in FIG. \ref{fig01} where a geometric representation of the quantity $\omega_\text{eff}+1$ in the $xu$-plane is shown. Here we used the new (bounded) variable: 

\bea u=\frac{y}{y+1},\;0\leq u\leq 1.\label{u-var}\eea This choice makes possible to fit the whole (semi-infinite) phase plane $xy$ into a finite size box: $\{(x,u):0\leq x\leq 1/3,0\leq u\leq 1\}$. The red-colored regions are the ones where $\omega_\text{eff}+1<0$, i. e., where the scalar field behaves like phantom matter. It is appreciated that, for negative-slope potentials (the decaying exponential and the inverse power-law in the figure), both the phantom region with $\omega_\text{eff}+1<0$ and the region where $\omega_\text{eff}+1>0$ (gray color) coexist, so that the crossing of the phantom divide is possible.

%-----------------------------------

\begin{figure*}
\includegraphics[width=4.2cm]{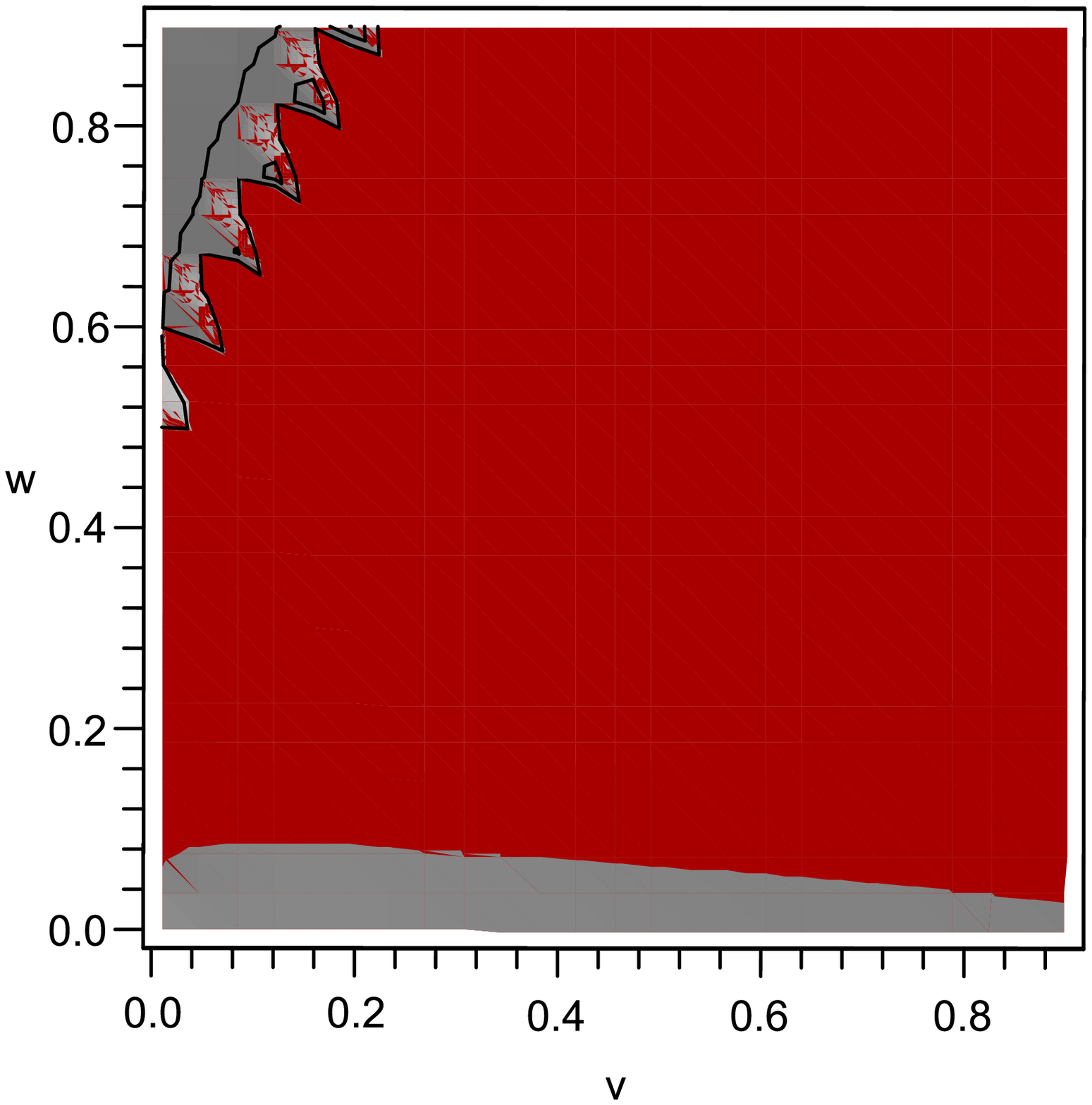}
\includegraphics[width=4.2cm]{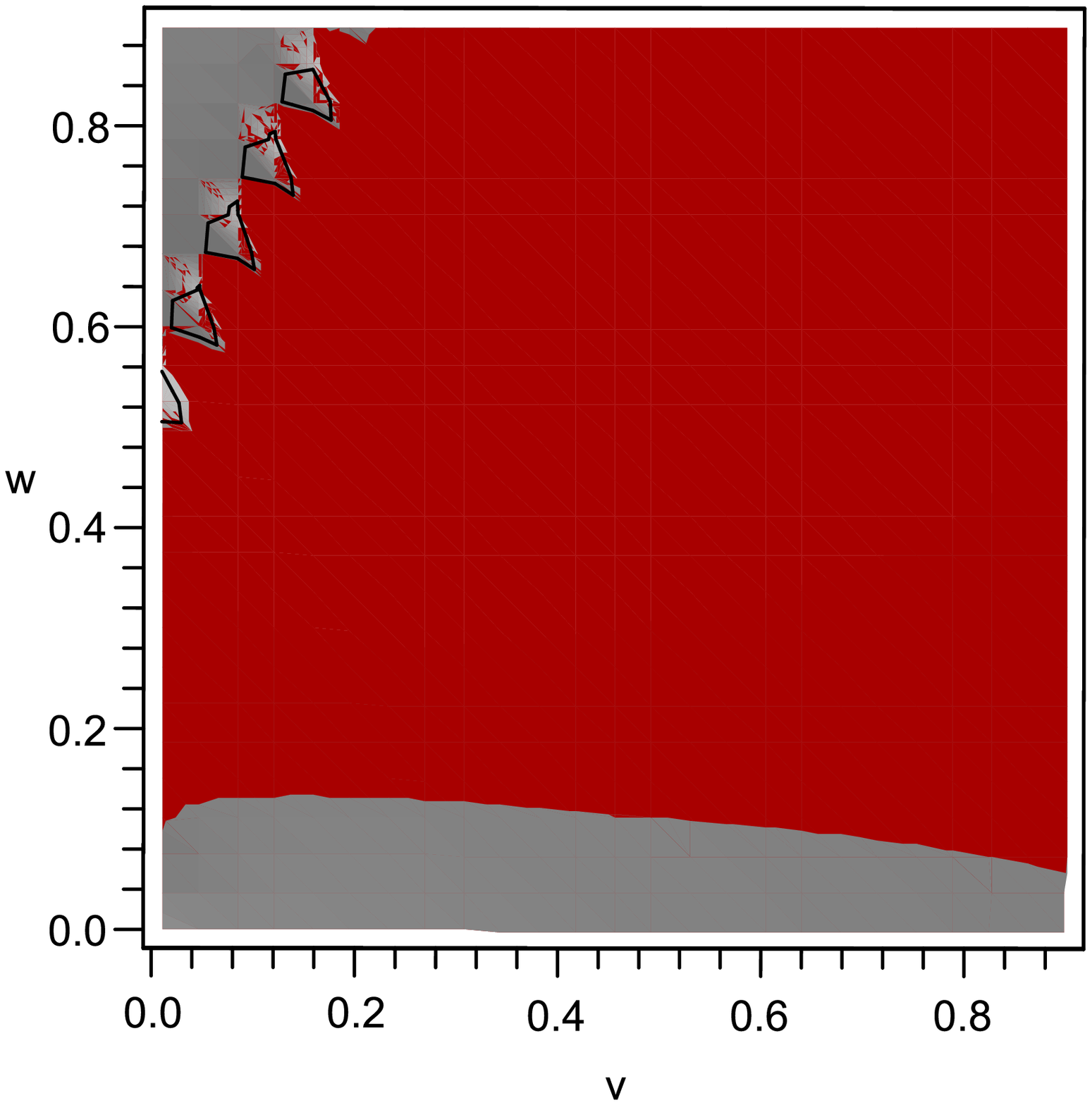}
\includegraphics[width=4.2cm]{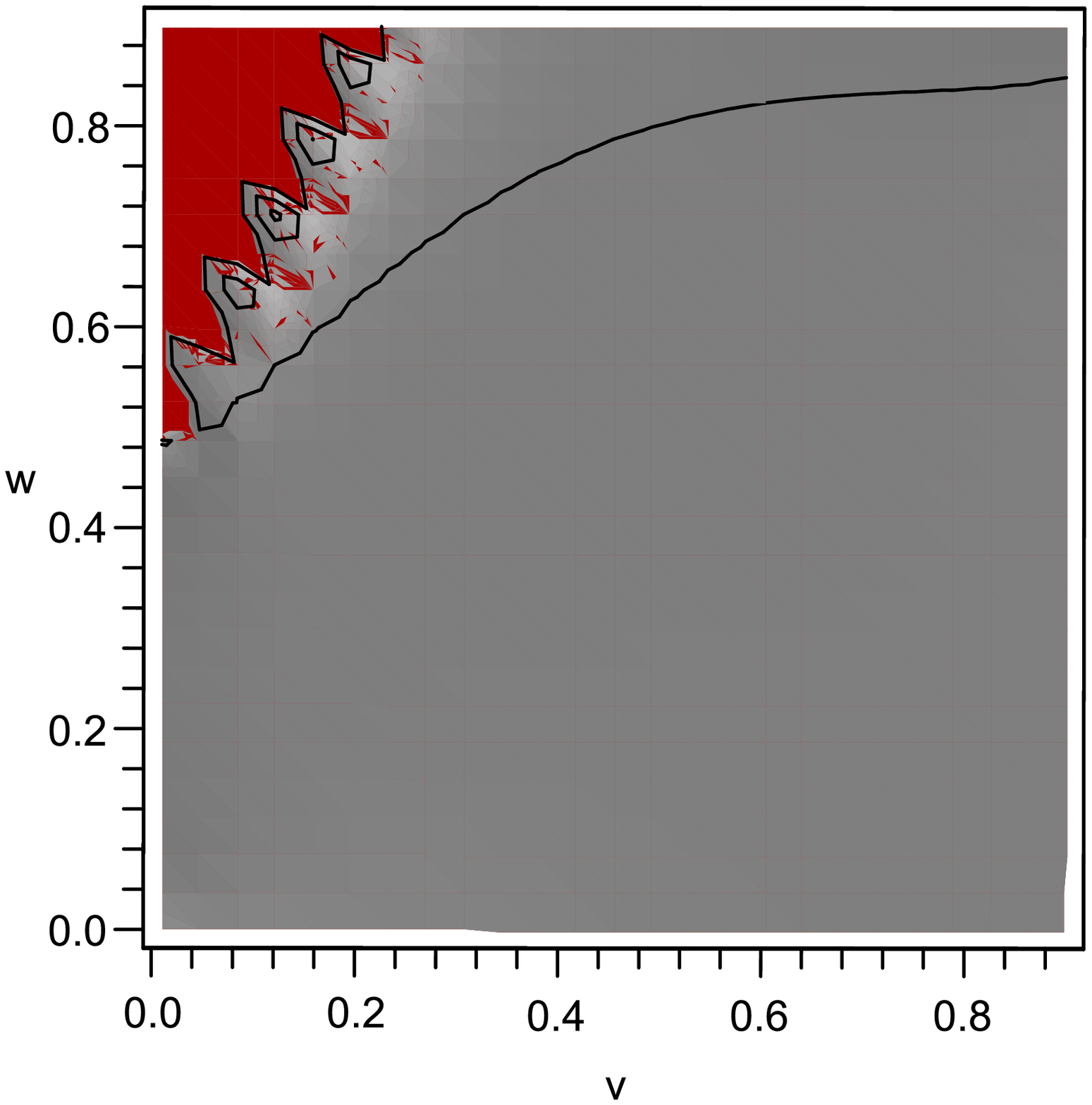}
\includegraphics[width=4.2cm]{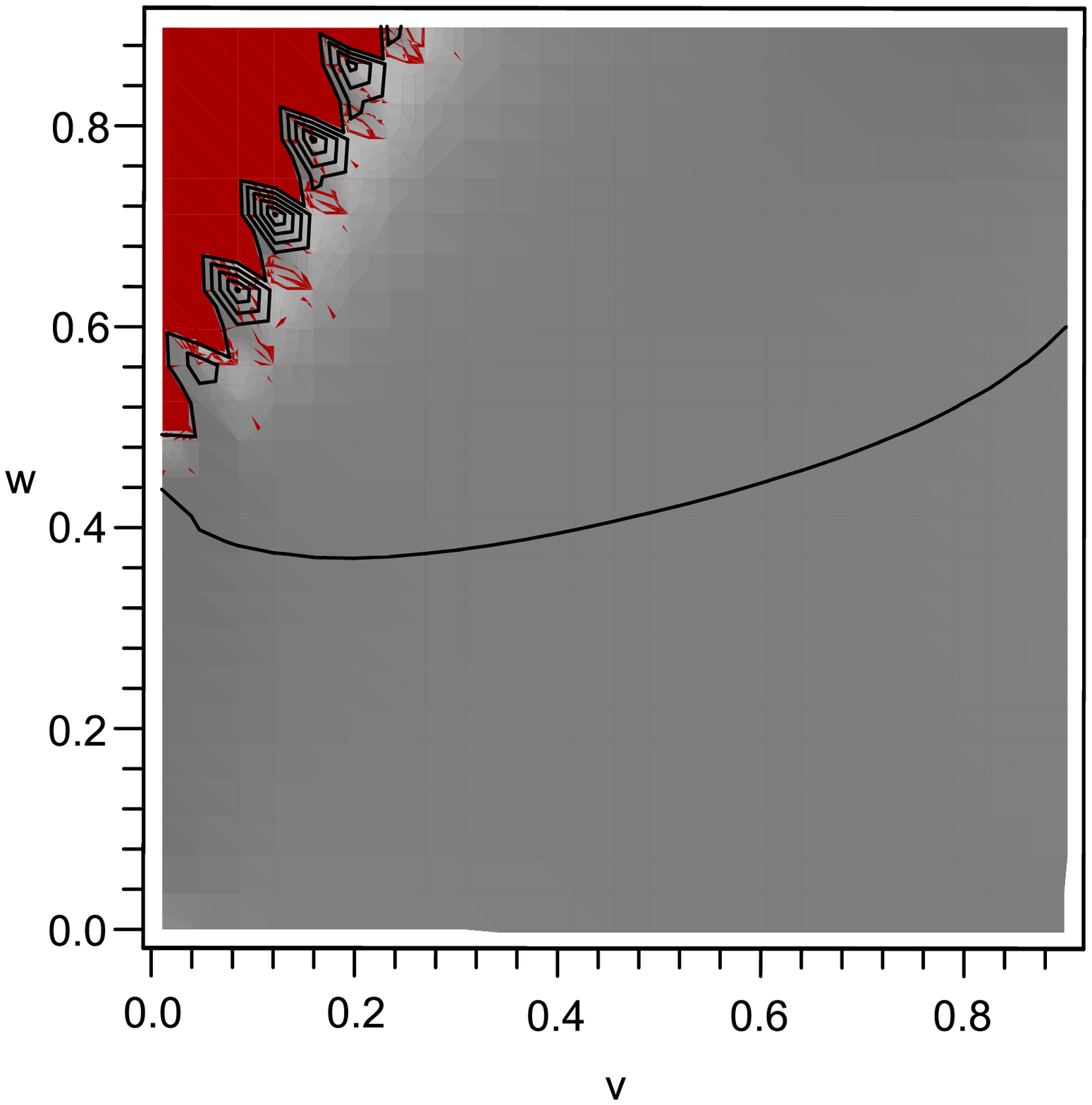}
\includegraphics[width=4.2cm]{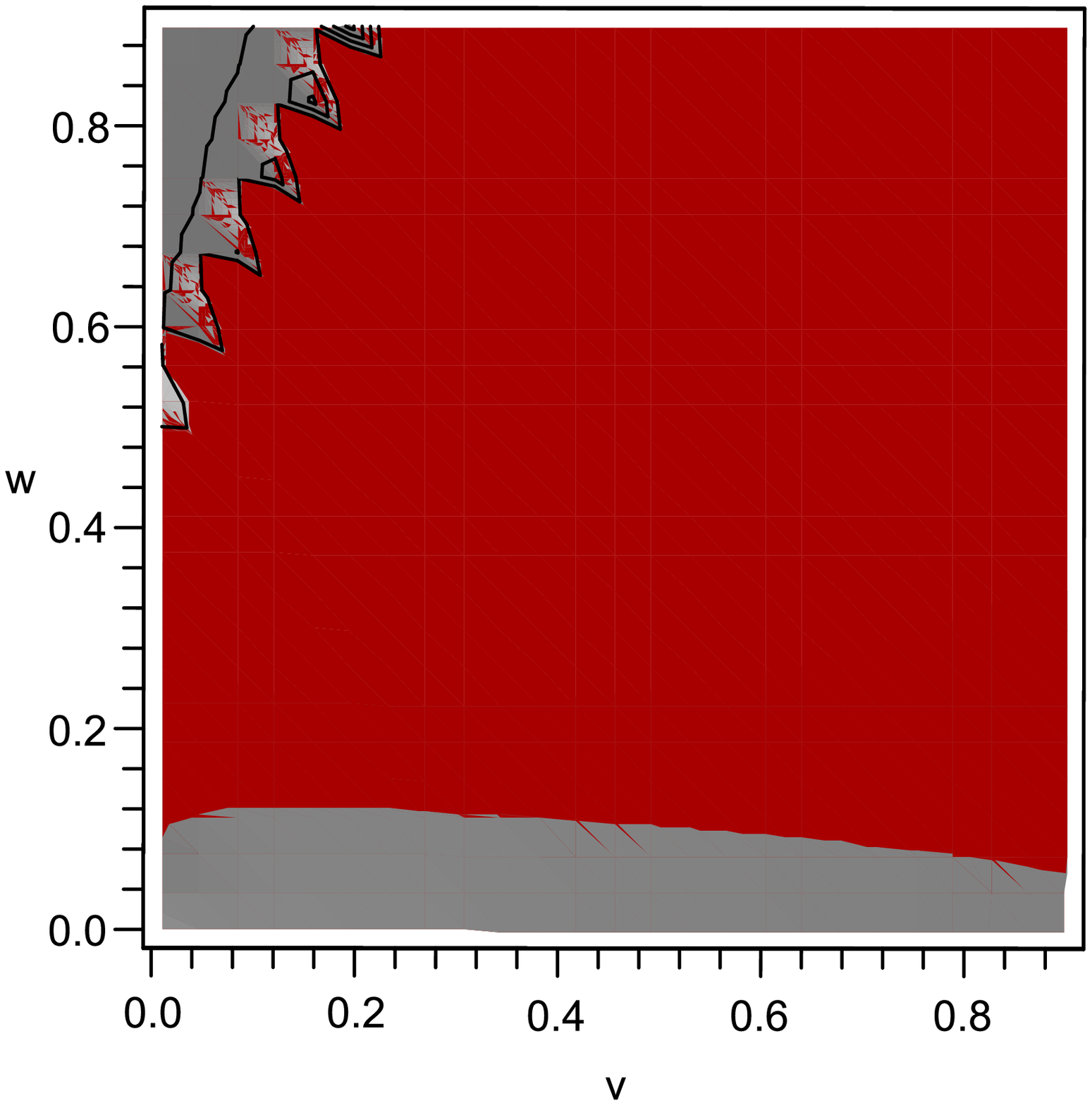}
\includegraphics[width=4.2cm]{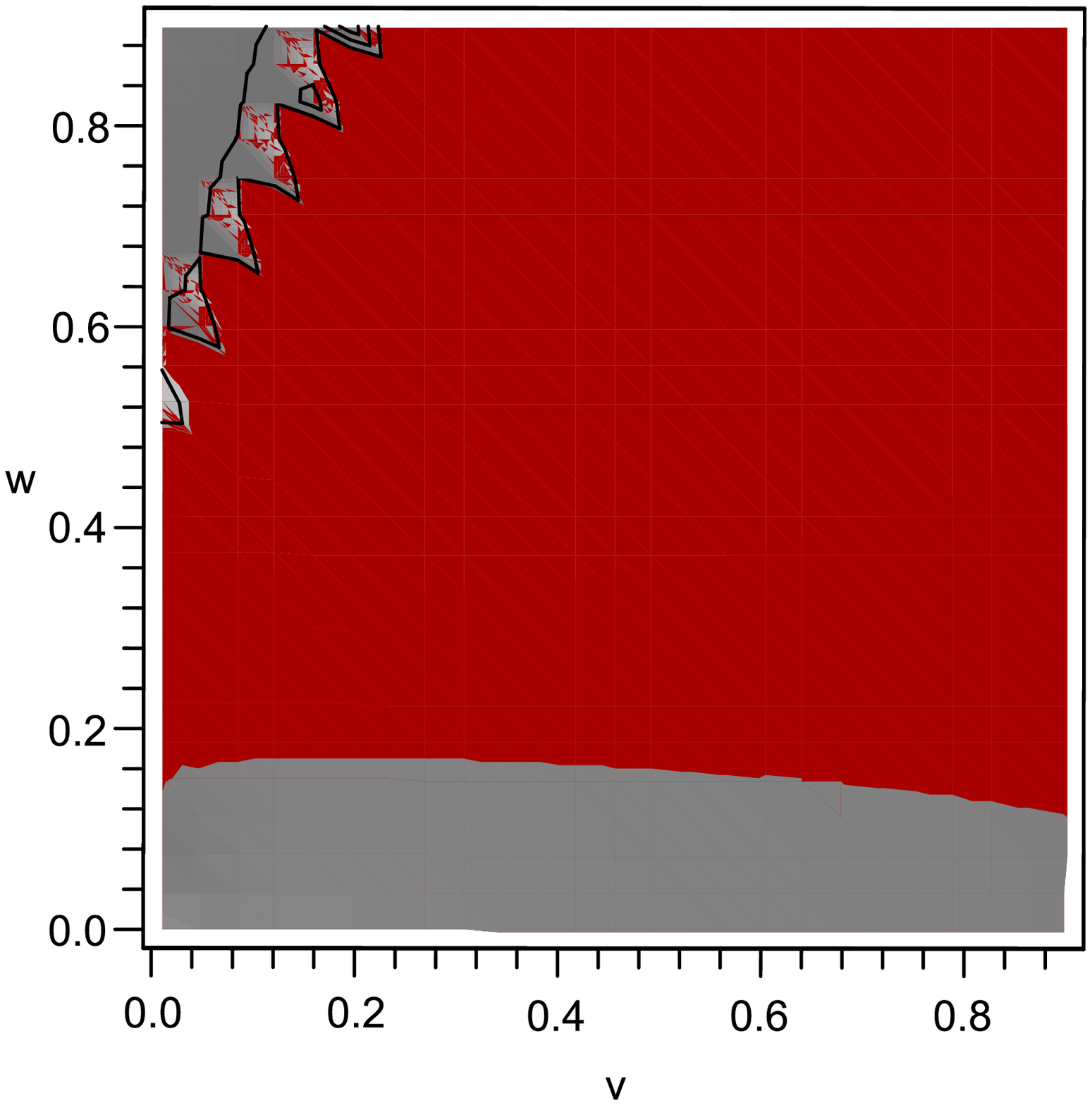}
\includegraphics[width=4.2cm]{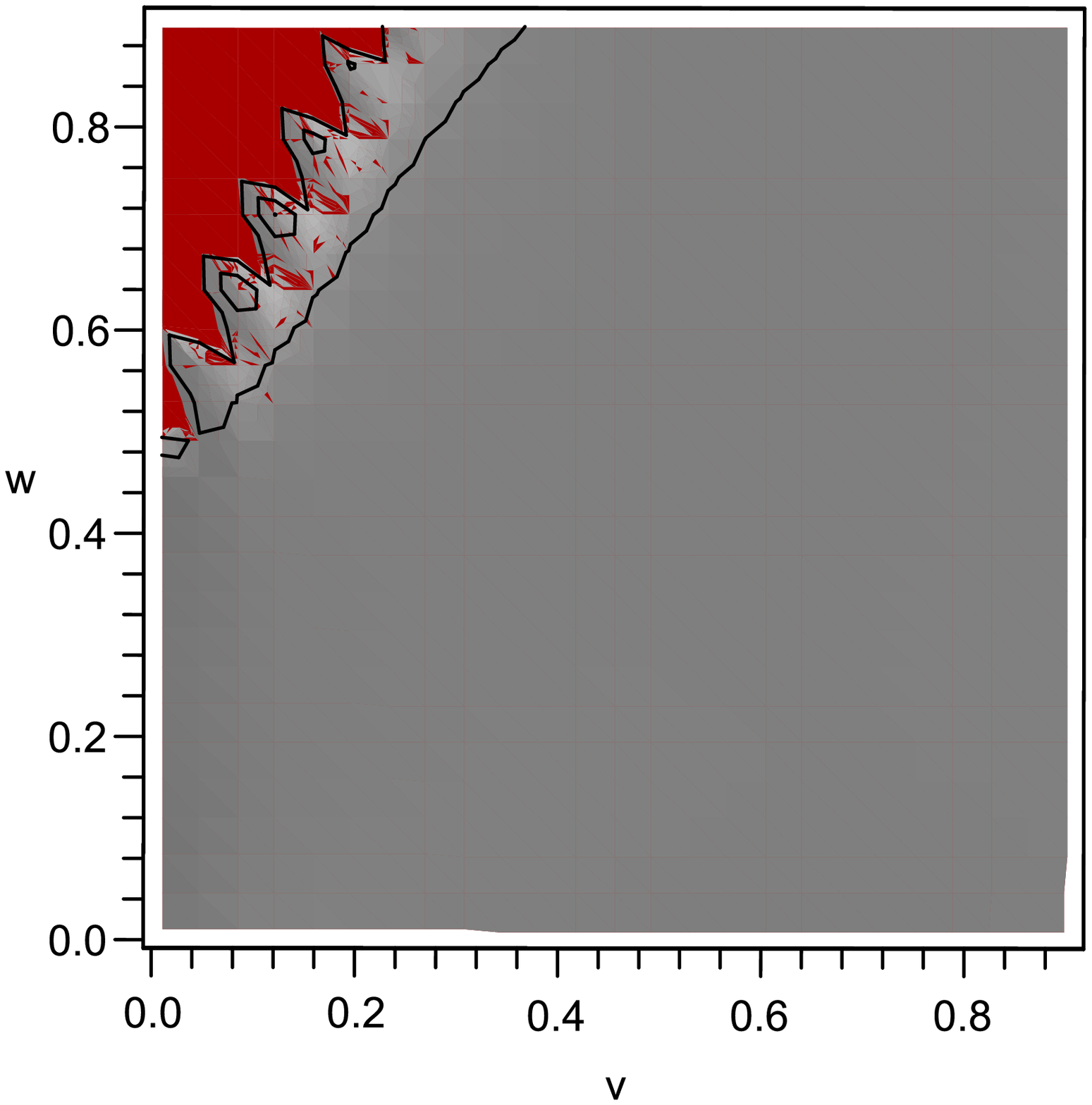}
\includegraphics[width=4.2cm]{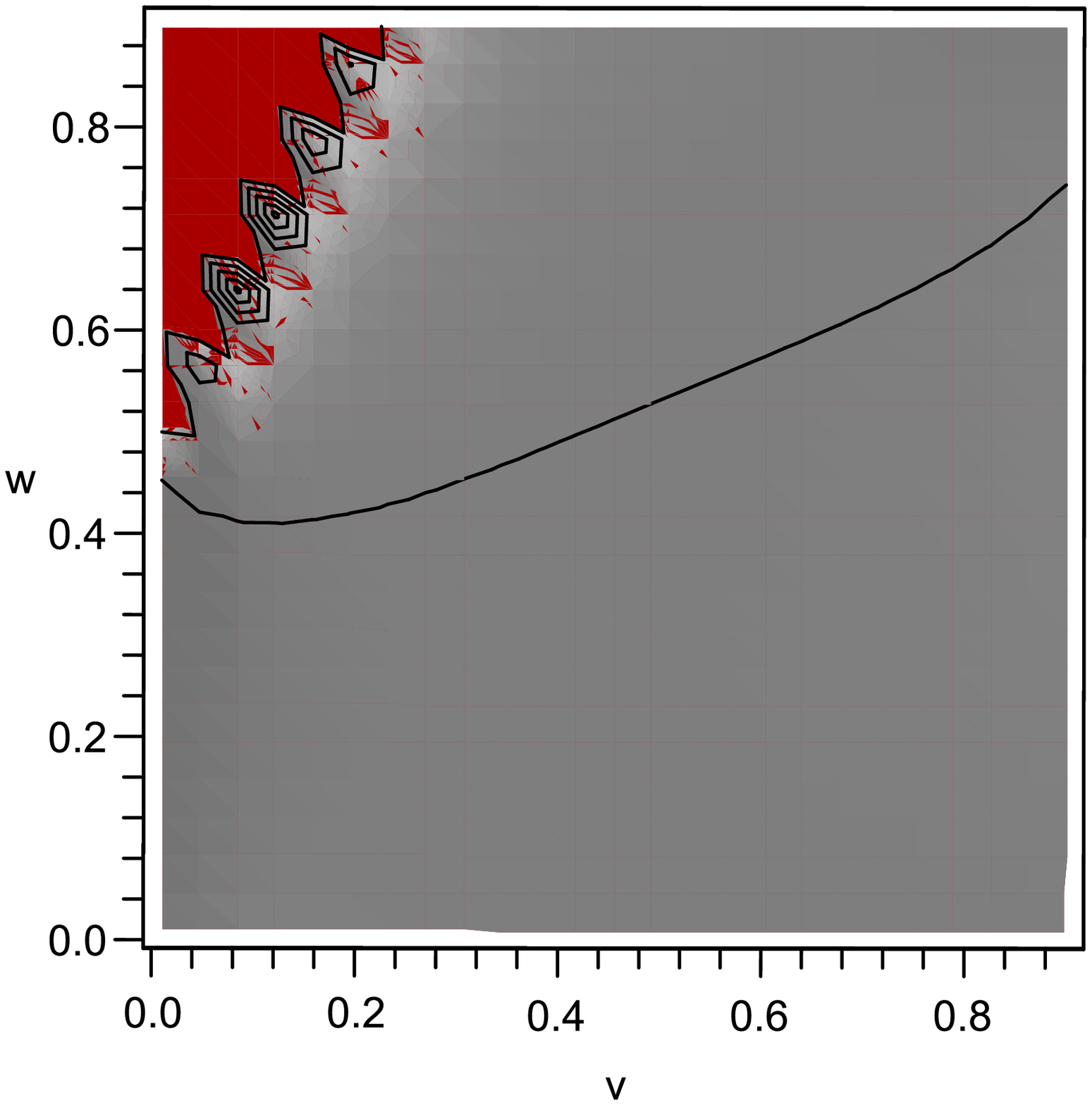}\vspace{0.7cm}
\caption{Region in the $vw$-plane where the scalar field behaves like phantom matter, i. e., where $\omega_\text{eff}<-1$ (red-colored regions) for the negative coupling case ($\alpha<0$). Here, in order to fit the whole phase plane into a finite size box -- recall that for the negative coupling both variables $x<0$ and $y<0$ are unbounded from below -- we have chosen the bounded variables $v=x/x-1$ ($0\leq v\leq 1$) and $w=y/y-1$ ($0\leq w\leq 1$). Under this choice, the phase plane $vw$ is the unit square. In the figure the exponential potential -- top panels -- and the power-law potential -- bottom panels -- are chosen for different values of the parameters $\lambda$ and $n$ respectively. In the top panels, from left to the right: $\lambda=-5$, $\lambda=-2$, $\lambda=2$ and $\lambda=5$, while in the bottom panels: $n=-2$, $n=-1$, $n=1$ and $n=2$, respectively. The zigzagging (slanted) curves in the top-left corners in the figures represent asymptotic separatrices in the $vw$-plane. This means that the domains lying at each side of the slanted zigzagging curves are disconnected (no continuous curve can join them). It is seen that in the right-hand figures (growing potentials) the regions where $\omega_\text{eff}+1\geq 0$ (gray color) and where the scalar field behaves like phantom matter: $\omega_\text{eff}+1<0$ (red color), are disconnected in the $vw$-plane, so that the crossing is not possible. Meanwhile, in the left-hand figures there exists a third region where $\omega_\text{eff}\geq-1$ (gray-colored region in the bottom of the figures) that can be joined to the phantom domain by continuous curves in the $vw$-plane, so that the crossing may eventually happen.}\label{fig02}\end{figure*}

%-----------------------------------

\subsection{Negative coupling ($\alpha<0$)}

For negative coupling, i. e., for $-\infty<x\leq 0$, $-\infty<y\leq 0$, the situation is a bit more complex. In this case it is more appropriate to go to a bounded set of variables:

\bea v=\frac{x}{x-1},\;w=\frac{y}{y-1},\label{vw-var}\eea where $0\leq v\leq 1$, $0\leq w\leq 1$. In terms of the latter variables the whole plane $xy$ fits into the unit square $\{(v,w):0\leq v\leq 1,0\leq w\leq 1\}$. We have that\footnote{For definiteness here we set $\epsilon=1$. The pure derivative coupling case $\epsilon=0$ will be discussed separately in section \ref{sec-e0}.}

\bea \omega_\text{eff}=-1+\frac{2v(4w-1)(1+v-2w)}{(1-v)(1-w)(v+w-2vw)F_1}+\frac{2}{F_1}\sqrt\frac{2v(1+2v)^3(1-w)}{3(1-v)^3(v+w-2vw)}\,y_\phi,\label{weff-vw}\eea where $y_\phi=-w_\phi/(1-w)^2$, and

\bea F_1\equiv F_{\epsilon=1}=\frac{1+v+4v^2-2(1-2v+4v^2)w}{(1-v)^2(1-w)}.\label{fe-vw}\eea This latter quantity, as well as the second and the third terms in the RHS of \eqref{weff-vw}, in general have not definite sign. As a consequence, there can be curves $w_c=w_c(v)$ where $F_1$ vanishes, meaning that the surface $\omega_\text{eff}=\omega_\text{eff}(v,w)$ tends to asymptotic large values, so that the surface literally ''breaks off'' (slanted zigzagging curves in the top left hand corners in the figures in FIG. \ref{fig02}). The curves $w_c=w_c(v)$ that annihilate the function $F_1$: $F_1(w_c(v),v)=0$, are in fact asymptotic separatrices in the $vw$-plane (the unit square). This means that any other curve in the unit square can only asymptotically approach to -- or leave off -- the curve $w_c$.

The competition between the second and the third terms in the RHS of \eqref{weff-vw} does not depend only on the slope of the potential $y_\phi$ (or $-w_\phi$), but also on whether the given $vw$-region is located in respect to the asymptote at $w_c=w_c(v)$ where $F_1=0$. From the plots in FIG. \ref{fig02} it is seen that, the only continuous regions in the $vw$-plane where the crossing of the phantom divide is possible, are those located below and to the right of the separatrices (the slanted zigzagging curves in the top-left corner) for the monotonically decreasing potentials: the decaying exponential and the inverse power-law in the left-hand panels. In these regions there can be curves in the $vw$-plane that continuously joint the domains where $\omega_\text{eff}>-1$ (gray color region at the bottom of the figures) with those where $\omega_\text{eff}<-1$ (red color). Hence, these curves can continuously cross the phantom divide: $\omega_\text{eff}+1=0$.

One may conclude that, independent of the sign of the coupling constant $\alpha$, the crossing of the phantom divide can happen only for negative-slope potentials: $y_\phi<0$ ($V_\phi<0$).

%-----------------------------------

\begin{figure*}
\includegraphics[width=4.2cm]{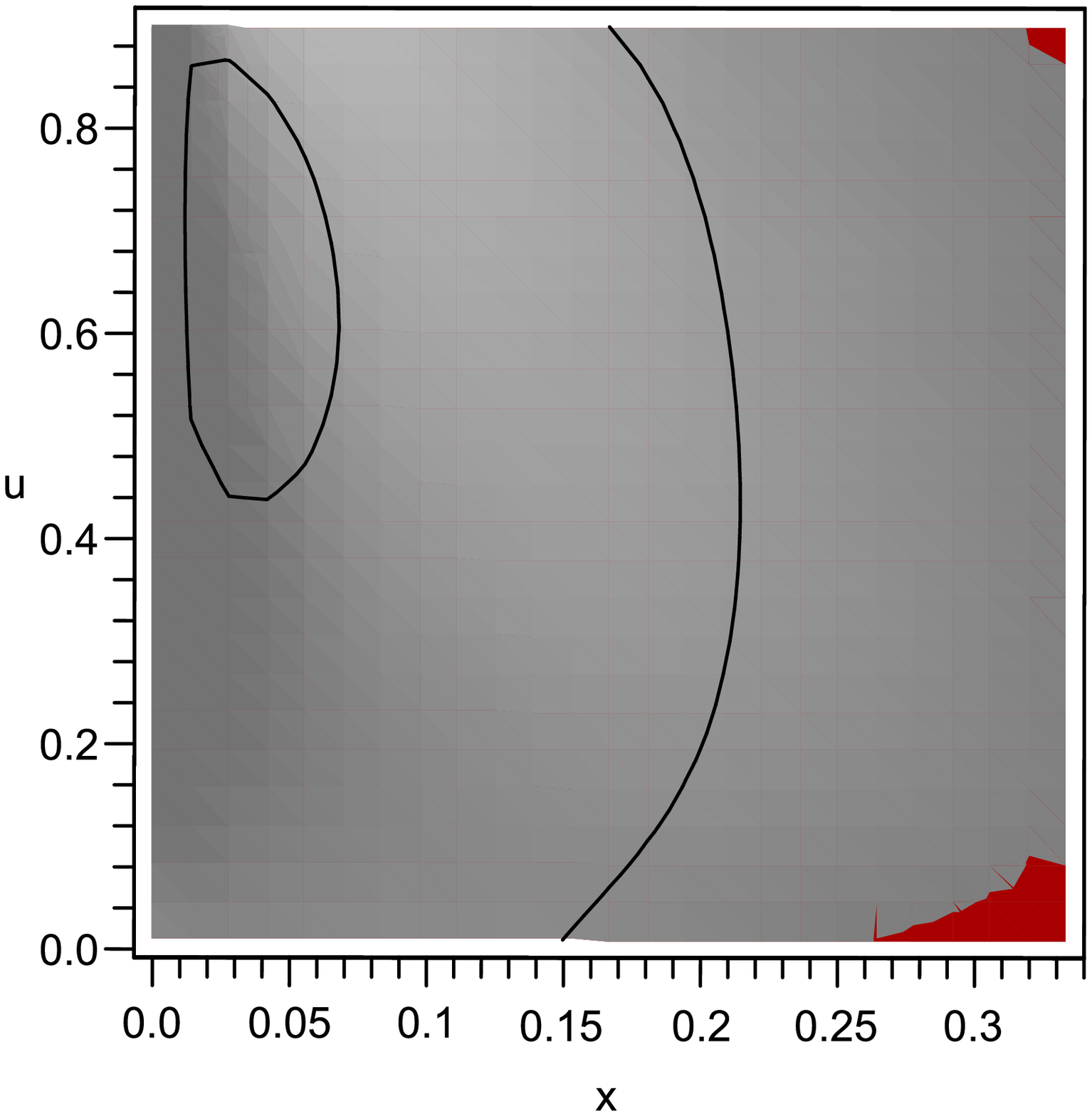}
\includegraphics[width=4.2cm]{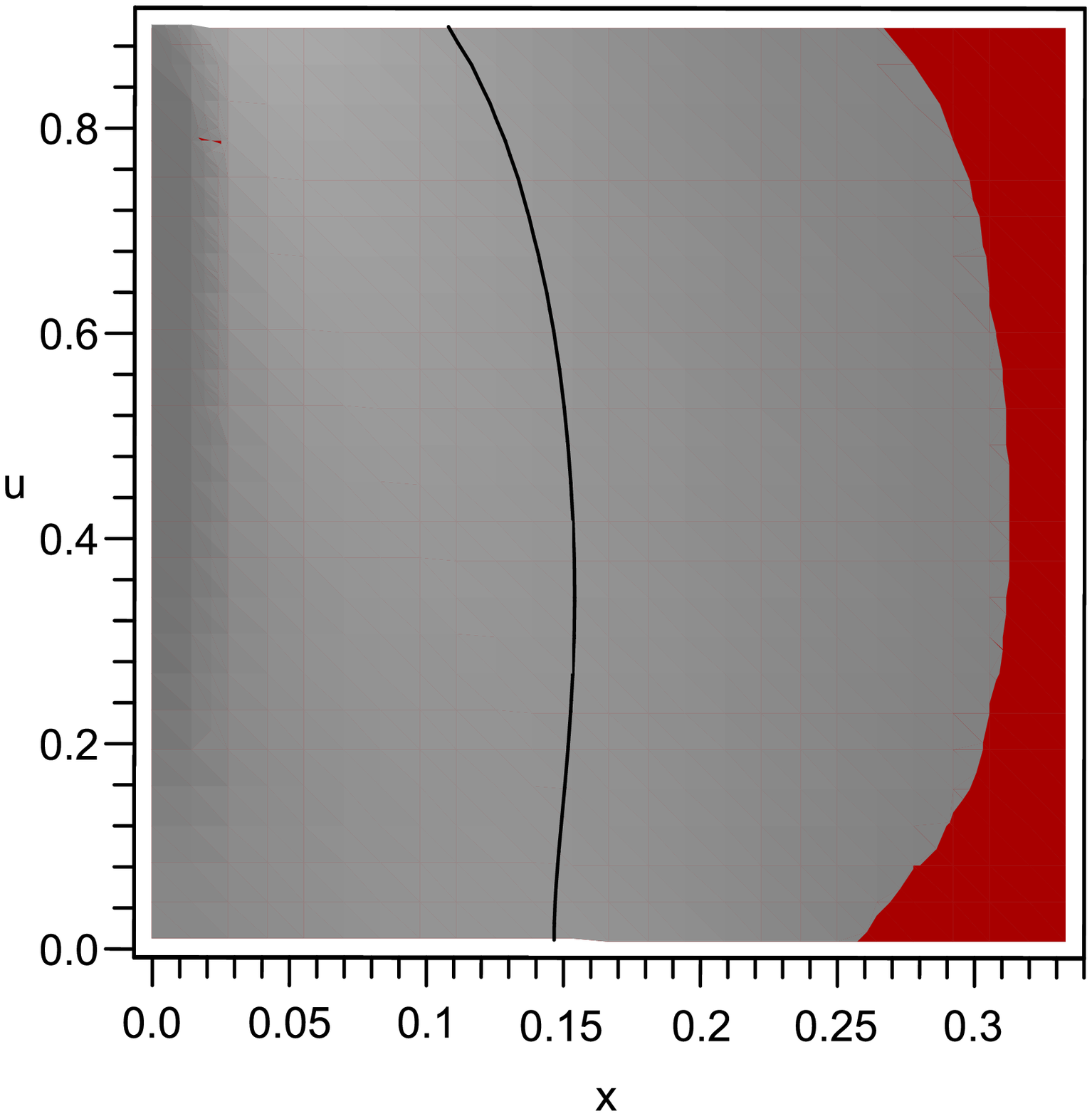}
\includegraphics[width=4.2cm]{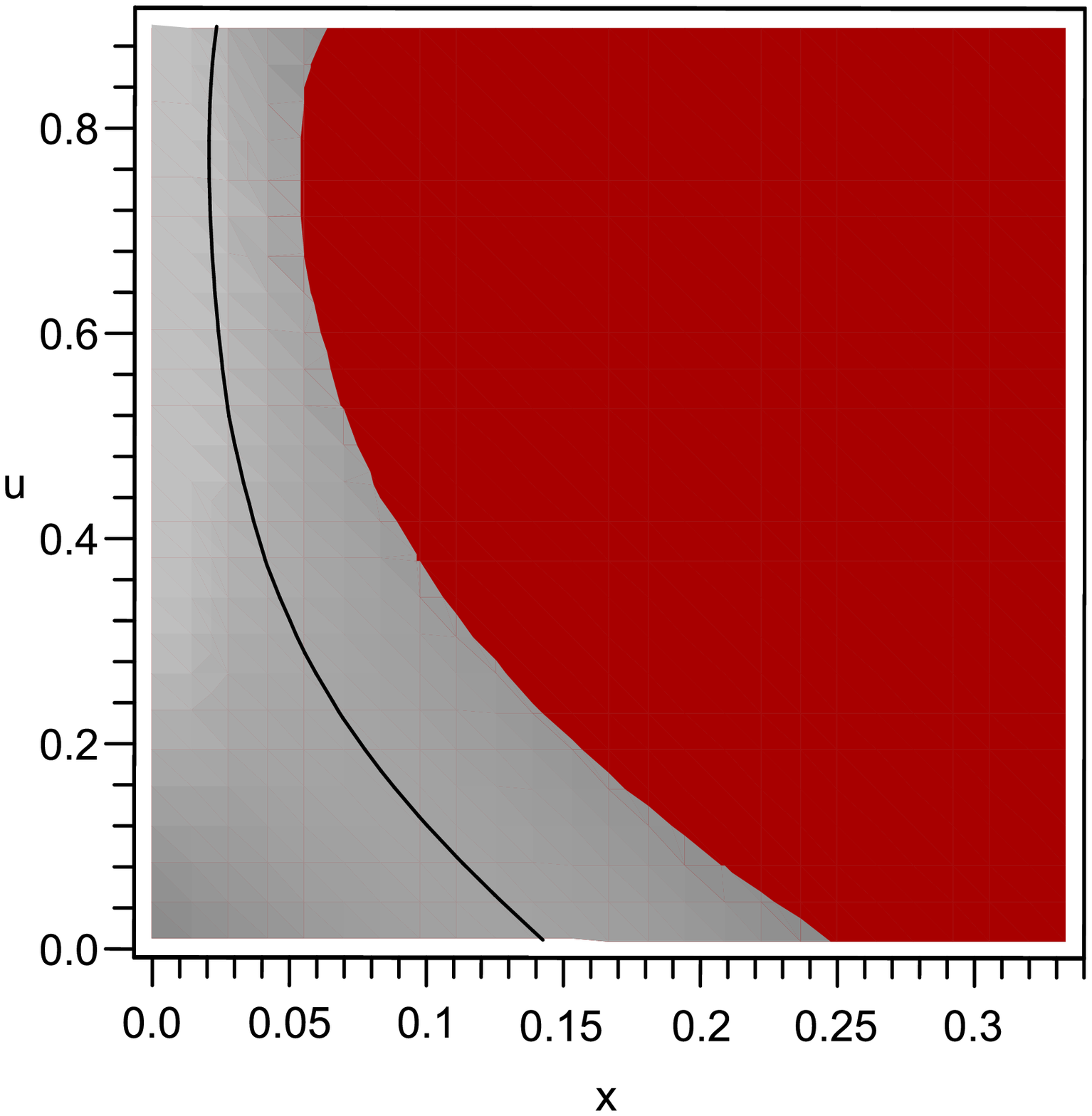}
\includegraphics[width=4.2cm]{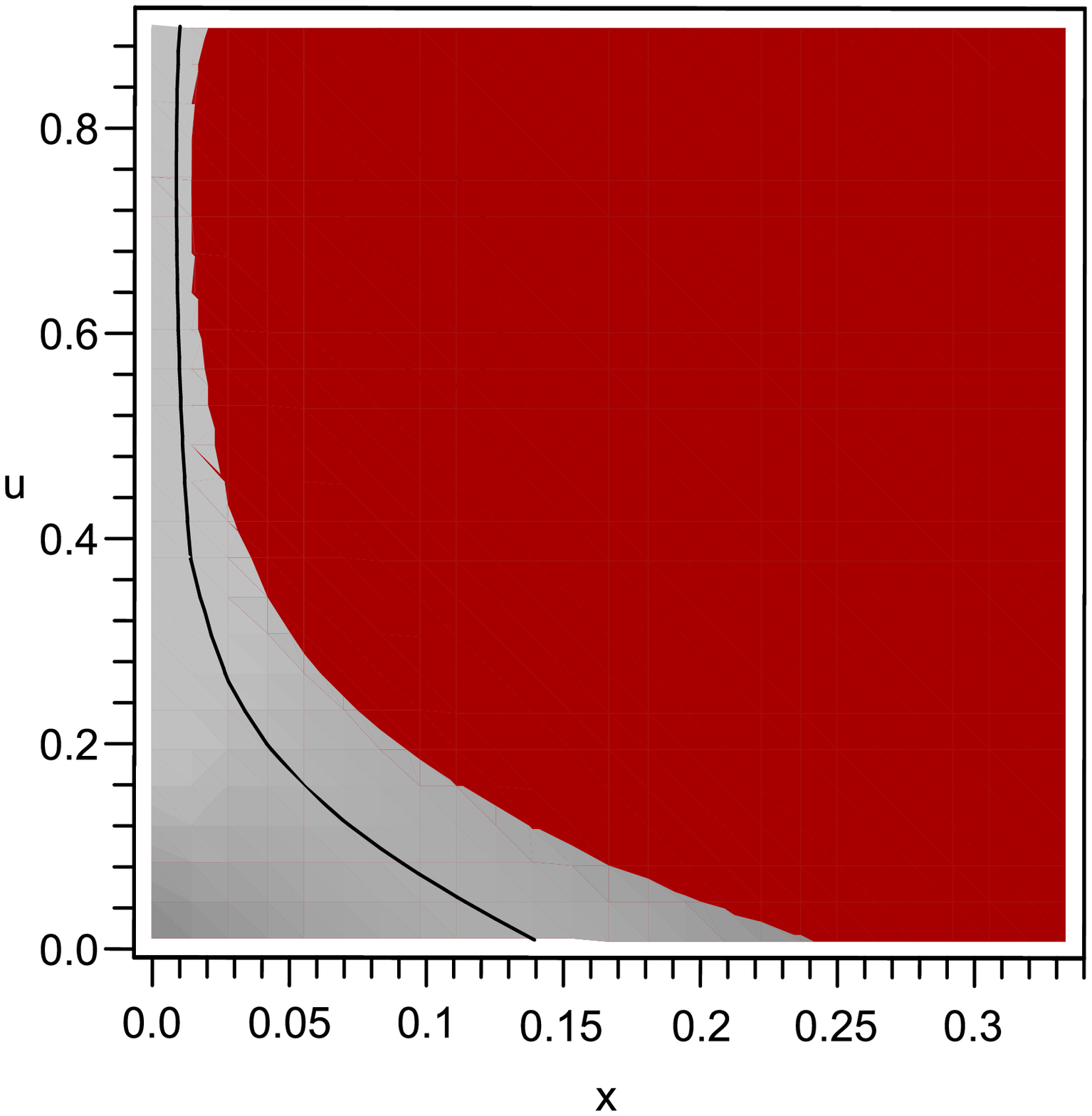}
\includegraphics[width=4.2cm]{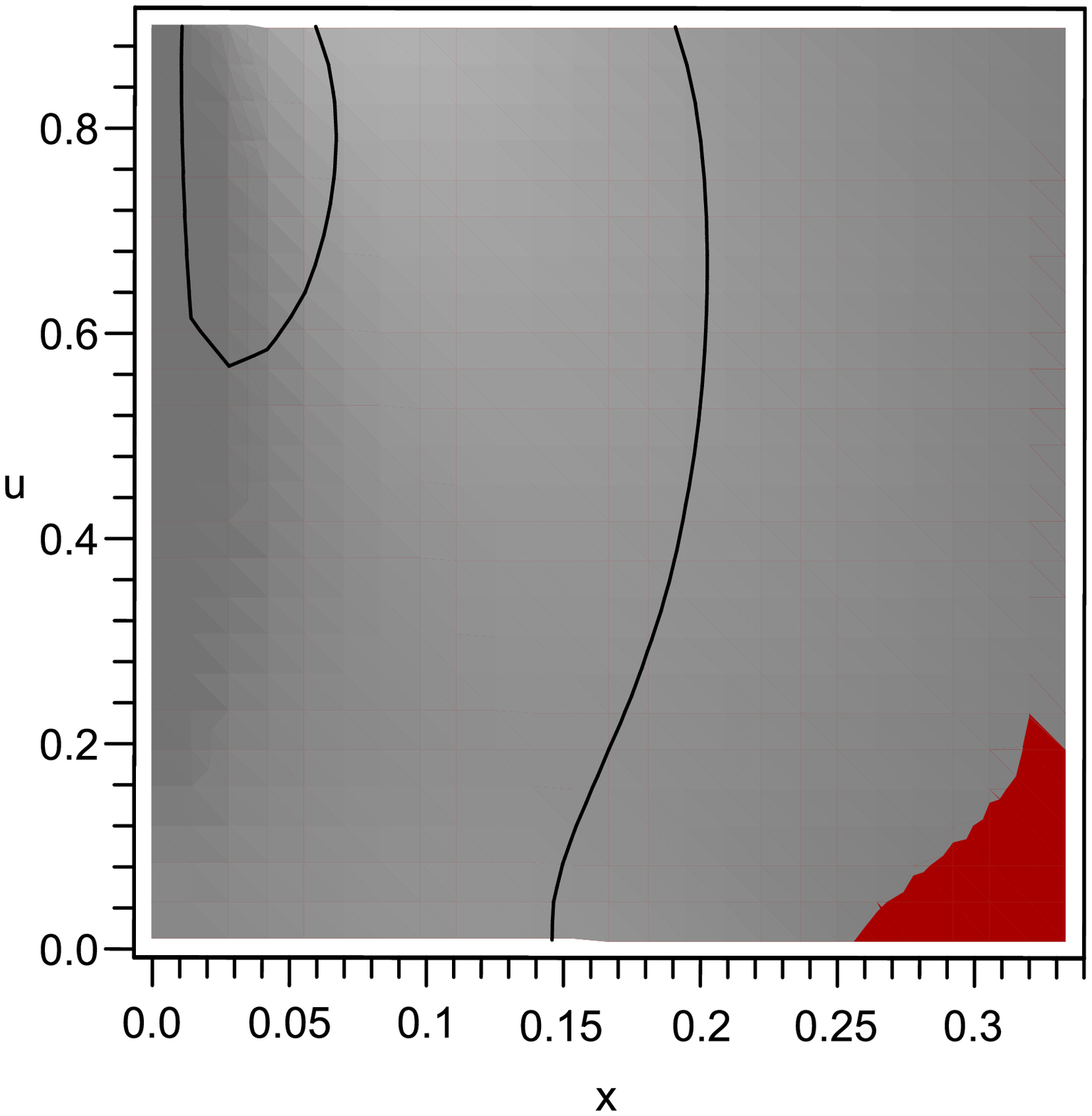}
\includegraphics[width=4.2cm]{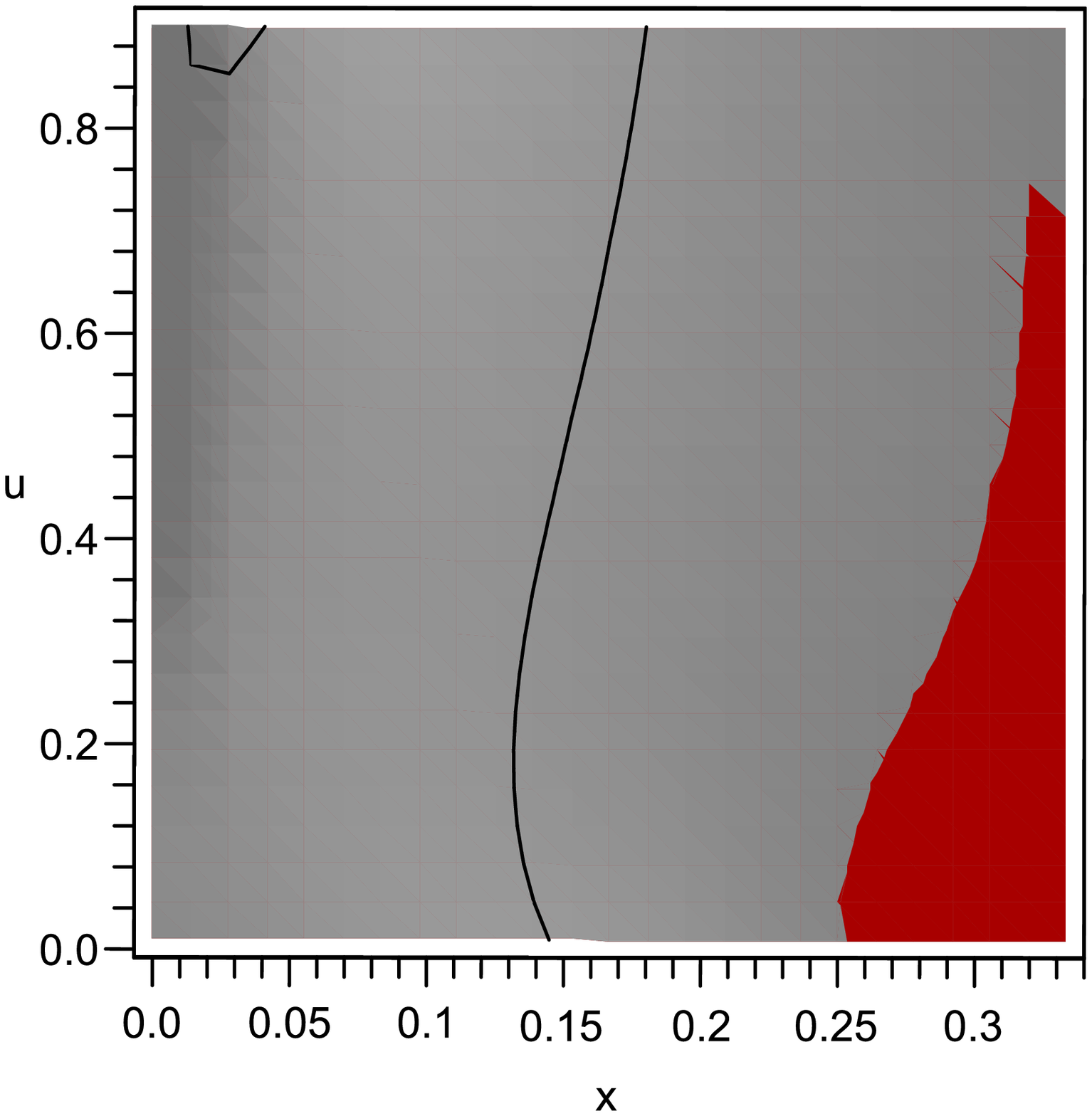}
\includegraphics[width=4.2cm]{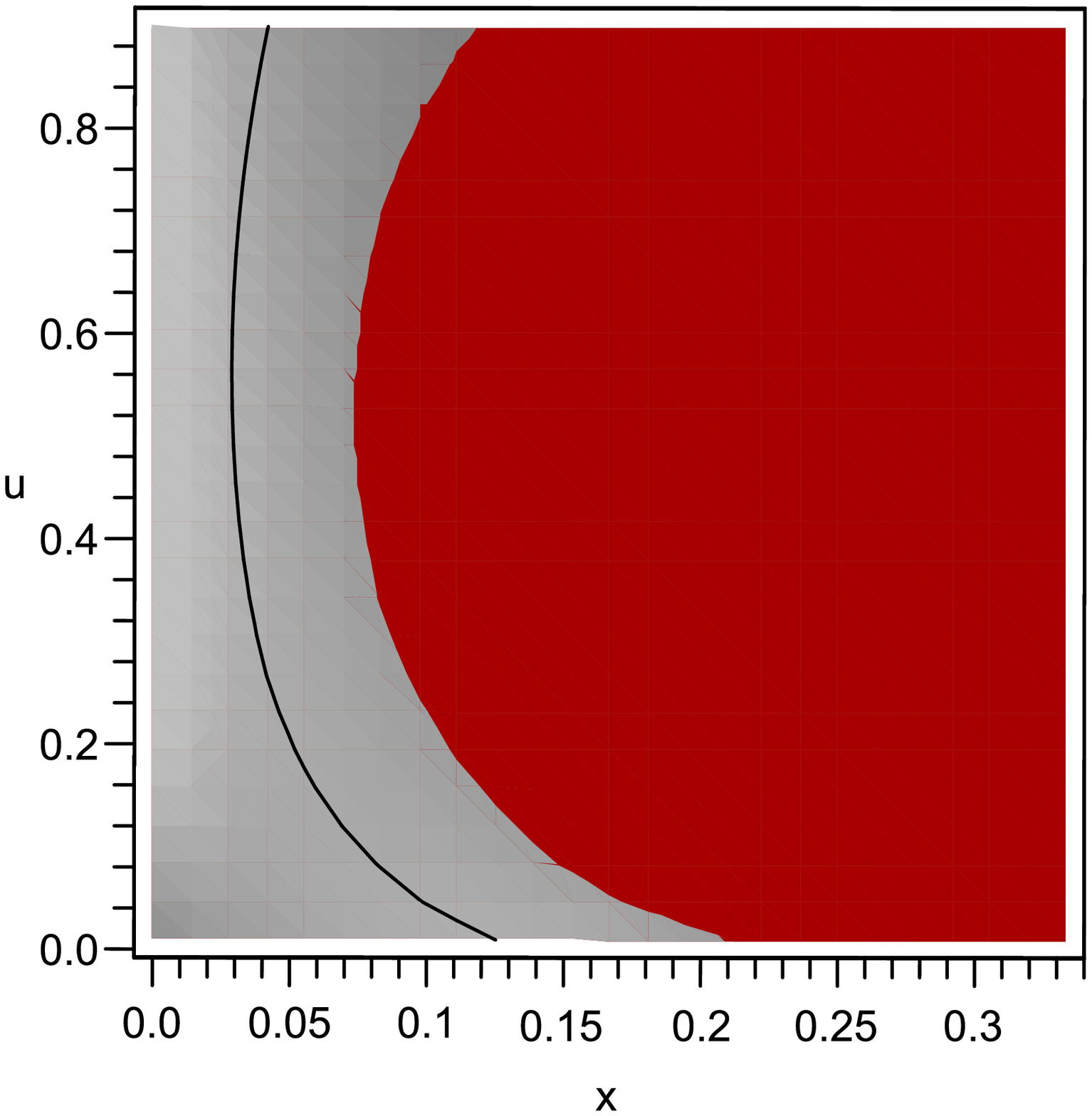}
\includegraphics[width=4.2cm]{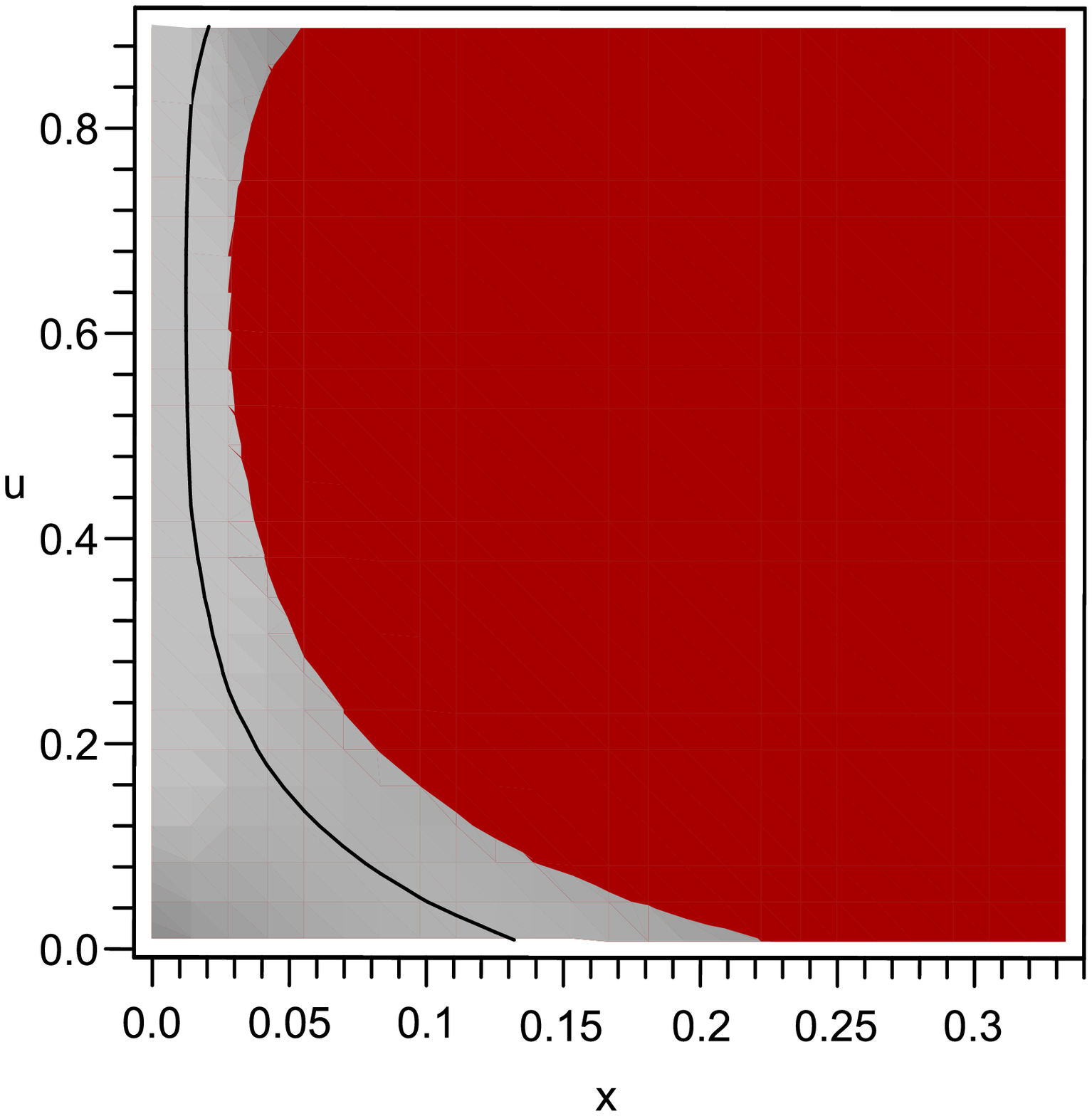}\vspace{0.7cm}
\caption{Geometric representation of the bound $c^2_s\geq 0$ in the $xu$-plane for positive coupling $\alpha>0$. For illustrative purposes we consider the exponential potential $V=V_0\exp{(\lambda\phi)}$ -- top panels -- and the power-law potential $V=V_0\phi^{2n}$ -- bottom panels -- for different values of the parameters $\lambda$ and $n$ respectively. As in FIG. \ref{fig01}, here we use the bounded variables $x=\alpha\dot\phi^2/2$ ($0\leq x\leq 1/3$) and $u=y/y+1$ ($0\leq u\leq 1$) where $y=\alpha V$, so that the whole phase plane $xu$ fits into a finite size box. In the top panels, from left to the right: $\lambda=-5$, $\lambda=-2$, $\lambda=2$ and $\lambda=5$, while in the bottom panels: $n=-2$, $n=-1$, $n=1$ and $n=2$, respectively. The red-colored regions are the ones where the squared sound speed is negative ($c^2_s<0$), i. e., where the Laplacian instability eventually develops. It is seen that, although the bound $c^2_s<0$ is always met in some -- even small -- region in the $xu$-plane, for monotonically growing potentials ($\lambda>0|n>0$), i. e. for potentials that do not allow the crossing of the phantom divide, the region of the phase plane where the Laplacian instability arises is appreciably larger.}\label{fig1}\end{figure*}

%-----------------------------------

%-----------------------------------

\begin{figure*}
\includegraphics[width=5cm]{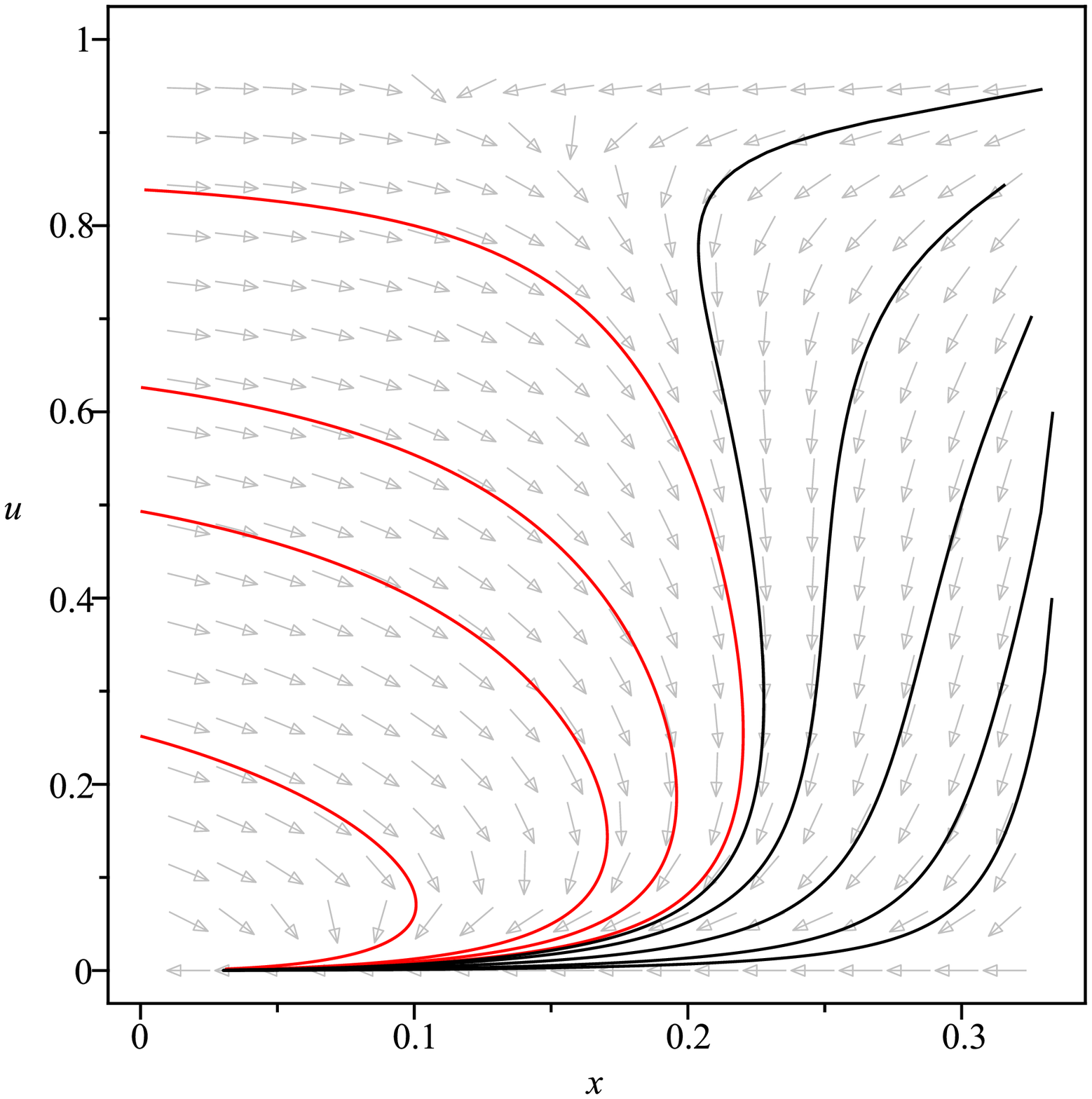}
\includegraphics[width=6cm]{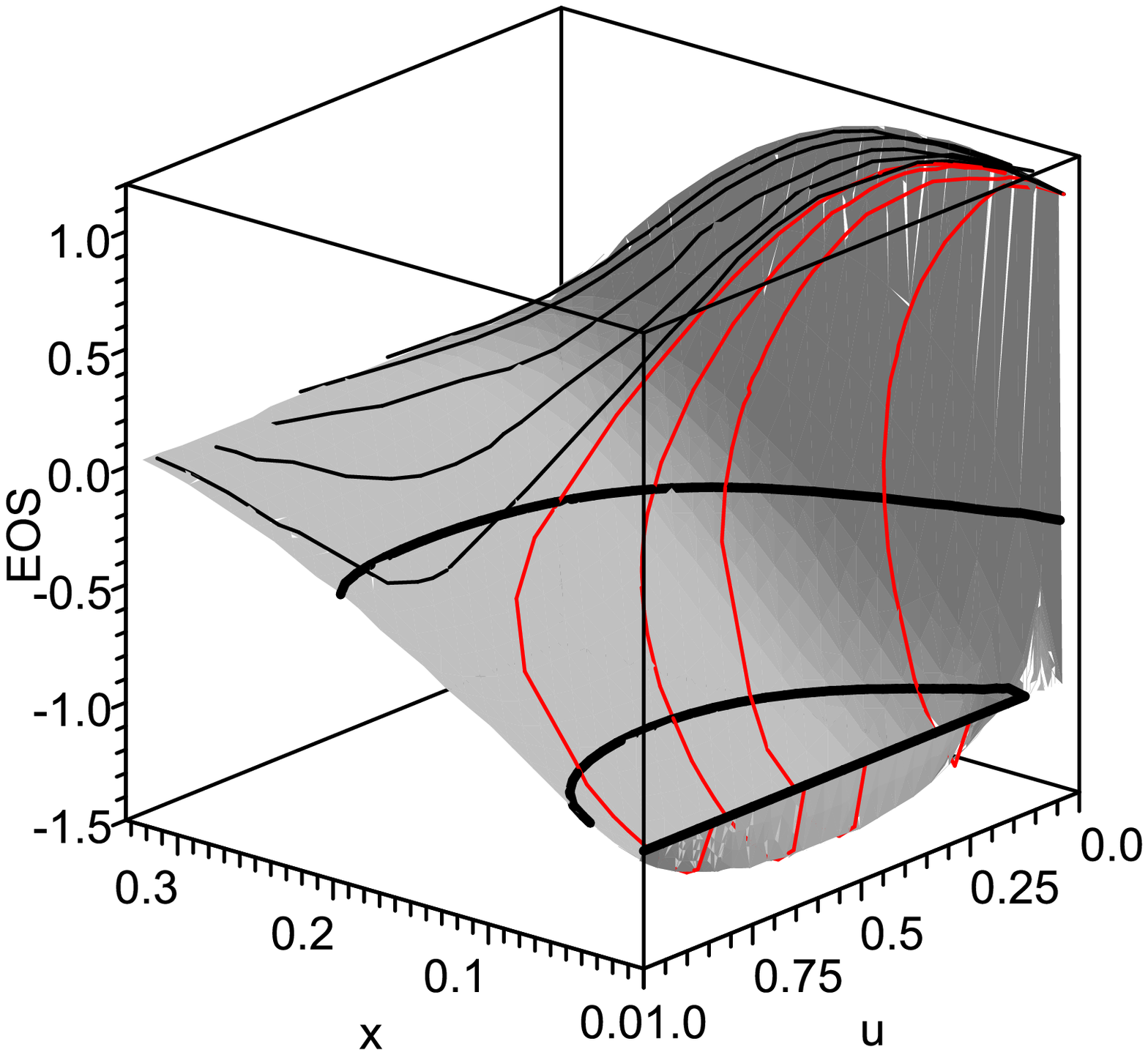}
\includegraphics[width=6cm]{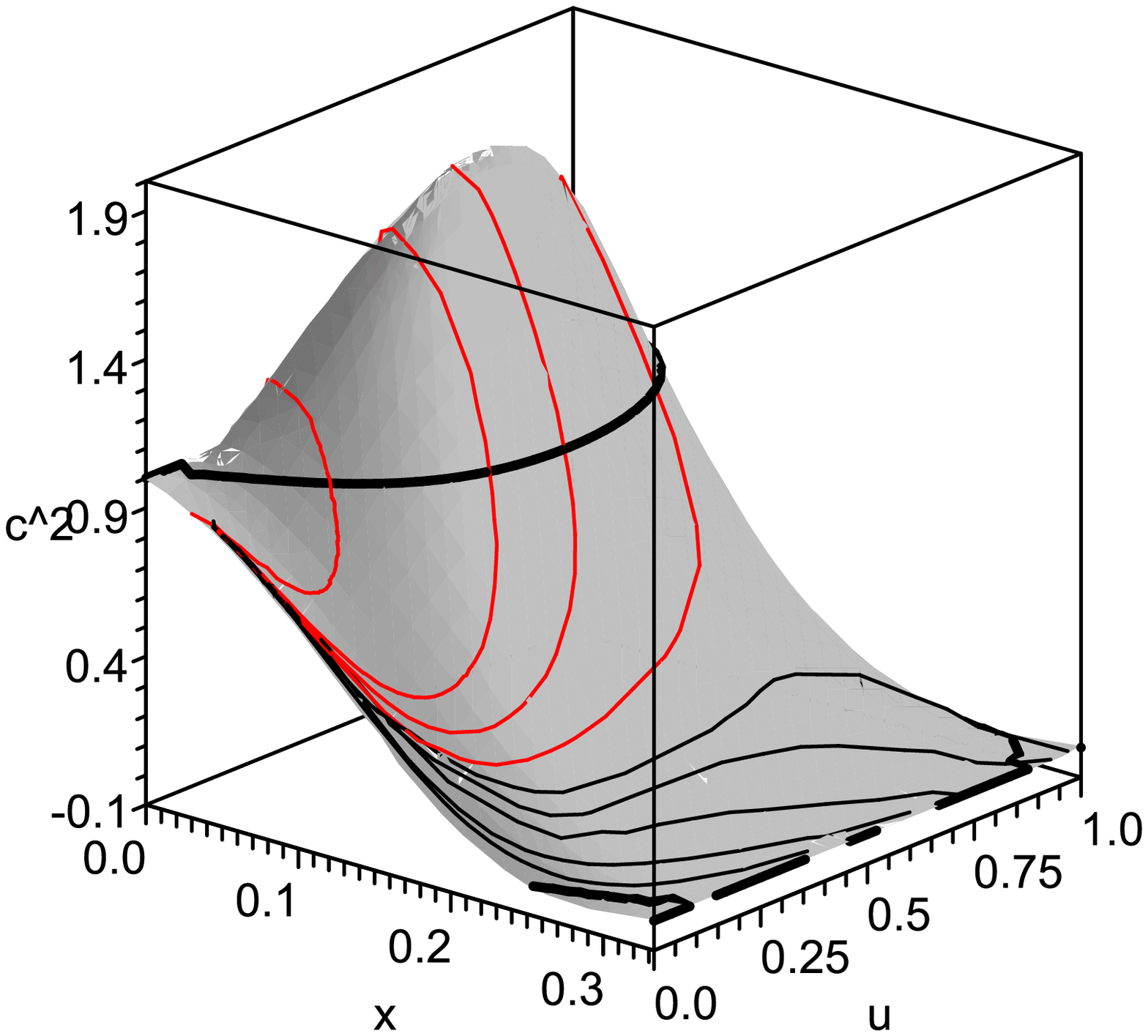}
\includegraphics[width=5cm]{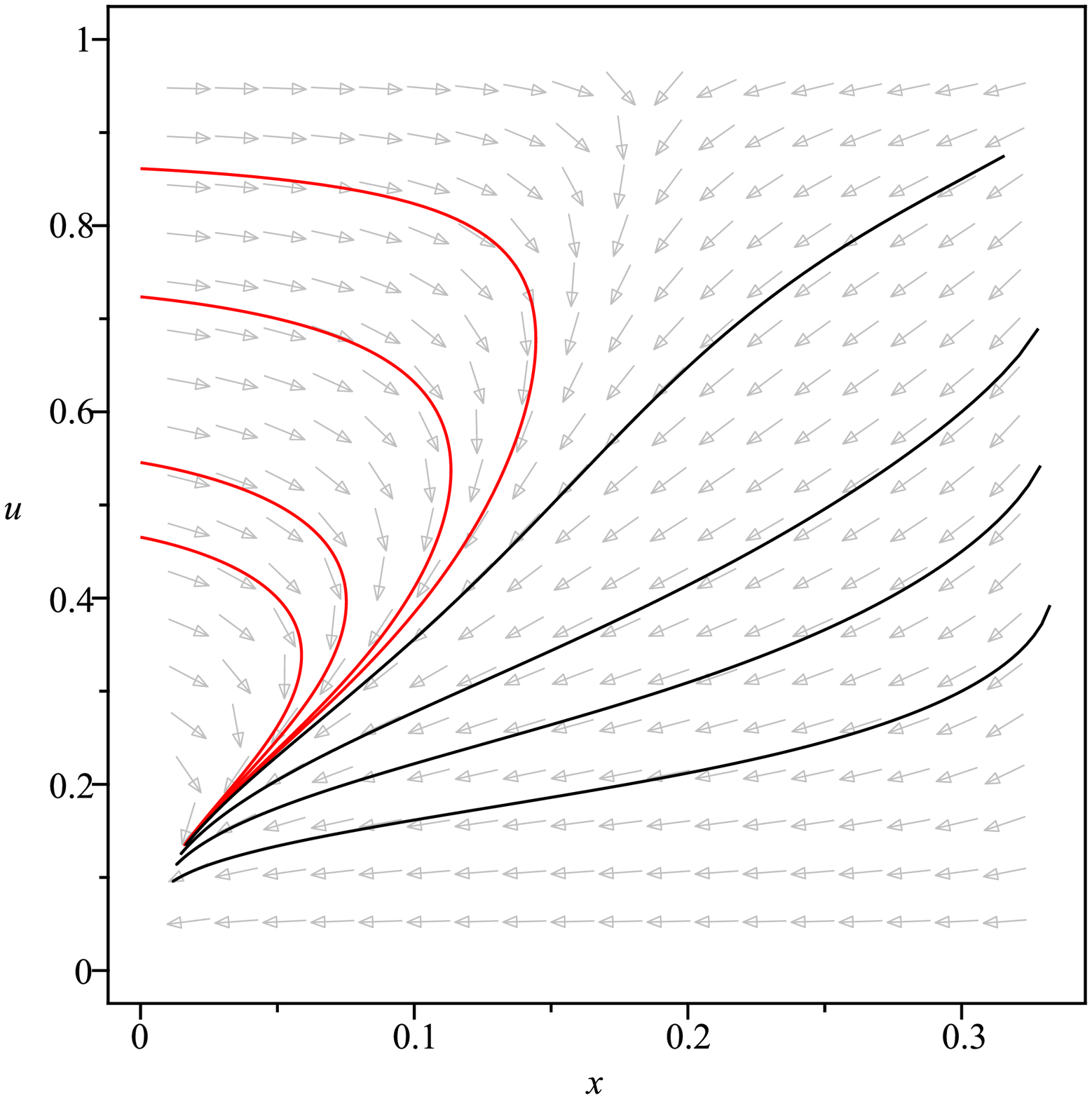}
\includegraphics[width=6cm]{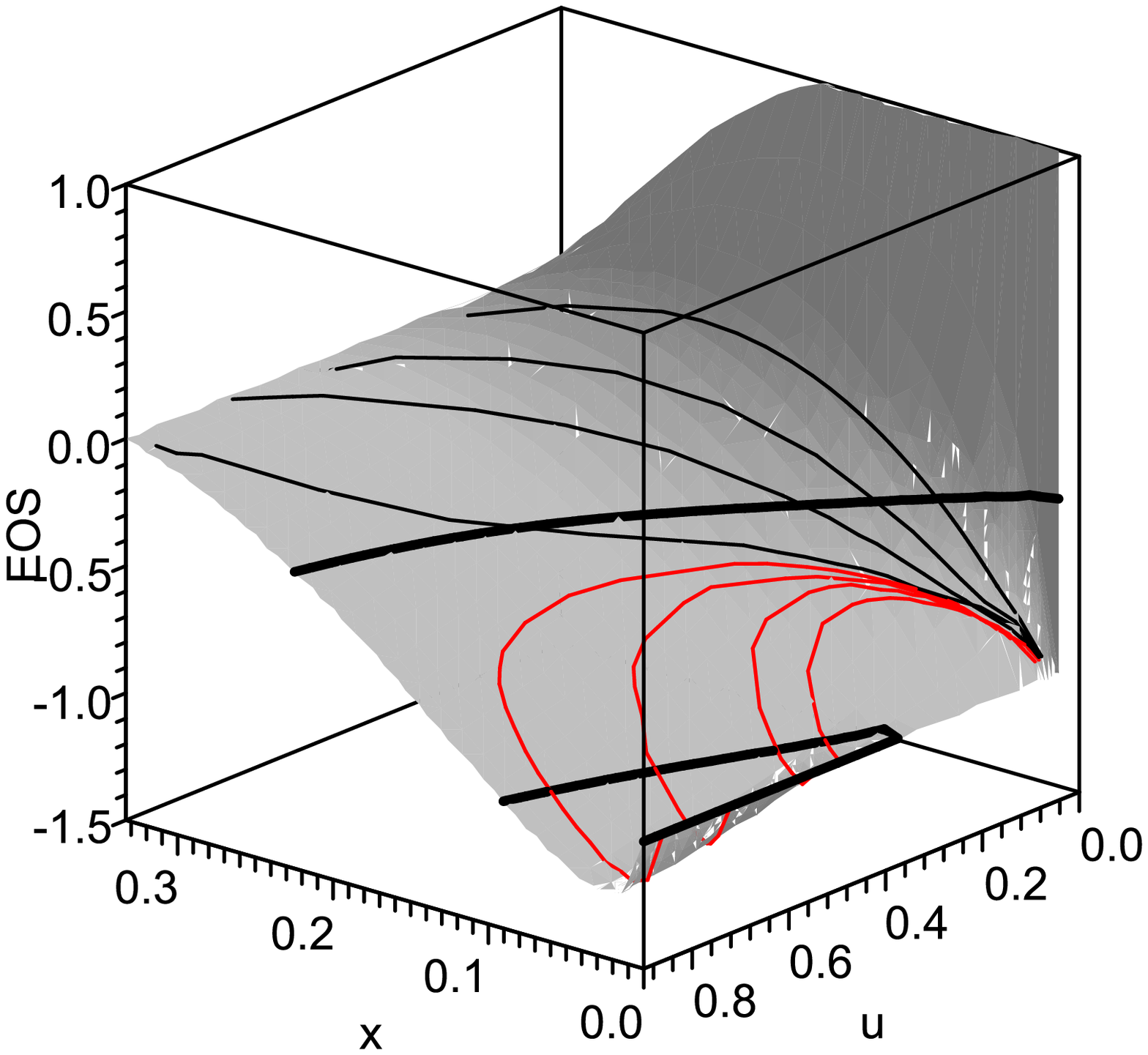}
\includegraphics[width=6cm]{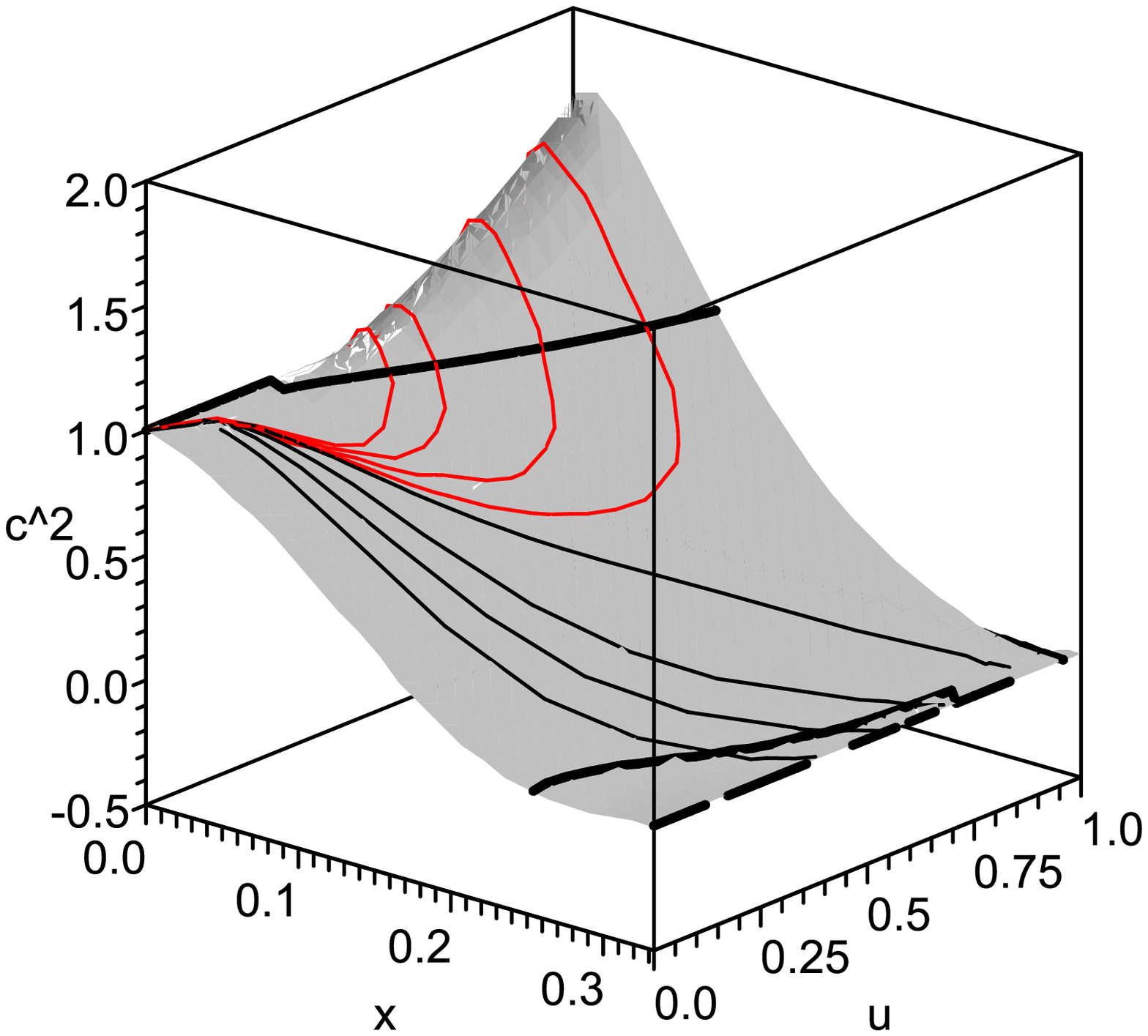}\vspace{0.7cm}
\caption{Phase portraits (left) of the dynamical system \eqref{ode-xu}, EOS-embedding diagrams (middle) and $c^2_s$-embedding diagrams (right) corresponding to the cosmological model \eqref{action} with positive coupling ($\alpha>0$). In the top panels the decaying exponential potential \eqref{exp-pot} ($\lambda=-5$) has been chosen, while in the bottom panels the inverse power-law potential \eqref{pow-law-pot} ($n=-1$) is considered. In the EOS-embedding diagrams the contours -- thick horizontal curves -- are drawn for $\omega_\text{eff}=-1/3$ (upper contour) and for $\omega_\text{eff}=-1$ (lower contour), while in the $c^2_s$-embeddings the drawn (quite irregular) contours are for $c^2_s=1$ (upper contour) and for $c^2_s=0$ (lower contour). It is seen that the red-colored orbits do the crossing of the phantom divide (middle panels) since these cross through the $\omega_\text{eff}=-1$ contour, and also violate causality since in the right-hand panels these orbits come from domains on the surface $c^2_s=c^2_s(x,u)$ that lie above the contour $c^2_s=1$, representing the local speed of light.}\label{fig3}\end{figure*}

%-----------------------------------

%%%%%%%%%%%%%%%%%%%%%%%%%%%%%%%%%%%%%%%%%%%%%

\section{Squared Sound Speed}\label{sec-cs2}

In \cite{cartier} the authors derived the evolution equations for the most general cosmological scalar, vector and tensor perturbations in a class of non-singular cosmologies derived from higher-order corrections to the low-energy bosonic string action:

\bea {\cal L}=\frac{1}{2}f(\phi,R)-\frac{1}{2}\omega(\phi)\nabla^\mu\phi\nabla_\mu\phi-V(\phi)+{\cal L}_q,\label{cartier-lag}\eea where $f(\phi,R)$ is an algebraic function of the scalar field $\phi$ and of the curvature scalar $R$, while $\omega(\phi)$ and $V(\phi)$ are functions of the scalar field. For our purposes it is enough to consider $f(\phi,R)=R$ and $\omega(\phi)=1$. Through ${\cal L}_q$ the inclusion of higher order derivative terms is allowed:

\bea {\cal L}_q=-\frac{\lambda}{2}\xi\left[c_1R^2_\text{GB}+c_2 G^{\mu\nu}\der_\mu\phi\der_\nu\phi+c_3\nabla^2\phi\der^\mu\phi\der_\mu\phi+c_4\left(\der^\mu\phi\der_\mu\phi\right)^2\right],\label{lag-q}\eea where $\xi=\xi(\phi)$ is a function of the scalar field, $R^2_\text{GB}\equiv R_{\mu\nu\tau\lambda}R^{\mu\nu\tau\lambda}-4R_{\mu\nu}R^{\mu\nu}+R^2$ is the Gauss-Bonnet combination, $\lambda$, $c_1,\ldots,c_4$ are constants and we have chosen the units where $\alpha'=1$. In this paper, without loss of generality we set $\xi=1$.

Our action \eqref{action} is a particular case of \eqref{cartier-lag}, so that the results of \cite{cartier} are easily applicable to the present model (see, for instance, \cite{gao}). The Einstein's field equations that are derived from the Lagrangian \eqref{cartier-lag} read: 

\bea G_{\mu\nu}=T^\text{eff}_{\mu\nu}=T^{(\phi)}_{\mu\nu}+T^{(q)}_{\mu\nu},\;\nabla^2\phi-T^{(q)}=V',\nonumber\eea where the comma stands for derivative with respect to $\phi$, $$T^{(\phi)}_{\mu\nu}=\der_\mu\phi\der_\nu\phi-\frac{1}{2}g_{\mu\nu}\left(\der^\tau\phi\der_\tau\phi\right)-g_{\mu\nu}V,$$ is the standard stress-energy tensor of a scalar field, while $$T^{(q)}_{\mu\nu}=-2\frac{\der{\cal L}_q}{\der g^{\mu\nu}}-g_{\mu\nu}{\cal L}_q,$$ and $T^{(q)}$ represent the contributions derived from the next to leading order corrections given by ${\cal L}_q$ in equation \eqref{lag-q} (equation (2) of \cite{cartier}). These contribute towards the effective stresses and energy. 

The perturbed line-element reads \cite{cartier, hwang}:

\bea ds^2=-a^2(1+2\psi)d\eta^2-2a^2\left(\beta_{,i}+B_i\right)d\eta dx^i+a^2\left[g_{ij}(1+2\vphi)+2\gamma_{,i|j}+2C_{(i|j)}+2C_{ij}\right]dx^idx^j,\label{ds-pert}\eea where $d\eta=dt/a$. Latin letters denote space indices while $\psi=\psi(t,{\bf x})$, $\beta=\beta(t,{\bf x})$, $\vphi=\vphi(t,{\bf x})$ and $\gamma=\gamma(t,{\bf x})$ characterize the scalar-type perturbations. The traceless modes $B_i$ and $C_i$ ($B^i_{|i}=C^i_{|i}=0$) represent the vector-type perturbations, meanwhile, $C_{ij}=C_{ij}(t,{\bf x})$ are trace free and transverse: $C^j_{i|j}=C^i_i=0$, and correspond to the tensor-type perturbations. The vertical bar denotes covariant derivative defined in terms of the space metric $g_{ij}$. Following \cite{hwang} in \cite{cartier} the uniform-field gauge ($\delta\phi=0$) is chosen since this gauge admits the simplest analysis. In this case each variable is replaced by its corresponding gauge-invariant combination with $\delta\phi$, for instance, for the scalar perturbation the gauge-invariant combination $$\vphi_{\delta\phi}\equiv\vphi-H\frac{\delta\phi}{\dot\phi},$$ is considered (in the uniform-field gauge $\vphi_{\delta\phi}$ is identified with $\vphi$ since $\delta\phi=0$). The second-order differential (wave) equation for the scalar-metric perturbation $\vphi_{\delta\phi}$ in closed form reads \cite{cartier}: 

\bea \frac{1}{a^3Q_s}\frac{\der}{\der t}\left(a^3Q_s\frac{\der}{\der t}\,\vphi_{\delta\phi}\right)-c^2_s\frac{\nabla^2}{a^2}\,\vphi_{\delta\phi}=0,\label{s-waveq}\eea where $$Q_s=\frac{\dot\phi^2+\frac{3Q_a^2}{2+Q_b}+Q_c}{\left(H+\frac{Q_a}{2+Q_b}\right)^2},$$ and the squared speed of propagation of the scalar perturbation is given by

\bea c^2_s=1+\frac{(2+Q_b)Q_d+Q_aQ_e+\frac{Q_a^2Q_f}{2+Q_b}}{(2+Q_b)(\dot\phi^2+Q_c)+3Q_a^2},\label{c2s-eff}\eea with

\bea &&Q_a=\lambda\dot\phi^2\left(2c_2H+c_3\dot\phi\right),\;Q_b=\lambda c_2\dot\phi^2,\;Q_c=-3\lambda\dot\phi^2\left(c_2 H^2+2c_3H\dot\phi+2c_4\dot\phi^2\right),\nonumber\\
&&Q_d=-2\lambda\dot\phi^2\left[c_2\dot H+c_3\left(\ddot\phi-H\dot\phi\right)\right],\;Q_e=4\lambda\dot\phi\left[c_2\left(\ddot\phi-H\dot\phi\right)-c_3\dot\phi^2\right],\;Q_f=2\lambda c_2\dot\phi^2=2Q_b.\label{Q-s}\eea

For the linearized tensor-type perturbations we obtain the following second order equation of motion \cite{cartier}: 

\bea \frac{1}{a^3Q_T}\frac{\der}{\der t}\left(a^3Q_T\frac{\der}{\der t}C^i_{\;\;j}\right)-c^2_T\frac{\nabla^2}{a^2}C^i_{\;\;j}=\frac{1}{Q_T}\delta T^i_{\;\;j},\label{t-waveq}\eea where $\delta T^i_{\;\;j}$ includes contributions to the tensor-type energy-momentum tensor, $$Q_T=1+\frac{\lambda}{2}\,c_2\dot\phi^2,$$ and

\bea c^2_T=\frac{2-\lambda c_2\dot\phi^2}{2+\lambda c_2\dot\phi^2},\label{c2t}\eea is the squared speed of propagation of the gravitational waves perturbation. Notice that for $c^2_s>0$ and $c^2_T>0$ the wave equations \eqref{s-waveq} and \eqref{t-waveq}, respectively, are hyperbolic differential equations -- the Cauchy problem is well posed -- meanwhile for negative $c^2_s<0$ and $c^2_T<0$, these equations are elliptic so there is not propagating mode (the Cauchy problem is not well posed). In this later case a Laplacian instability develops (see the appendix).

In the present cosmological model based in \eqref{action} the Lagrangian \eqref{lag-q} can be written in the following way: $${\cal L}_q=\frac{3\alpha}{2}\dot\phi^2H^2,$$ where we have set $\xi=1$, $\lambda c_2=-\alpha$ (the remaining constants in \eqref{lag-q} vanish). Hence, we obtain that

\bea Q_a=-2\alpha H\dot\phi^2,\;Q_b=-\alpha\dot\phi^2,\;Q_c=3\alpha H^2\dot\phi^2,\;Q_d=2\alpha\dot H\dot\phi^2,\;Q_e=-4\alpha\dot\phi(\ddot\phi-H\dot\phi).\label{qs}\eea For the squared speed of propagation of the gravitational waves perturbation \eqref{c2t} it is found that, for the present cosmological model:

\bea c^2_T=\frac{1+\alpha\dot\phi^2/2}{1-\alpha\dot\phi^2/2},\label{c2t'}\eea where it is appreciated that, for the positive coupling $\alpha>0$, the tensor perturbations propagate superluminally. A similar result has been formerly reported in \cite{germani-1} for the same model but under the slow-roll approximation, i. e., valid for primordial inflation. For negative coupling $\alpha<0$, provided that $\dot\phi^2>2/|\alpha|$ the squared sound speed of the tensor perturbations becomes negative, signaling to the eventual occurrence of a Laplacian instability. For a detailed derivation of \eqref{c2s-eff} and of \eqref{c2t} within the perturbative approach we recommend the reference \cite{cartier}.

Equation \eqref{c2s-eff} with the substitution of the quantities \eqref{qs} will be our master equation for determining the (squared) speed of propagation of the scalar perturbations of the energy density. In terms of the field variables $x=\alpha\dot\phi^2/2$ and $y=\alpha V(\phi)$ we have that:

\bea c^2_s=1+\frac{4x[\epsilon(3-11x+6x^2)+(1-3x)y]}{3(1-x)F_\epsilon}-\frac{3(1-x)(\epsilon x+y)(\omega_\text{eff}+1)}{F_\epsilon},\label{c2s-master-eq}\eea where $\omega_\text{eff}$ is given by \eqref{eos-master-eq} and the funciton $F_\epsilon$ has been defined in \eqref{f-eps}.

\subsection{Positive coupling}

In this case we have that $0\leq x\leq 1/3$ and $0\leq y<\infty$. This means that $F_\epsilon$ is always a positive function. Besides, both the numerator and the denominator in the second term in the right-hand side (RHS) of equation \eqref{c2s-master-eq} are positive quantities. The same is true for the factor $(1-x)(\epsilon x+y)/F_\epsilon$ in the third term in the RHS of the mentioned equation. Hence, while the second term always contributes towards superluminality of propagation of the scalar perturbations, the contribution of the third term depends on the sign of $\omega_\text{eff}+1$. For $\omega_\text{eff}>-1$ the superluminal contribution of the second term in the RHS of \eqref{c2s-master-eq} may be compensated by the third term. However, when $\omega_\text{eff}<-1$, both terms in the RHS of \eqref{c2s-master-eq} contribute towards superluminality of the propagation of the scalar perturbations of the energy density. This means that, whenever the crossing of the phantom divide is allowed, then $\omega_\text{eff}+1$ becomes necessarily negative during a given stage of the cosmic evolution and, consequently, causality violations are inevitable. This result is independent on the specific functional form of the self-interaction potential. 

In general, from \eqref{c2s-master-eq} it follows that whenever the condition 

\bea \frac{4x[\epsilon(3-11x+6x^2)+(1-3x)y]}{9(1-x)^2(\epsilon x+y)}>\omega_\text{eff}+1,\label{cond}\eea is fulfilled, the squared sound speed is superluminal ($c^2_s>1$). The latter condition may be satisfied only if $\omega_\text{eff}+1<0$, i. e., if $\omega_\text{eff}<-1$. For positive $\omega_\text{eff}+1>0$, the inequality \eqref{cond} is never satisfied. 

We want to point out that, although the condition $\omega_\text{eff}<-1$ boosts further superluminality of the propagation of the scalar perturbations, in general the $\omega=-1$ crossing is not required for the superluminality to arise in the present model. Actually, as seen from \eqref{c2s-master-eq}, given that the second term in the RHS of \eqref{c2s-master-eq} is always a postive quantity, superluminality arises even if $\omega_\text{eff}+1=0$.  

The potential situation where $\omega_\text{eff}+1>0$, i. e., where $\omega_\text{eff}>-1$, leads to another interesting and disturbing possibility, namely that 

\bea \omega_\text{eff}+1>\frac{(1-x)F_\epsilon}{3(1-x)^2(\epsilon x+y)}+\frac{4x[\epsilon(3-11x+6x^2)+(1-3x)y]}{9(1-x)^2(\epsilon x+y)},\label{laplac-instab}\eea that is, that $c^2_s<0$. Fulfillment of this latter bound leads to the development of the Laplacian/gradient instability. This is a classical instability associated with the uncontrolled growth of the amplitude of the scalar perturbations of the background density (see the appendix \ref{app}). 

In the FIG. \ref{fig1} we have geometrically represented the bound $c^2_s\geq 0$ for the exponential \eqref{exp-pot} and for the power-law \eqref{pow-law-pot} potentials, for different values of the free parameters $\lambda$ and $n$, respectively.\footnote{In this section we focus in the quintessence case $\epsilon=1$ exclusively.} Meanwhile, in FIG. \ref{fig2} we have drawn the surfaces $\omega_\text{eff}=\omega_\text{eff}(x,u)$ and $c^2_s=c^2_s(x,u)$ for the growing exponential potential with $\lambda=5$. In these figures we have used the bounded coordinate in \eqref{u-var}: $$u=\frac{y}{y+1},\;0\leq u\leq 1,$$ instead of $y$ ($0\leq y<\infty$), in order to comprise the whole phase plane into a finite-size region. In the right-hand panel of FIG. \ref{fig2} different orbits of the dynamical system corresponding to the present cosmological model, have been mapped into the surface $c^2_s=c^2_s(x,u)$ in order to show geometrically, that the choice of free parameters that is not compatible with the crossing of the phantom divide -- in this case the growing exponential (positive slope) -- leads eventually to the development of the Laplacian instability. 

As already shown, the potentials that allow for the crossing of the phantom divide -- potentials with the negative slope -- can lead also to causality problems. This finding is geometrically illustrated in the figure FIG. \ref{fig3}, where the EOS-embedding and $c^2_s$-embedding diagrams are shown for potentials with the negative slope: (i) decaying exponential potential \eqref{exp-pot} with $\lambda=-5$ (top panels) and (ii) inverse power-law potential \eqref{pow-law-pot} with $n=-1$ (bottom panels), respectively.

\subsection{Negative coupling}

For $\alpha<0$ we have that $-\infty<x\leq 0$, $-\infty<y\leq 0$, so that it is recommended to use the bounded variables $v$, $w$ in \eqref{vw-var}. Under the latter choice the whole of $xy$-plane is comprised within the unit square: $\{(v,w):0\leq v\leq 1,0\leq w\leq 1\}$. In this case the analysis of the bounds on the squared sound speed: $0\leq c^2_s\leq 1$, is a very complicated task and one has to heavily rely in the numeric investigation. 

\begin{figure*}
\includegraphics[width=4.2cm]{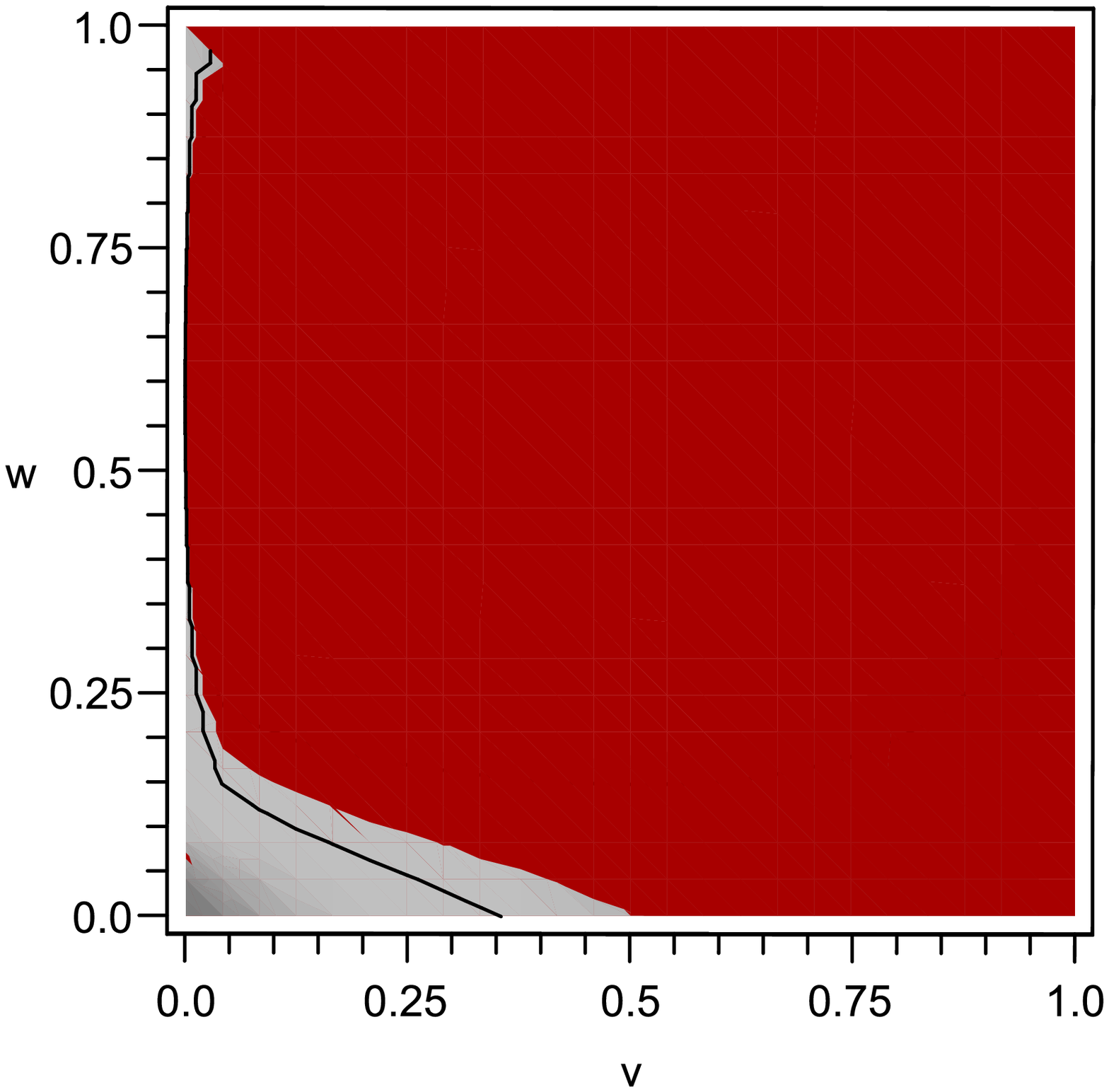}
\includegraphics[width=4.2cm]{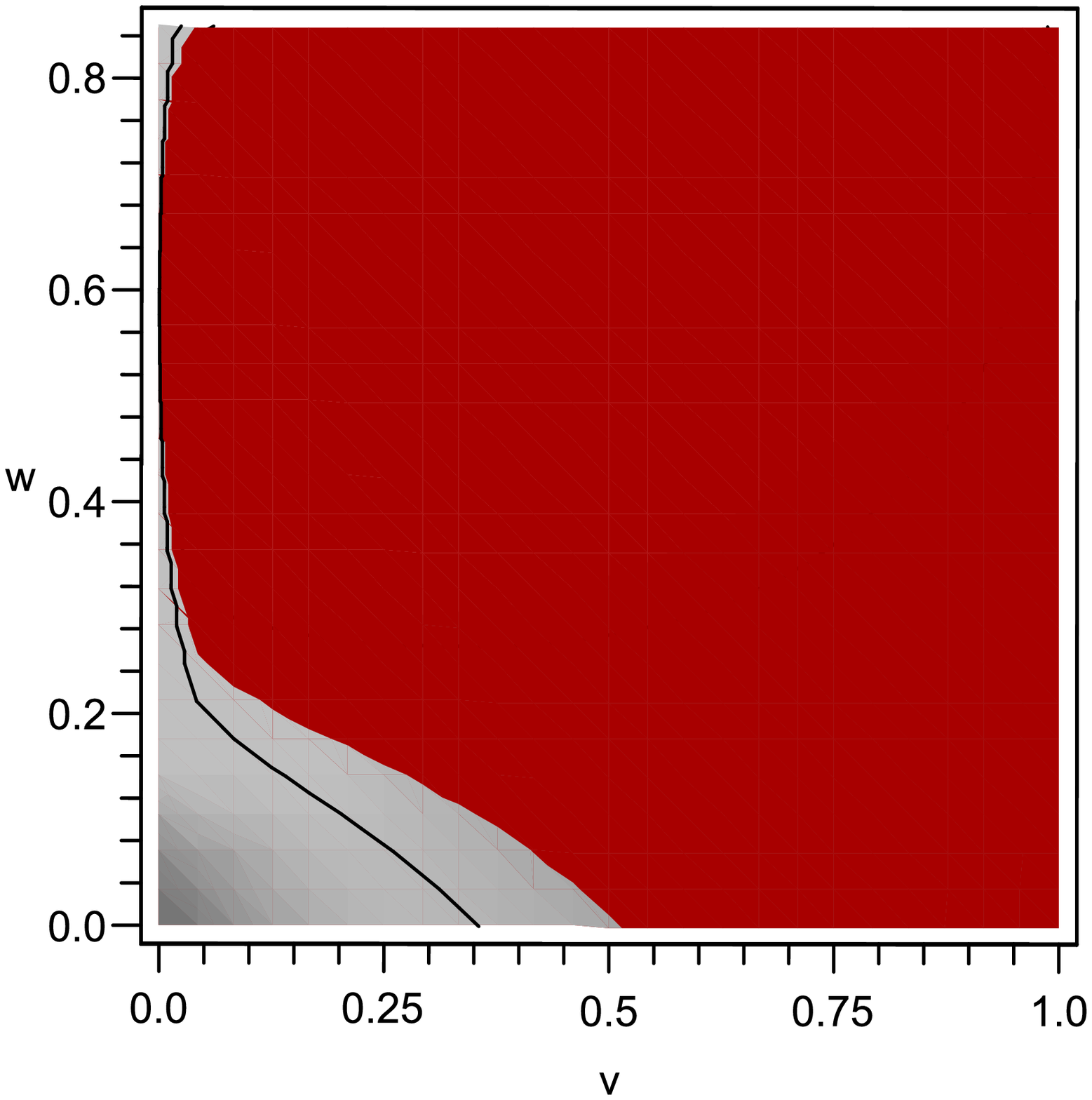}
\includegraphics[width=4.2cm]{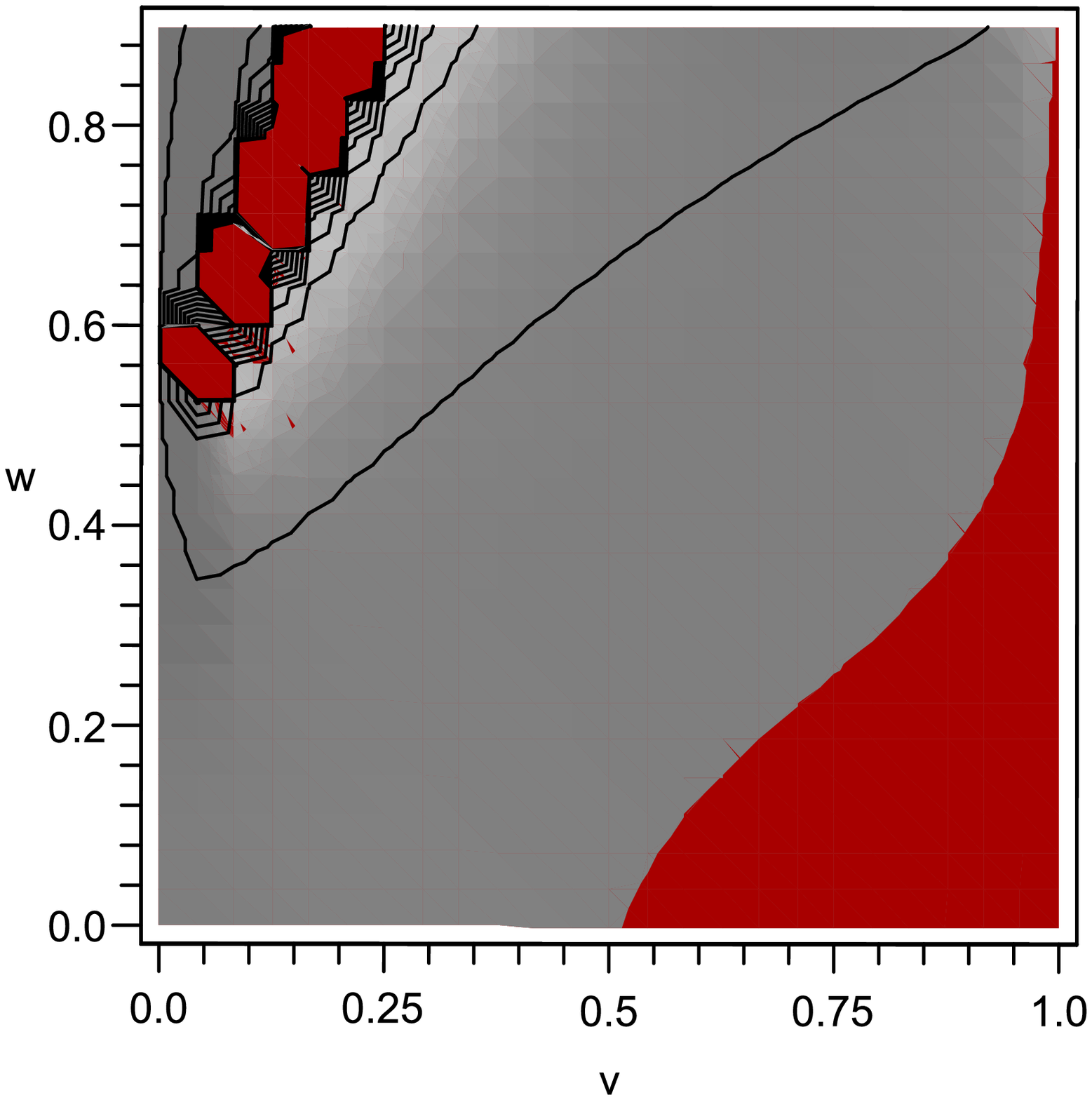}
\includegraphics[width=4.2cm]{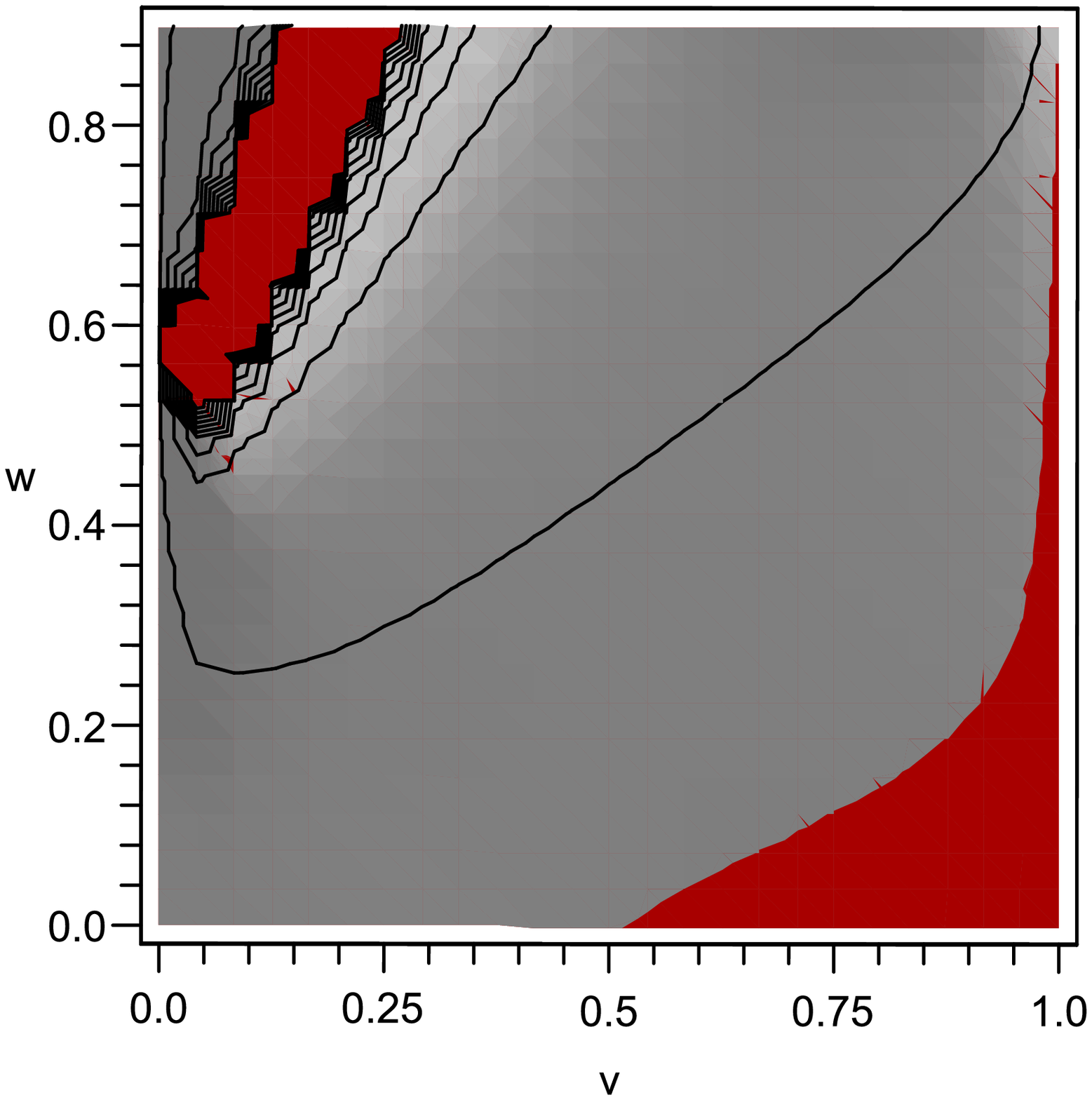}
\includegraphics[width=4.2cm]{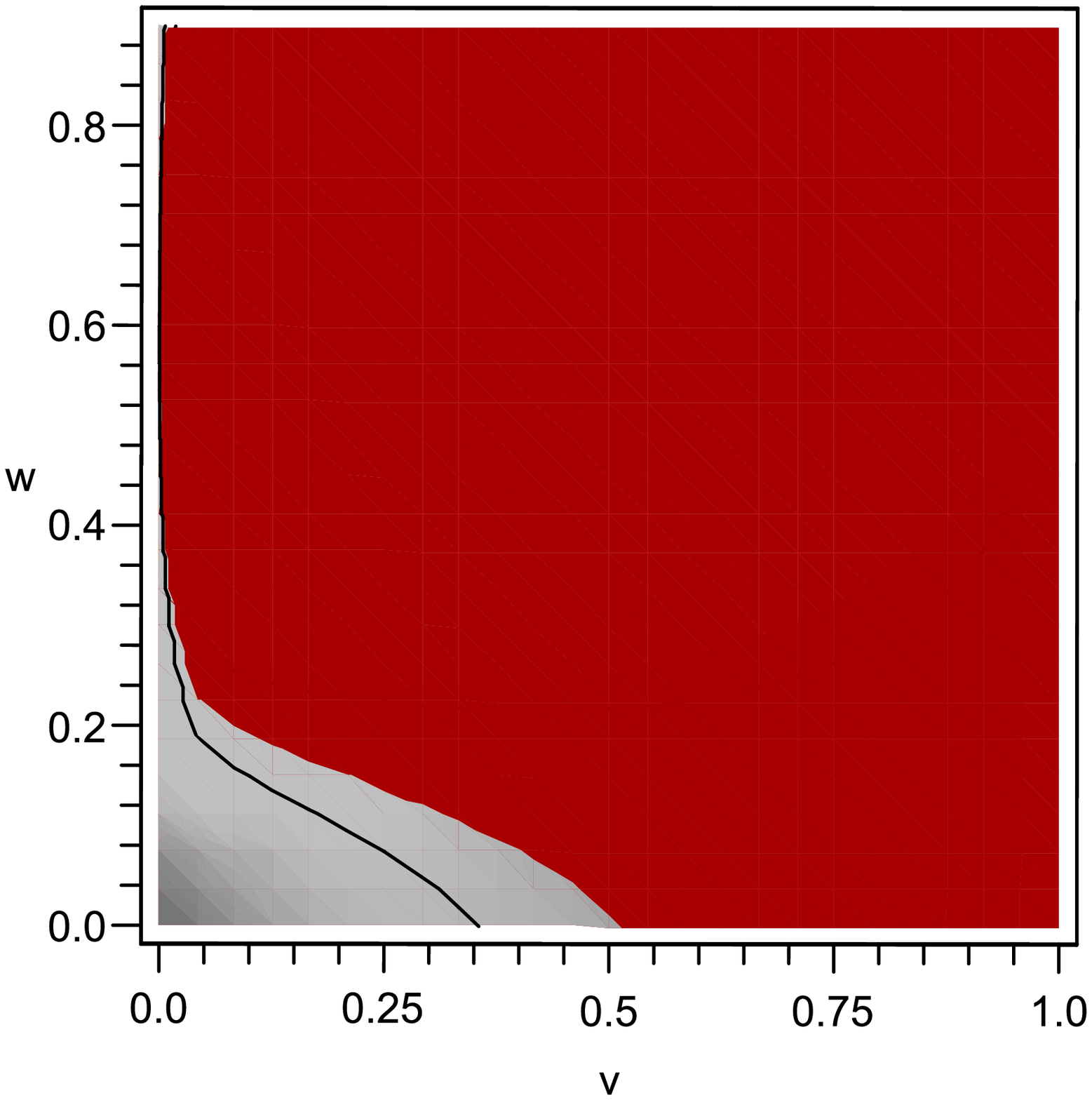}
\includegraphics[width=4.2cm]{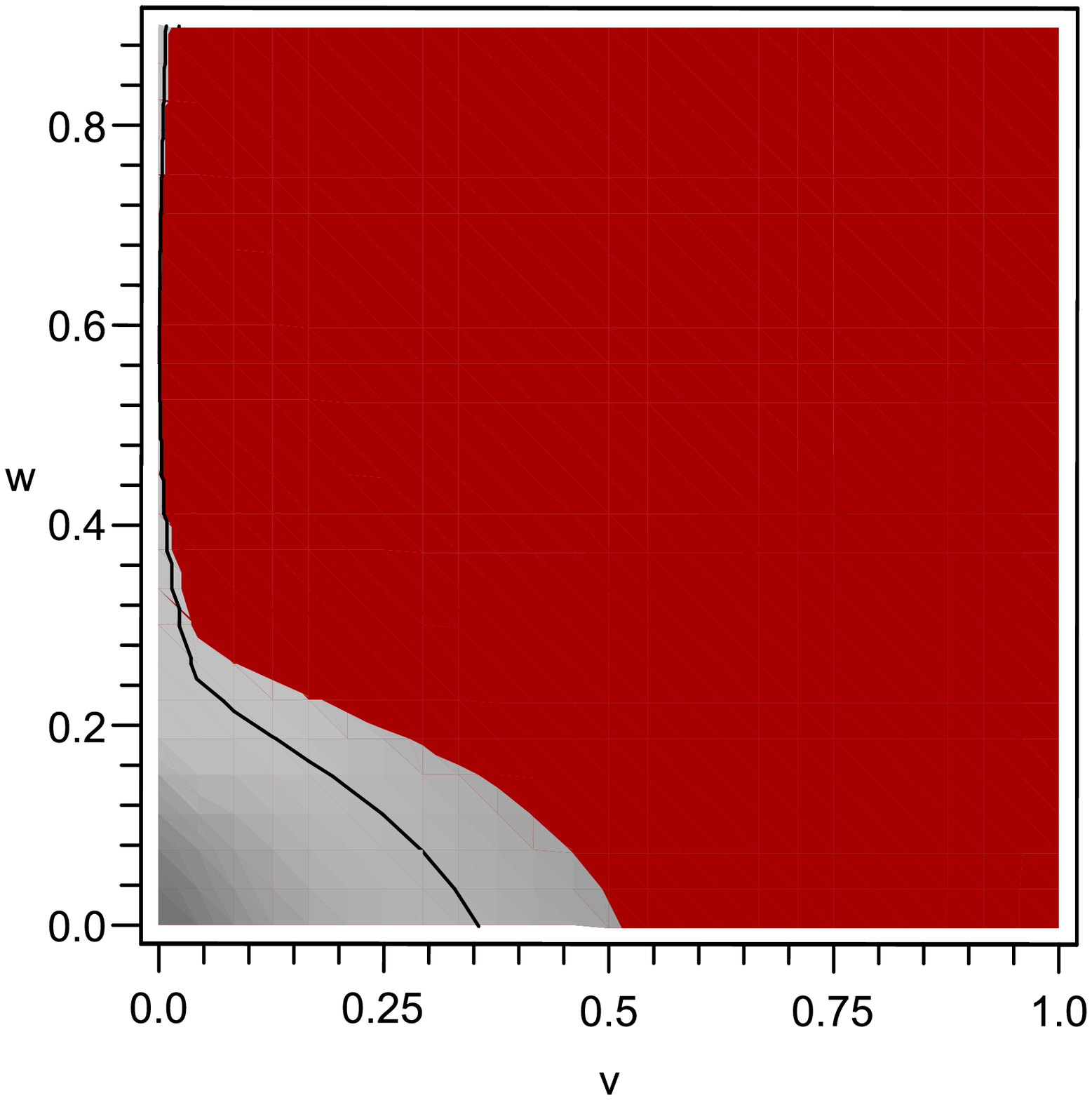}
\includegraphics[width=4.2cm]{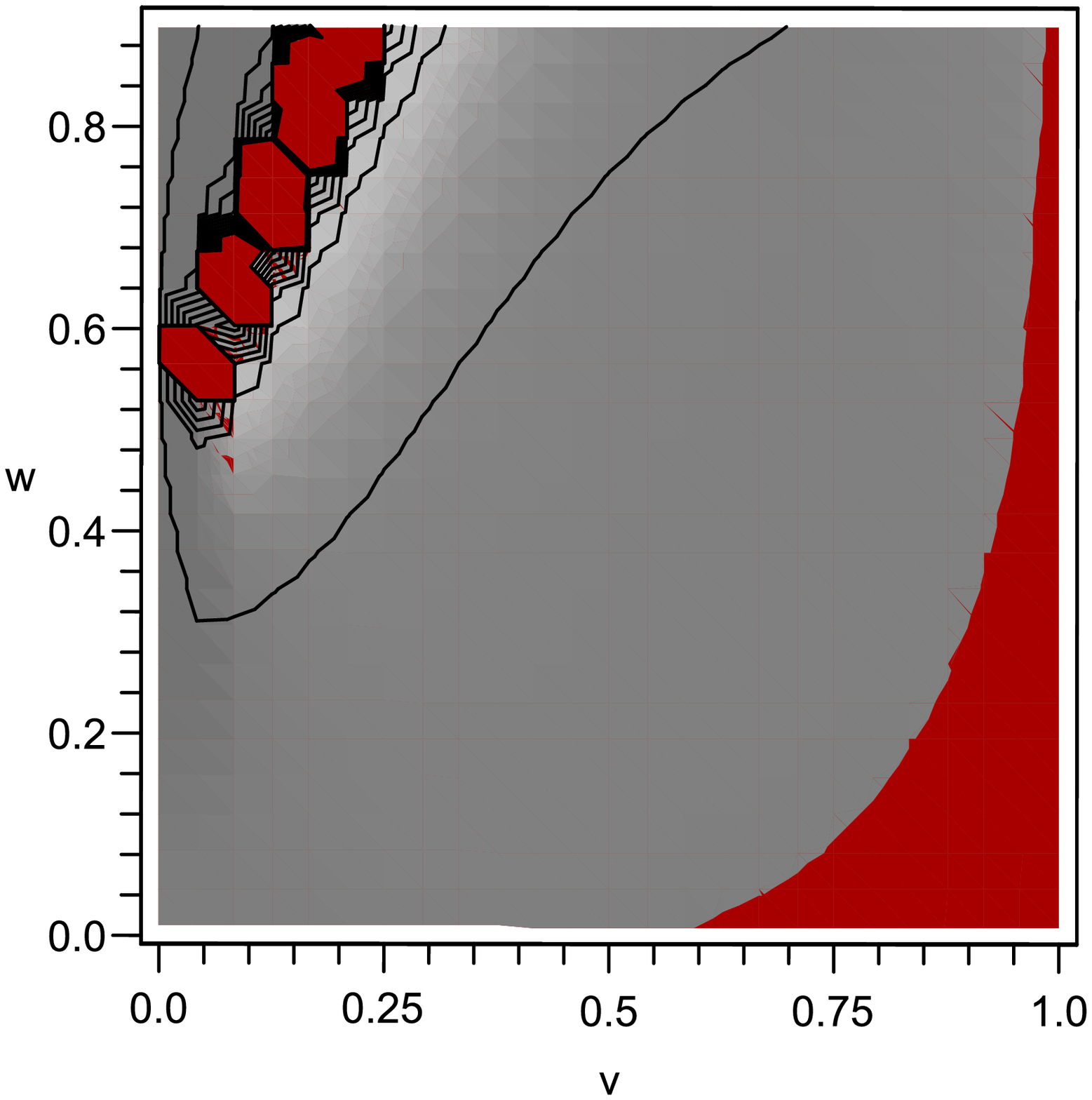}
\includegraphics[width=4.2cm]{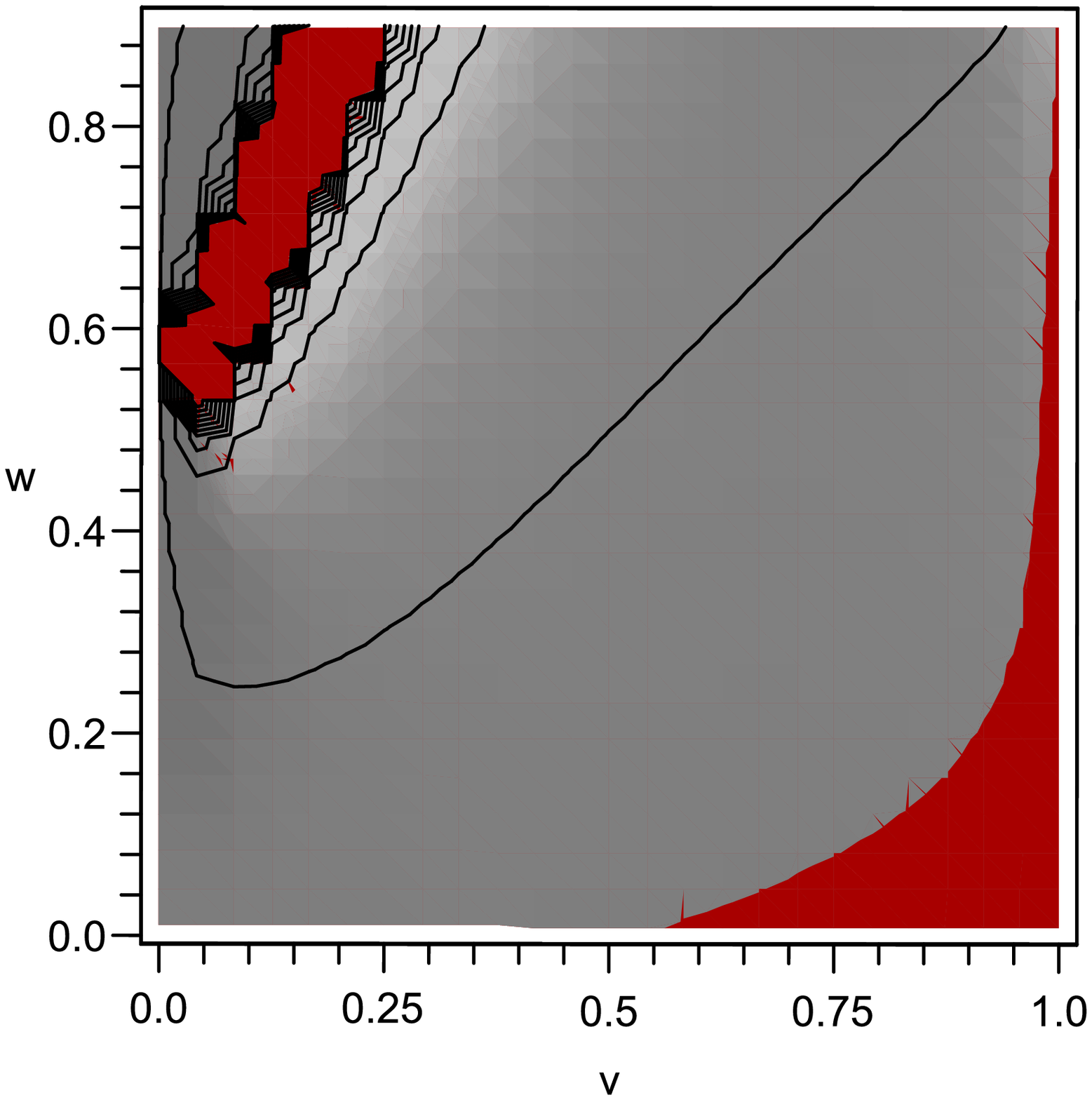}\vspace{0.7cm}
\caption{Region in the $vw$-plane where the squared sound speed is negative (red-colored regions) i. e. where the Laplacian instability eventually develops, for the negative coupling case ($\alpha<0$). As in FIG. \ref{fig02}, in order to fit the whole phase plane into a finite-size box, we have chosen the bounded variables $v=x/x-1$ ($0\leq v\leq 1$) and $w=y/y-1$ ($0\leq w\leq 1$), so that the phase plane $vw$ is the unit square. In the figure the exponential potential (top panels) and the power-law potential (bottom panels) are chosen for different values of the parameters $\lambda$ and $n$ respectively. In the top panels, from left to the right: $\lambda=-5$, $\lambda=-2$, $\lambda=2$ and $\lambda=5$, while in the bottom panels: $n=-2$, $n=-1$, $n=1$ and $n=2$, respectively. The slanted irregular (cell-shaped) regions in the top-left corner of the right-hand figures, represent asymptotic regions so that the red colored cells in these regions should not be related to the Laplacian instability.}\label{fig4}\end{figure*}

\begin{figure*}
\includegraphics[width=5cm]{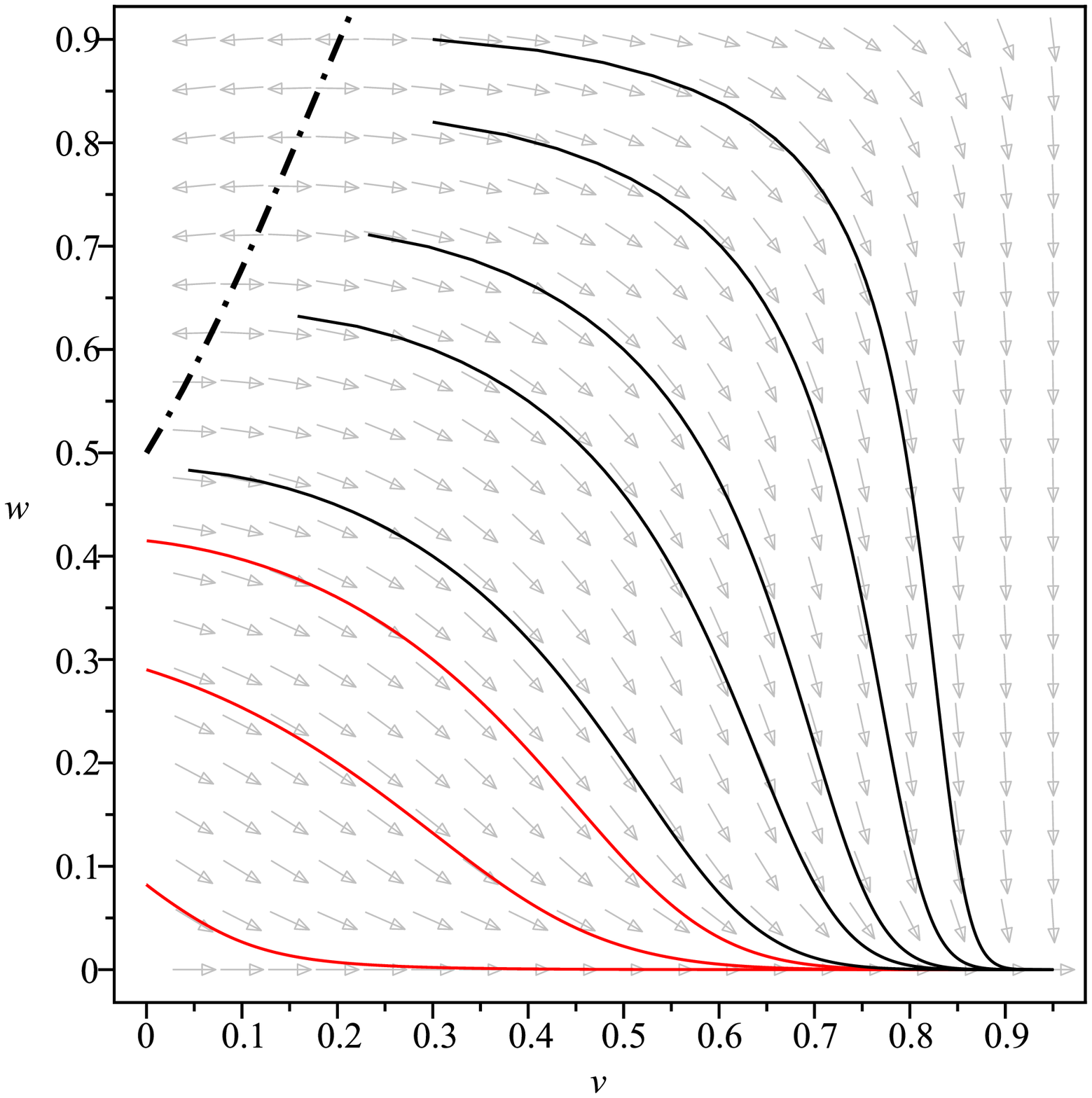}
\includegraphics[width=6cm]{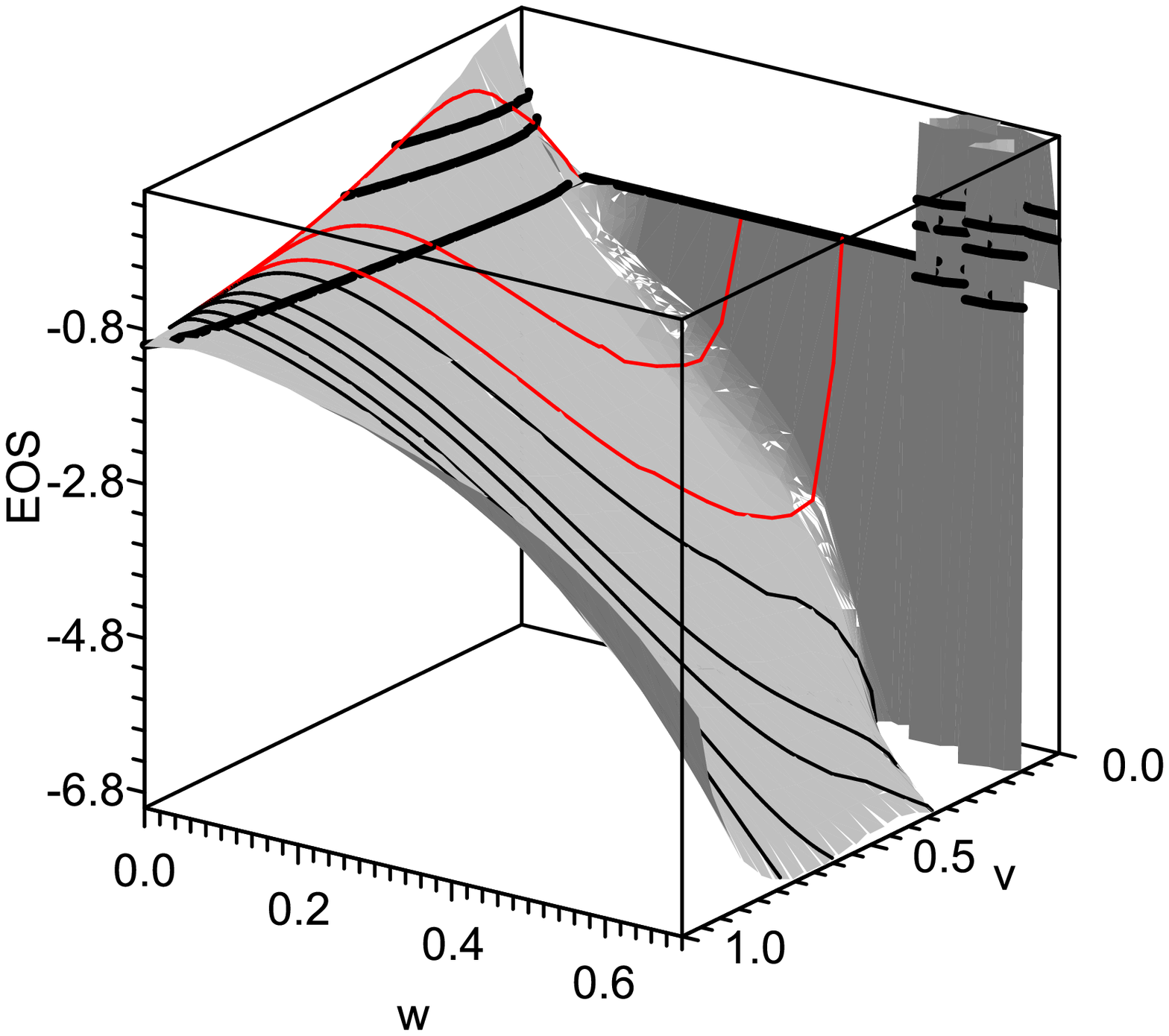}
\includegraphics[width=6cm]{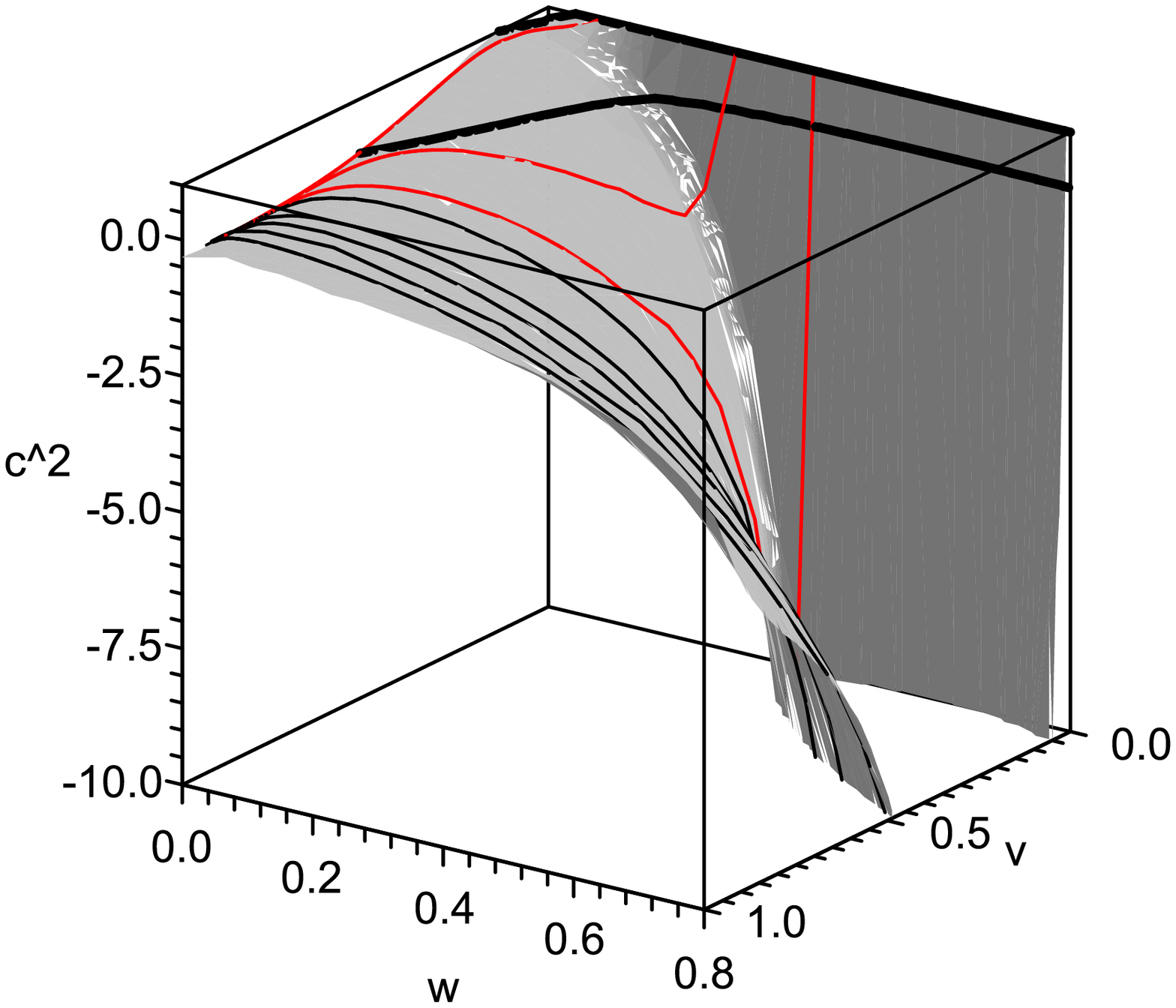}
\includegraphics[width=5cm]{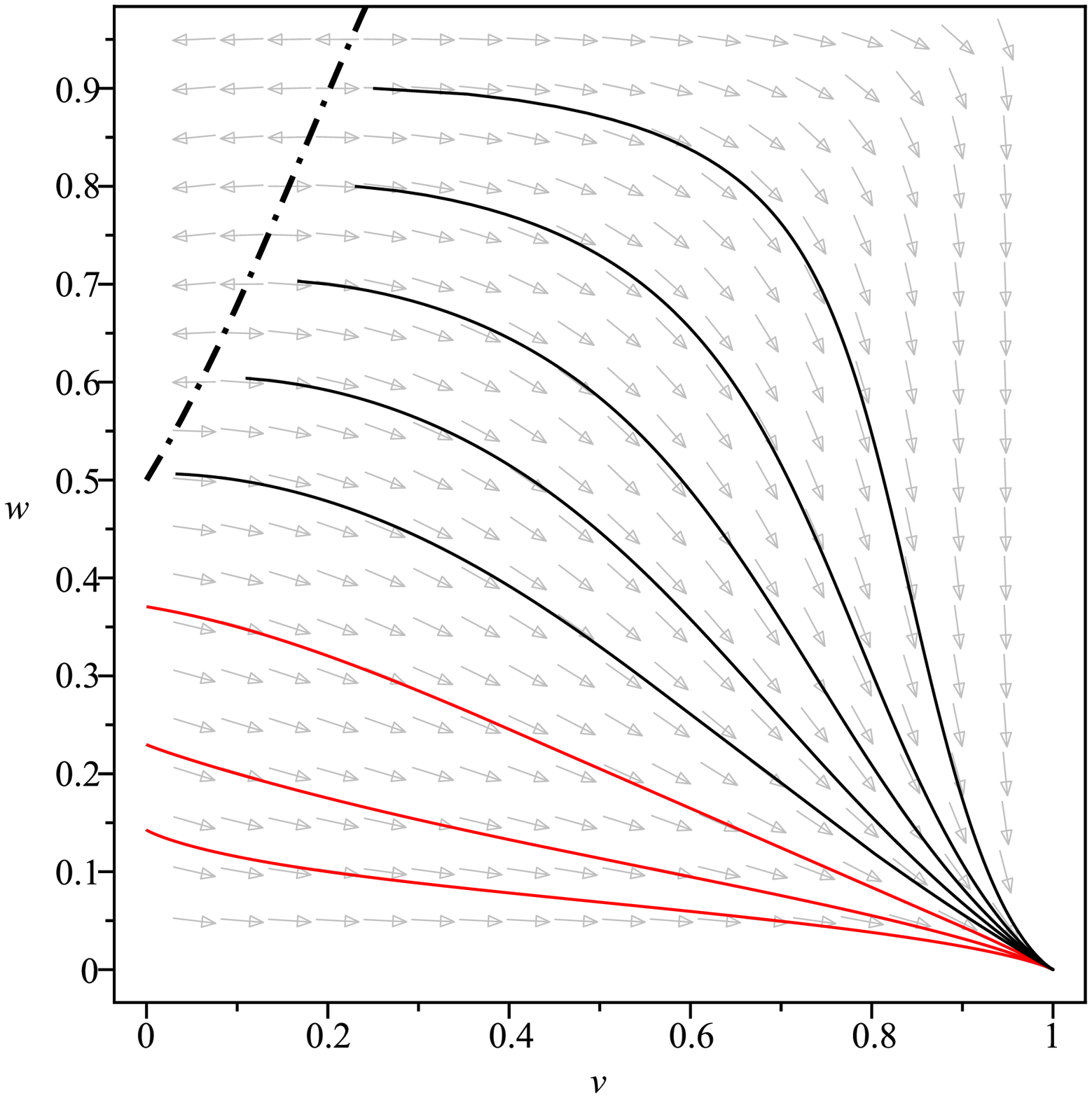}
\includegraphics[width=6cm]{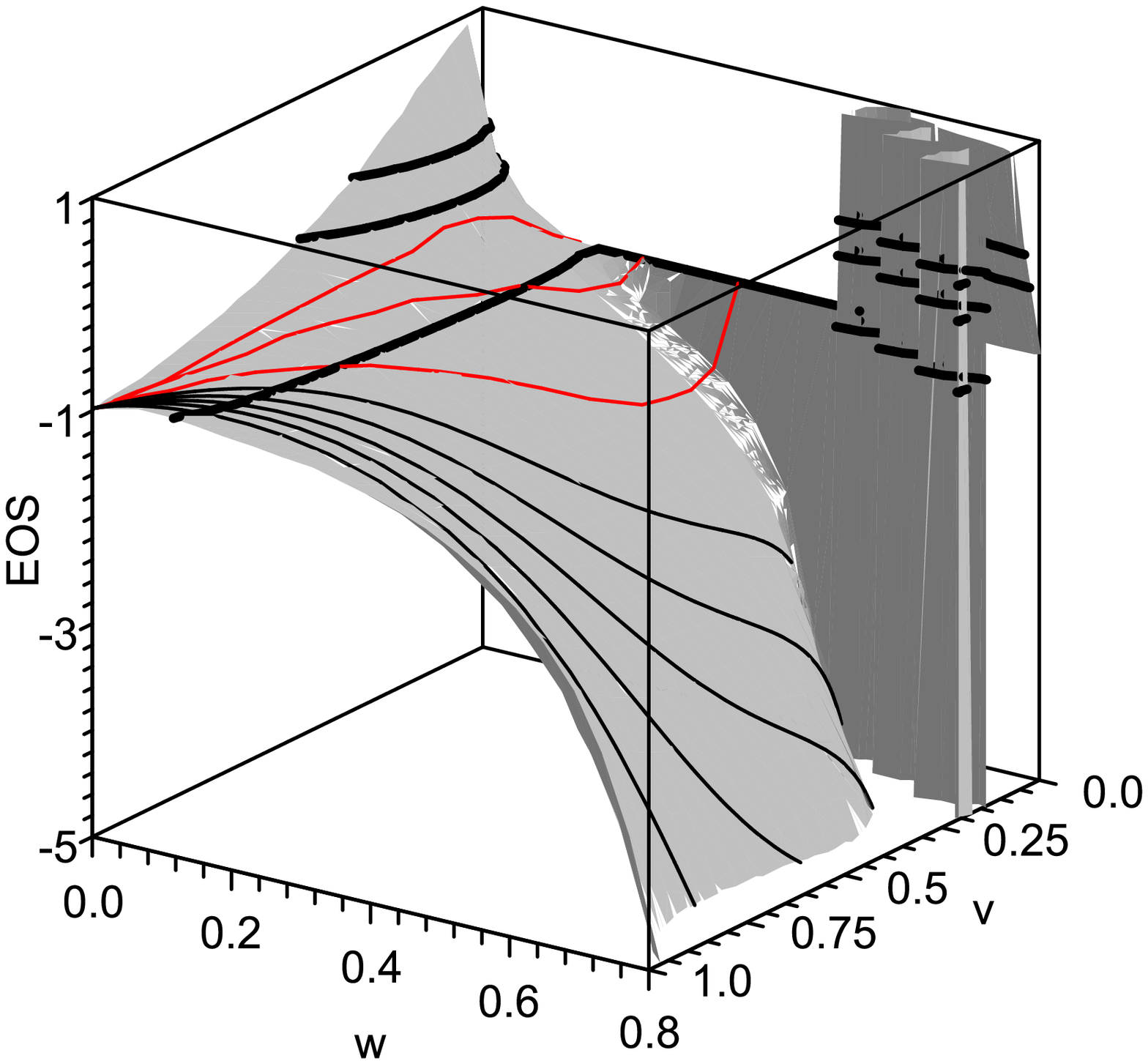}
\includegraphics[width=6cm]{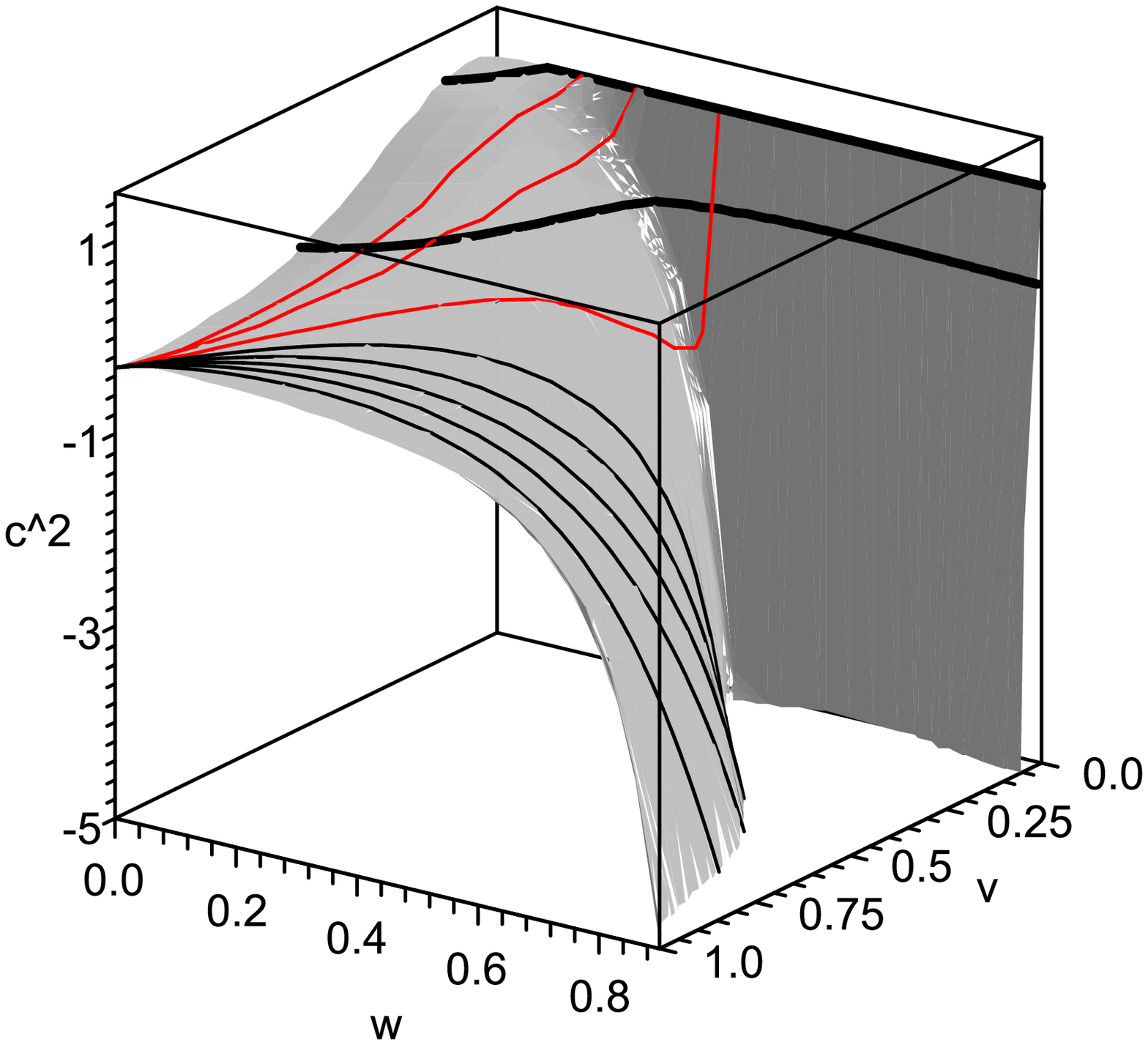}\vspace{0.7cm}
\caption{Phase portraits (left) of the dynamical system \eqref{ode-vw}, EOS-embedding diagrams (middle) and $c^2_s$-embedding diagrams (right) corresponding to the cosmological model \eqref{action} with the negative coupling ($\alpha<0$). In the top panels the decaying exponential potential \eqref{exp-pot} ($\lambda=-5$) has been chosen, while in the bottom panels the inverse power-law potential \eqref{pow-law-pot} ($n=-1$) is considered. As in FIG. \ref{fig3}, in the EOS-embedding diagrams the thick contours are drawn for $\omega_\text{eff}=-1/3$ (upper contour) and for $\omega_\text{eff}=-1$ (lower contour), while in the $c^2_s$-embeddings the drawn thick contours are for $c^2_s=1$ (upper contour) and for $c^2_s=0$ (lower contour). It is evident from the $c^2_s$-embedding diagrams that as the orbits approach to the global attractor ${\cal P}_A:(1,0)$, these enter a domain on the surface $c^2_s=c^2_s(v,w)$ where the squared sound speed becomes negative, signaling the eventual development of a Laplacian instability.}\label{fig5}\end{figure*}

In the figures in FIG. \ref{fig4} the red-colored regions in the unit square are the ones where $c^2_s<0$, i. e., where the Laplacian instability develops. We concentrate in the first and second figures from left to the right -- the plots corresponding to the decaying exponential (top) and to the inverse power-law (bottom) potentials, respectively -- since only for these choices the crossing of the phantom divide may happen. It is obvious from the $c^2_s$-embedding diagrams in FIG. \ref{fig5}, that independent of the choice of the self-interaction potential (either the decaying exponential or the inverse power-law) and of the initial conditions, the development of a gradient instability is inevitable since, as the orbits in the unit (phase) square $\{(v,w):0\leq v\leq 1,0\leq w\leq 1\}$ approach to the global attractor, these necessarily enter the region where $c^2_s<0$. Besides, at the global attractor itself the squared sound speed is negative. We shall come back to this issue again in the next section where the basic properties of the corresponding dynamical system are discussed in connection with the bounds on the squared sound speed.

%-----------------------------------

%-----------------------------------

%-----------------------------------

%-----------------------------------

%%%%%%%%%%%%%%%%%%%%%%%%%%%%%%%%%%%%%%%%%%%%%%%%%%%%%%%%%%%%%%%%%%%%%%%%%%

\section{Squared Sound Speed and the Dynamical System}\label{sec-ds}

Given that in order to illustrate the main results of the present investigation we heavily rely on the properties of the dynamical system corresponding to the cosmological model of interest, here we give a compact exposition of the most elementary of these properties in connection with the bounds on the squared sound speed.

We want to underline that here we do not care about a detailed study of the critical points of the dynamical system and their stability. A detailed dynamical systems study of the present model can be found in \cite{huang}. Different orbits in the given phase space will correspond to possible patterns of cosmological evolution that are sustained by the dynamical system and, consequently, by the cosmological equations \eqref{feqs}. Moreover, every possible orbit that can be generated by every possible choice of the initial conditions, represents a potential cosmic history for our universe. The critical points of the dynamical system correspond to ``outstanding'' or generic cosmological solutions of \eqref{feqs}.

\subsection{Positive coupling $\alpha>0$}

Let us investigate the asymptotic properties of the dynamical system corresponding to the cosmological equations \eqref{feqs} in the phase plane $$\psi=\{(x,y):0\leq x\leq 1/3,y\geq 0\}.$$ It can be demonstrated that the second order cosmological field equations \eqref{feqs} can be traded by the following system of 2 ordinary differential equations on the variables $x$, $y$:

\bea &&x'=\frac{x[\epsilon(1-2x)+y]}{1-3x}-\frac{(1-x)(\epsilon x+y)(\omega_\text{eff}+1)}{2(1-3x)},\nonumber\\
&&y'=y_\phi\sqrt\frac{2x(\epsilon x+y)}{3(1-3x)},\label{ode-xy}\eea where the comma means derivative with respect to the time variable $d\tau=\alpha Hdt$. The problem with \eqref{ode-xy} is that the phase plane is unbounded ($0\leq y<\infty$) so that it may happen that one or several critical points of the dynamical system at infinity are unseen in a finite region of the phase plane. This is why in \eqref{u-var} we introduced the bounded variable $u=y/y+1$ ($0\leq u\leq 1$). After this choice the whole phase plane is shrunk into the phase rectangle:

\bea \psi_{\alpha>0}=\{(x,u):0\leq x\leq 1/3,0\leq u\leq 1\},\label{bound-rect}\eea and the ODE system \eqref{ode-xy} is rewritten as:

\bea &&x'=\frac{x[\epsilon(1-2x)(1-u)+u]}{(1-3x)(1-u)}-\frac{x(1-x)[\epsilon(1-u)+3u][\epsilon(1-2x)(1-u)+u]}{(1-3x)(1-u)^2F_\epsilon}\nonumber\\
&&\;\;\;\;\;\;\;\;\;\;\;\;\;\;\;\;\;\;\;\;\;\;\;\;\;\;\;\;\;\;\;\;\;\;\;\;\;\;\;\;\;\;\;\;\;-\sqrt\frac{2x(1-x)^2(1-3x)[\epsilon x(1-u)+u]}{3(1-u)}\frac{u_\phi}{(1-u)^2F_\epsilon},\nonumber\\
&&u'=u_\phi\sqrt\frac{2x[\epsilon x(1-u)+u]}{3(1-3x)(1-u)},\label{ode-xu}\eea where $$F_\epsilon=\frac{\epsilon(1-3x+6x^2)(1-u)+(1+3x)u}{1-u}.$$ In the left-hand figures in FIG. \ref{fig3} the phase portraits of the dynamical system \eqref{ode-xu} are drawn for the decaying exponential with $\lambda=-5$ (top) and for the inverse power-law with $n=-1$ (bottom), for a set of 9 and 8 different initial conditions respectively.

A crude inspection of \eqref{ode-xu} reveals that, independent of the specific functional form of the self-interaction potential, among the equilibrium configurations of the dynamical system in the phase rectangle \eqref{bound-rect}, there is a critical manifold: ${\cal M}_0=\{(0,u):0\leq u\leq 1\}.$ Equilibrium points in this manifold have different stability properties. The origin ${\cal P}_0:(0,0)$ is a stable critical point. Moreover, it is the global future attractor. The remaining points ${\cal P}_i\in{\cal M}_0$ represent unstable equilibrium configurations and can be only local sources. In the phase portraits (left-hand figures) in FIG. \ref{fig3} the red-color orbits start at local sources in ${\cal M}_0$ and end up at the global attractor ${\cal P}_0$. For points ${\cal P}_\delta:(\delta,u)$ in the neighborhood of ${\cal M}_0$, where $\delta\ll 1$ is a small parameter, we have that: $$c^2_s\approx 1-6\sqrt{2}\,\delta^{1/2}\sqrt\frac{u}{1-u}\,u_\phi,$$ where the terms $\propto\delta$ and of higher orders in the small parameter have been omitted. Hence, if we assume that $u\neq 0$ -- i. e., if exclude the global attractor at the origin -- assuming potentials with the negative slope: $$V_\phi<0\Rightarrow y_\phi<0\Rightarrow u_\phi<0,$$ for points in the neighborhood of the critical manifold ${\cal M}_0$, the speed of sound becomes superluminal $c^2_s>1$. This behavior is illustrated in the $c2_s$-embedding diagrams in FIG. \ref{fig3}, where it is appreciated that as the red-colored orbits leave the source points the speed of sound becomes superluminal.\footnote{At the source points, as well as at the global attractor at the origin, where $\delta=0$, we have that $c^2_s=1$.} For orbits that start at points to the right of the phase rectangle ($x=1/3-\delta$), it is found that there are regions in the phase plane where the squared sound speed becomes negative, signaling the development of Laplacian instability. This is illustrated in the first and second figures (from left to the right) in FIG. \ref{fig1} where the small red-colored regions in the $xu$-plane represent the domains in the phase rectangle where $c^2_s<0$. In the $c^2_s$-embedding diagrams in FIG. \ref{fig3} it is appreciated that several of the mentioned orbits (continuous black curves) indeed meet the gradient instability regions.

\subsection{Negative coupling $\alpha<0$}

In terms of the variables $v$, $w$ in \eqref{vw-var} the autonomous system of ODE \eqref{ode-xy} can be written in the following way:

\bea &&v'=\left(\frac{1-v}{1-w}\right)\left(\frac{1+v-2w}{1+2v}\right)-\frac{(1-v)(v+w-2vw)(\omega_\text{eff}+1)}{2(1+2v)(1-w)},\nonumber\\
&&w'=w_\phi\sqrt\frac{2v(v+w-2vw)}{3(1-v)(1-w)(1+2v)},\label{ode-vw}\eea where $\omega_\text{eff}$ is given by \eqref{weff-vw}. The phase portraits of the dynamical system \eqref{ode-vw} are shown in the left-hand figures in FIG. \ref{fig5} for the decaying exponential with $\lambda=-5$ (top) and for the inverse power-law potential with $n=-1$ (bottom).\footnote{Recall that we are interested in potentials that allow for the crossing of the phantom divide exclusively.} The global (future) attractor at ${\cal P}_A:(1,0)$ is sharply appreciated. If we make the replacement of $x\rightarrow v/v-1$ and of $y\rightarrow w/w-1$ in \eqref{c2s-master-eq}, and then we evaluate the squared sound speed at the attractor, we get: $$\lim_{(v,w)\rightarrow(1,0)}c^2_s(v,w)=-\frac{1}{3}.$$ This means that, at least at the attractor $c^2_s<0$, so that a Laplacian instability eventually develops. In the $c^2_s$-embedding diagrams in FIG. \ref{fig5} it is seen that, as a matter of fact, to a large extent the embedded phase space orbits lie in domains on the surface $c^2_s=c^2_s(v,w)$ that are below the contour corresponding to $c^2_s=0$. Moreover, there are orbits that entirely lie in domains below the mentioned contour in the extended phase space, which means that the corresponding whole cosmic histories are classically unstable under scalar perturbations of the cosmic background.

%%%%%%%%%%%%%%%%%%%%%%%%%%%%%%%%%%%%%%%%%%%%%%%%%%%%%%%%%%%%%%%%

\section{Pure derivative coupling ($\epsilon=0$)}\label{sec-e0}

The interest in the case where the kinetic coupling is exclusive to the Einstein's tensor, i. e., where the term $g_{\mu\nu}\der^\mu\phi\der^\nu\phi$ is removed from \eqref{action}, is due to the significant simplification of the equations of the resulting cosmological model that allows one to discuss in a fully analytical way on the phantom crossing and the bounds on the squared sound speed.

Actually, in this particular case where $\epsilon=0$, the expressions for the effective EOS parameter and for the squared sound speed become

\bea \omega_\text{eff}+1=\frac{6x}{1+3x}+\frac{\sqrt{8x(1-3x)^3}}{3(1+3x)^2y^3}\;y_\phi,\label{weff-e0}\eea and 

\bea c_s^2=1-\frac{2x(45x^2-54x+29)}{3(1-x)(1+3x)^2}-\frac{6(1-x)}{(1+3x)^2}\sqrt\frac{2x(1-3x)^3}{3y^3}\,y_\phi,\label{c2s-e0}\eea respectively. The analysis of the behavior of the above quantities is straightforward.

\subsection{Positive coupling}

From equation \eqref{weff-e0} it is seen that at the upper boundary: $x=1/3$, the effective (background) fluid behaves like pressureless dust. It is seen also that, provided the slope of the potential is negative: $y_\phi<0$, the second term in the RHS of \eqref{weff-e0} may compensate the contribution of the first-one. For the exponential potential $y_\phi=\lambda y$, for instance, for $$y<\frac{2\lambda^2(1-3x)^3}{27x},$$ the crossing of the phantom divide may happen since $\omega_\text{eff}+1$ becomes negative. For monotonically growing potentials the crossing is not possible.

Causality violations and the development of Laplacian instability in this case are apparent. Even for the constant potential $y_\phi=0$ (this includes the vanishing potential case $V=0$) the instability issue is apparent. Actually, in this case \eqref{c2s-e0} simplifies even more: 

\bea c_s^2=1-\frac{2x(45x^2-54x+29)}{3(1-x)(1+3x)^2}.\label{pot-0}\eea It is straightforward to show that the squared sound speed above is a monotone decreasing function of $x$, and that it vanishes at $x=0.0897$. In the interval $0.0897<x\leq 1/3$, $c^2_s$ is negative. In particular at $x=1/3$ the squared sound speed $c^2_s=-1/3$. The violation of causality in connection with superluminal propagation of the scalar perturbations may happen only for potentials with the negative slope $y_\phi<0$. Only in this case the third term in the RHS of \eqref{c2s-e0} may compensate the contribution from the second one, and may contribute towards superluminality ($c^2_s>1$).

\subsection{Negative coupling}

In this case $-\infty<x\leq 0$, $-\infty<y\leq 0$, so that both variables are unbounded. In terms of the bounded variables $v$, $w$ in \eqref{vw-var}, for the simplest case when the potential is a constant ($y_\phi=0$), the squared sound speed \eqref{pot-0} can be written as: 

\bea c^2_s=1+\frac{2v(29+54v-9v^2)}{3(1-4v)^2},\label{pot-0'}\eea while the corresponding autonomous ODE is

\bea v'=-\frac{2y_0\,v(1-v)^2}{1-4v},\label{ode-pot-0}\eea where $y_0=\alpha V_0$ is a constant. The squared sound speed blows up at the asymptote $v\rightarrow 1/4$, so that a coarse violation of causality eventually occurs. In the phase line $0\leq v\leq 1$ the asymptote $v=1/4$ represents a separatrix, since the orbits of \eqref{ode-pot-0} can not cross from the left to the right of $v=1/4$ and vice versa.

%%%%%%%%%%%%%%%%%%%%%%%%%%%%%%%%%%%%

\section{Discussion}\label{sec-disc}

Our results in the previous sections are clear and convincing. These show that in general terms, without specifying the functional form of the self-interacting potential, the cosmological models based in the theory \eqref{action} -- where the scalar field is kinetically coupled to the curvature -- are unsatisfactory due to the occurrence of causality violations and -- what is more problematic -- of classical Laplacian instabilities, for a non-empty set of initial conditions. These results do not depend on the sign of the coupling constant $\alpha$ in \eqref{action}. We have shown this analytically and also numerically by specifying the form of the potential; we have done this for the exponential and for the power-law potentials. There is, however, a particular class of such models without the potential ($V=0$) and with the constant potential ($V=V_0$) that deserve separate comments since these can be treated in a fully analytical way (see below). 

In general terms theories with the kinetic coupling of the scalar field to the Einstein's tensor -- this is true also for more general Horndeski theories -- all possess some configurations with a superluminal propagation. Besides, these theories have also the speed of propagation of the gravity waves different from the speed of light. In particular, the speed of sound for the scalar perturbations can be subluminal while, simultaneously, the speed of propagation for the gravity waves can be superluminal \cite{germani-1}. In the later reference this has been shown for the theory \eqref{action} with the positive coupling, for the quartic potential during inflation. In \eqref{c2t} the squared speed of propagation of the gravity waves perturbations is given independent of the self-interaction potential: 

\bea c^2_T=\frac{1+x}{1-x}.\label{c2t-x}\eea This confirms that the speed of the gravitational waves is always superluminal if assume the positive coupling $\alpha>0$. For the negative coupling, in terms of the bounded variable $v$ ($0\leq v\leq 1$) we have that: 

\bea c^2_T=1-2v.\label{c2t-v}\eea This means that for $0\leq v\leq 1/2$ the speed of propagation of the gravitational waves meets the bounds: $0\leq c^2_T\leq 1$, meanwhile, for $v>1/2$, the squared sound speed of the tensor perturbations is a negative quantity that leads eventually to the development of a Laplacian instability.

%-----------------------------------

\begin{figure}
\includegraphics[width=6cm]{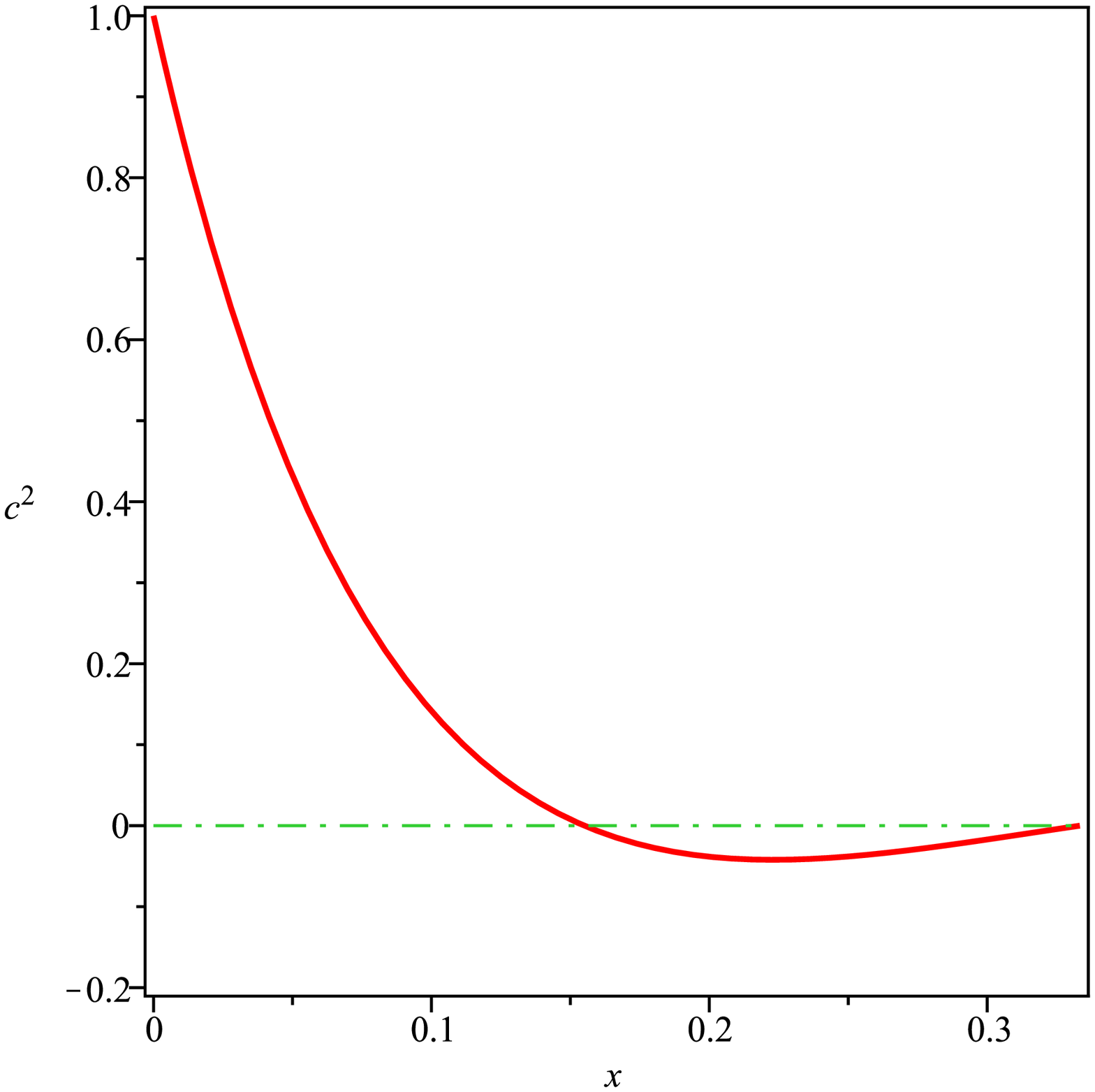}
\includegraphics[width=6cm]{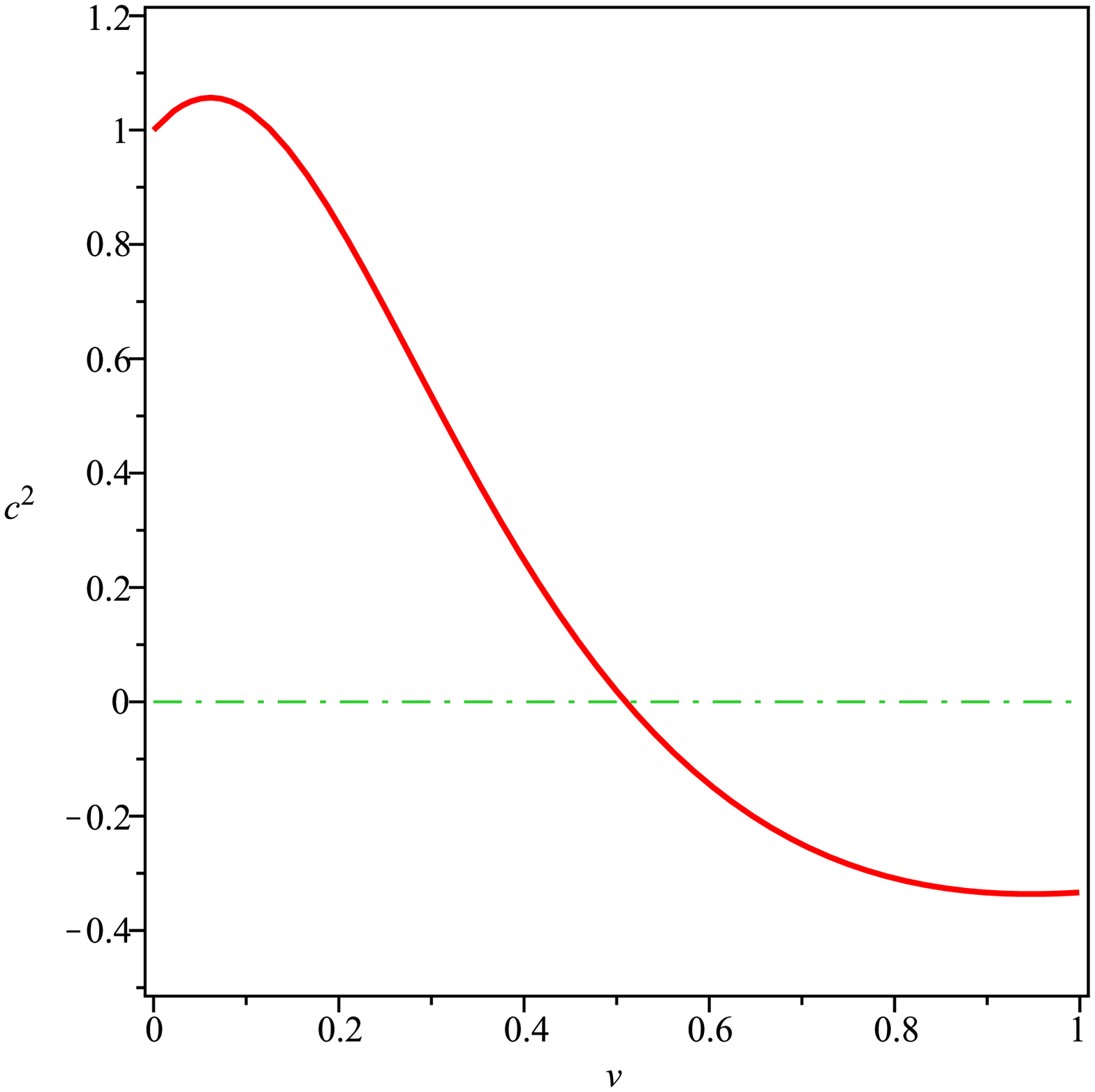}\vspace{0.7cm}
\caption{Plot of $c^2_s$ vs $x$ (top) and of $c^2_s$ vs $v$ (bottom) for the model \eqref{action} with the constant potential ($y_0=\alpha V_0$). The top figure is for the positive coupling case $\alpha>0$ ($0\leq x\leq 1/3$), while the bottom figure is for the negative coupling case $\alpha<0$ ($0\leq v\leq 1$). In the top we have arbitrarily set $y_0=10$, while in the bottom $y_0=-0.01$. The dash-dot horizontal line marks the lower bound $c^2=0$ on the squared speed of sound. It is appreciated that, independent of the sign of the coupling, there always exist an interval in the $x$/$v$-coordinate where $c^2_s<0$, meaning that a Laplacian instability may eventually arise.}\label{figfin}\end{figure}

%-----------------------------------

In order to further illustrate our results, let us to discuss in detail the constant potential case: $$V=V_0\Rightarrow y=y_0=\alpha V_0,$$ with the vanishing potential as the particular case when $y_0=0$, that can be studied analytically. We have that (for definiteness we consider $\epsilon=1$):

\bea 3\alpha H^2=\frac{x+y_0}{1-3x}.\label{3h2-y0}\eea Since for the positive coupling $0\leq x\leq 1/3$, from \eqref{3h2-y0} it follows that for $\alpha>0$ the Hubble rate is unbounded from above and bounded from below: $\sqrt{y_0/3\alpha}\leq H<\infty$. 

For the negative coupling $\alpha<0$ ($-\infty<x\leq 0$) the Hubble rate is bounded (in this case the constant $y_0$ should be a negative quantity as well): 

\bea &&\frac{1}{3\sqrt{-\alpha}}\leq H\leq\sqrt\frac{y_0}{3\alpha}=\sqrt\frac{V_0}{3}\;(V_0>1/|3\alpha|),\nonumber\\
&&\sqrt{V_0/3}\leq H\leq 1/3\sqrt{-\alpha},\;(V_0<1/|3\alpha|).\label{h-bound}\eea

For the constant potential the dynamical system \eqref{ode-xy} reduces to a single ordinary differential equation (ODE):

\bea x'=-\frac{2x(1-2x+y_0)}{1-3x}\left[\frac{y_0+(1-3y_0)x-3x^2}{1+y_0+3(y_0-1)x+6x^2}\right].\label{ode-apos}\eea For the positive coupling ($0\leq x<1/3$), one of the critical points of the ODE \eqref{ode-apos} is at the origin $x=0$. This is a stable equilibrium point since linear perturbations $\delta$ around it ($x\rightarrow 0+\delta$) exponentially decay with the time $\tau=\alpha\ln a$: $\delta(\tau)\propto\exp(-2y_0\tau),$ or in terms of the scale factor of the Universe: $$\delta(a)\propto a^{-2\alpha y_0},$$ the perturbations decay as an inverse power-law. The above means that the cosmic dynamics ends up at the de Sitter attractor $x=0$, where $H=H_0=\sqrt{V_0/3}$. Consistently with the fact that, for the positive coupling, the late time dynamics is not modified by the kinetic coupling \cite{sushkov}, the above is the standard late time behavior expected in any scalar field model with a constant potential. For the vanishing potential the asymptotic late time dynamics corresponds to the empty static universe $H=0$, since for this particular case the origin ($x=0$) is the attractor equilibrium configuration as well: The small linear perturbations around the origin decay like $$\delta(\tau)\propto\tau^{-1}\Rightarrow\delta(a)\propto\frac{1}{\alpha\ln a}.$$ This model is plagued by the Laplacian instability as it can be seen from the top figure in FIG. \ref{figfin}, where the squared sound speed is plotted against $x$.

For the negative coupling ($-\infty<x\leq 0$) it is better to use the bounded variable $v=x/x-1$ ($0\leq v\leq 1$). In this case the autonomous ODE \eqref{ode-apos} transforms into:

\bea v'=-\frac{2v(1-v)[1+y_0+(1-y_0)v]}{1+2v}\left[\frac{y_0-(1-y_0)v-2(1+y_0)v^2}{1+y_0+(1-5y_0)v+4(1+y_0)v^2}\right].\label{ode-aneg}\eea Two of the critical points of the ODE \eqref{ode-aneg} are at the origin ($v=0$ $\Leftrightarrow x=0$), and at $v=1$ ($x\rightarrow\infty$). The dynamical equations for linear perturbations $\delta$ around these points read: $\delta'=-2y_0\delta$ and $\delta'=-\delta/3$, respectively. After integration, for perturbations around the origin $v=0$, we get that $\delta(a)\propto a^{-2\alpha y_0}$, i. e., given that both $\alpha$ and $y_0$ are negative for this case, then the perturbations decay with the cosmic expansion. Meanwhile, for perturbations around $v=1$ we get that $\delta(a)\propto a^{-\alpha/3}$ and, since $\alpha$ is negative, then the corresponding perturbation grows with the expansion of the Universe. Hence the point $v=1$ is unstable while the origin $v=0$ is the attractor. Since in this case: $$3\alpha H^2=\frac{y_0-(1+y_0)v}{1+2v},$$ in models with the constant potential (for the negative coupling) the Universe starts a the unstable de Sitter solution with $H=1/3\sqrt{-\alpha}$ and ends up its history at the late-time de Sitter solution with $$3\alpha H^2=y_0\Rightarrow H=H_0=\sqrt{V_0/3}.$$  The asymptotic de Sitter state at $v=1$: $H=1/3\sqrt{-\alpha}$, is to be associated with the primordial inflation \cite{sushkov} and the fact that it is a unstable equilibrium state warrants the natural (required) exit from the early times inflationary stage.\footnote{Transient quasi-de Sitter phases of the cosmic evolution can be found also for other potentials than the constant one.} Notice that for the above picture to make physical sense, in \eqref{h-bound} we have to choose the bottom-line bound, i. e., $V_0<1/|3\alpha|$. Otherwise the attractor would be at higher curvature than the starting point of the cosmic expansion, which is a non-sense from the point of view of the inflationary history of our Universe. 

In spite of the claims that this picture represents an appropriate description of the primordial inflation, according to \eqref{c2t-v} in the neighborhood of the inflationary equilibrium point: $v=1\mp\delta$ ($\delta\ll 1$), for the squared speed of propagation of tensor perturbations we have that: $c^2_T\approx -1\pm 2\delta$, so that the development of a Laplacian instability forbids the -- otherwise unphysical -- inflationary stage in the model.

The estimated value of the coupling constant in \cite{sushkov-a} is of about:

\bea |\alpha|\sim 10^{-74}\text{sec}^2\approx 10^{-24}\text{GeV}^{-2},\label{a-bound}\eea  where the authors chose the time at which inflation is assumed to start $t\approx 10^{-36}$sec. We may as well choose the time at which inflation is assumed to have ended: $t\approx 10^{-33}$sec. The estimated value for the coupling in this case is about 4 orders of magnitude larger: 

\bea |\alpha|\sim 10^{-70}\text{sec}^2\approx 10^{-20}\text{GeV}^{-2}.\label{a-bound'}\eea If combine the above estimates with the tight constraint on the difference in speed of photons and gravitons $|c^2_T-1|\leq 10^{-15}$ (in this paper we have chosen the units where $c^2=1$) implied by the announced detection of gravitational waves from the neutron star-neutron star merger GW170817 and the simultaneous measurement of the gamma-ray burst GRB170817A \cite{ligo}, since according to \eqref{c2t-x}: $$c^2_T-1=\frac{2x}{1-x}\Rightarrow 2x\leq 10^{-15},$$ we get that $\dot\phi^2\leq 10^5-10^9$GeV$^2$, i. e., $\dot\phi^2\leq 10^{-33}-10^{-29}M_\text{pl}$, where $M_\text{pl}\approx 10^{19}$GeV is the Planck mass. These estimates leave not much freedom for the scalar field to behave different from an effective cosmological constant. 

The above exposed -- quite simple -- picture is overshadowed by the stability problems associated with the scalar and tensor modes of the perturbations whose energy density grows without bound due to fact that, for these modes it may happen that $c^2_s<0$ ($c^2_T<0$). In the bottom figure in FIG. \ref{figfin} the plot of $c^2_s$ vs $v$ is drawn for $y_0=-0.01$. The conditions for the development of the Laplacian instability ($c^2_s<0$) are evident in the figure, in particular for points in the neighborhood of (including) the source equilibrium configuration that can be associated with the primordial inflation. Besides, in the neighborhood of this point we have also that $c^2_T<0$, so that the tensor modes are classical unstable as well.

%%%%%%%%%%%%%%%%%%%%%%%%%%%%%%%%%%%%%

\section{Conclusion}\label{sec-concl} 
 
In this paper we have investigated several problems: i) phantom barrier crossing, ii) causality and iii) classical Laplacian instability, and their potential interconnection in the model \eqref{action} where the scalar field has non-minimal derivative (kinetic) coupling to the Einstein's tensor. As far as we know this is the first time when the present model is checked in all detail against the physical bounds on the squared sound speed (see footnote 5 in the introductory section of this paper). We have also developed an illustrative procedure that allows to show geometrically the evolution of given physical parameters (the effective EOS and the squared sound speed in the present work) along given phase space orbits. The resulting procedure -- called here as ''embedding diagram'' -- geometrically illustrates the way these parameters of physical interest evolve along potential cosmic histories. The power of the procedure relies, precisely, on the fact that each phase space orbit entails a potential cosmic history that is sustained by the dynamical system corresponding to the cosmological field equations of the model \eqref{feqs}.

We have shown, both analytically and numerically, that violations of causality and -- what is more disturbing -- the occurrence of Laplacian instability during the propagation of the scalar and of the tensor perturbations, are distinctive features of the cosmological models based in the action \eqref{action} no matter what the sign of the coupling constant is.\footnote{For the tensor perturbations the violation of causality may happen only for the positive coupling case, while the Laplacian instability develops only for the negative coupling.} Moreover, even if the scalar perturbations can propagate subluminally, during inflation the gravitational waves travel with superluminal velocity (this is true for the positive coupling exclusively) as shown in \cite{germani-1} for the model \eqref{action} with the quartic potential $V\propto\phi^4$. In the general case -- see \eqref{c2t}, \eqref{c2t-x} or \eqref{c2t-v} -- the situation can not be more hopeless: Independent of the self-interaction potential, for the positive coupling the tensor perturbations propagate superluminally, while for the negative coupling a Laplacian instability arises. This latter instability invalidates the possibility for the model to describe the primordial inflation.

It has been shown also that, in the positive coupling case ($0\leq x\leq 1/3$), a sufficient (but not necessary) condition for superluminality to happen is that $\omega_\text{eff}+1<0$, since in this case the third term in the RHS of \eqref{c2s-master-eq} also adds to the unity. Since the crossing of the phantom barrier warrants that for some $x$-interval $\omega_\text{eff}+1<0$, then it also warrants that superluminality will happen. However, as mentioned before, it is not necessary that the crossing occurs in order to have superluminal propagation of the perturbations of the background. One trivial example can be the situation when $\omega_\text{eff}+1<0$ for all times. In this case violations of causality arise even when the crossing does not occur.

For the quintessence model with the kinetic coupling to the Einstein's tensor, in the particular case when the potential is a constant $V=V_0$, eventual violations of the physical bounds of the squared sound speed are evident as well: No matter whether the coupling is positive or negative, the asymptotic dynamics at early times develops a (classical) Laplacian instability that makes impossible the formation of cosmic structure. This makes very improbable that the primordial inflationary stage can be described by this cosmological model as suggested, for instance, in \cite{sushkov, matsumoto}. 

Although we lack a demonstration, we suspect that the violation of the bounds $0\leq c^2_s\leq 1$ on the squared sound speed are a feature of galileon models in general. In particular the cubic galileon model of \cite{q-gal, rob-q-gal} seems to suffer from the same problems. A demonstration of the latter assumption will be the subject of forthcoming work.

%%%%%%%%%%%%%%%%%%%%%%%%%%%%

\section*{ACKNOWLEDGMENTS}

Very interesting and useful comments by A. Vikman are sincerely acknowledged. The authors are grateful to SNI-CONACyT for continuous support of their research activity. The work of RG-S was partially supported by SIP20172234, SIP20160512, COFAA-IPN, and EDI-IPN grants. IQ, TG and FAH-R thank CONACyT of M\'exico for support of this research. UN also acknowledges PRODEP and CIC-UMSNH.

%%%%%%%%%%%%%%%%%%%%%%%%%%%%%%%%%%%%%%%%%%%%%%%%%%%%%%%%%%%%%%%%%%%%%%%%%%%%%%%%%%%%%

\section{Appendix: Classical instability due to imaginary sound speed}\label{app}

Even if the theory \eqref{action} is free of the Ostrogradsky instability (the equations of motion are second order in the derivatives), it may contain other kinds of instability since it is based in a non-standard Lagrangian. Here we shall discuss on one such kind of instability that may arise in the theory with non-minimal derivative coupling with the Einstein's tensor due to ``imaginary'' sound speed. 

Let $\rho_B$ and $p_B$ be the energy density and barotropic pressure of the FRW cosmological background. If consider small perturbations of the background energy density: $\rho_B(t)+\delta\rho_B({\bf x},t)$, the conservation of energy and stresses $\nabla^\mu T_{\mu\nu}=0$, leads to the wave equation \cite{peebles-rmp}: 

\bea \left(-\frac{\der^2}{\der t^2}+c^2_s\nabla^2\right)\delta\rho_B=0,\label{gen-waveq}\eea where $\nabla^2=\der^2/\der{\bf x}^2$ and $c^2_s=dp_B/d\rho_B$ is the speed of sound squared. The solution of the wave equation \eqref{gen-waveq} is given by $\delta\rho_B=\delta\rho_{B0}\exp(-i\omega t+i{\bf k}{\bf x})$, so that the standard dispersion relation is found: 

\bea \omega^2-c^2_sk^2=0.\label{dispersion-r}\eea For positive $c^2_s>0$ the solution is a free wave propagating with speed $c_s$, while for negative $c^2_s<0\Rightarrow c_s=i\bar c_s$, the frequency $\omega=\pm kc_s=\pm ik\bar c_s$ is imaginary, so that the solution of \eqref{gen-waveq} is not a propagating free wave but an exponentially growing spatial perturbation:

\bea \delta\rho_B=\delta\rho^+_{B0}\,e^{2\pi\bar c_s t/\lambda}\exp(i{\bf k}{\bf x})+\delta\rho^-_{B0}\,e^{-2\pi\bar c_s t/\lambda}\exp(i{\bf k}{\bf x}),\label{s-pert}\eea where we have taken into account that $k=2\pi/\lambda$ is the wave number of the perturbation ($a/k$ is the physical wavelength of the perturbation). Since the negative frequency part of the perturbation decreases with the time, eventually the energy density of the perturbations uncontrollably grows resulting in a classical instability of the cosmological model. As seen the increment of instability is inversely proportional to the wavelength of the perturbations and the models where $c^2_s<0$ are violently unstable so that these should be rejected \cite{chinos}.

The situation is a bit more complex for a scalar field \cite{ellis-roy, mukhanov} (see also \cite{noh}), which is the case considered in this paper. As an illustration, let us consider a general action of the form:

\bea S=\frac{1}{2}\int d^4x\sqrt{|g|}\,R+\int d^4x\sqrt{|g|}p_\phi(X,\phi),\label{action-mukhnov}\eea where $X\equiv(\der\phi)^2/2$, $p_\phi={\cal L}_\phi$ is the parametric pressure of the scalar field and $\rho_\phi=2X{\cal L}_{\phi,X}-{\cal L}_\phi$ is its energy density, with $Z_{,X}$ denoting the partial derivative with respect to $X$. Varying the scalar field Lagrangian ${\cal L}_\phi$ with respect to the metric one gets the stress-energy tensor for the scalar field: $$T^{(\phi)}_{\mu\nu}=\left(\rho_\phi+p_\phi\right)u_\mu u_\nu+p_\phi g_{\mu\nu},$$ where $u_\mu=\der_\mu\phi/\sqrt{2X}$. As stated in \cite{mukhanov}, the Lagrangian ${\cal L}_\phi$ can be used to draw a useful analogy with hydrodynamics. Indeed, if $p_\phi$ depends only on $X$, then $\rho_\phi=\rho_\phi(X)$. In many cases the equation $\rho_\phi=2X p_{\phi,X}-p$ can be solved giving the equation of state $p_\phi=p_\phi(\rho_\phi)$ for an ''isentropic'' fluid. In the general case, when $p_\phi=p_\phi(X,\phi)$, the pressure cannot be expressed only in terms of $\rho_\phi$. However, even in this case the hydrodynamical analogy is still useful.

If consider small perturbations of the scalar field: $\phi(t,x)=\phi_0(t)+\delta\phi(t,x)$, and recalling that $\delta T^i_k\propto \delta^i_k$, one can write the perturbed FRW metric in the longitudinal gauge: $$ds^2=-(1+2\Phi)dt^2+(1-2\Phi)a^2(t)g_{ik}dx^idx^k,$$ where $\Phi$ is the Newtonian gravitational potential. It is demonstrated in \cite{mukhanov} that the wave equation for the fluctuations of the scalar field in a spatially flat FRW background can be written as: 

\bea v''-c^2_s\nabla^2v-\frac{z''}{z}\,v=0,\label{wave-eq}\eea where $$v=z\left(\Phi+H\frac{\delta\phi}{\dot\phi}\right),$$ is the canonical quantization variable and 

\bea z\equiv\frac{a\sqrt{\rho_\phi+p_\phi}}{c_s H}.\label{z}\eea Besides, in \eqref{wave-eq} the comma denotes derivative with respect to the variable $\tau=\int dt/a$, while the quantity $$c^2_s=\frac{p_{\phi,X}}{\rho_{\phi,X}},$$ plays the role of the effective speed of sound (squared) for the perturbations of the scalar field. For negative $c^2_s<0$ the above equation \eqref{wave-eq} ceases to be a wave equation since it turns from hyperbolic $c^2_s>0$ (the Cauchy problem is well posed) into Elliptic. The imaginary effective sound speed ($c^2_s<0$) of the fluctuations of the scalar field is associated with the so called gradient instability. Notice that if set $v\propto v_k(\tau)\exp{(i{\bf k}{\bf x})}$ ($\nabla^2v=-k^2v$), the wave equation \eqref{wave-eq} can be written as 

\bea v''_k+\left(c^2_sk^2-\frac{z''}{z}\right)v_k=0.\label{waveq}\eea During slow-roll inflation the Hubble rate $H$, $c_s$ and $\rho_\phi+p_\phi$ change much slower than the scale factor $a$, so that, under the reasonable assumption that $(\rho_\phi+p_\phi)/\rho_\phi\ll 1$, from \eqref{z} it follows that $$\frac{z''}{z}\approx\frac{a''}{a}\approx 2(aH)^2.$$ For a given wave number $k$ the term $z''/z$ in \eqref{waveq} can be neglected at early times when the physical wavelength of the perturbations $a/k$ is much smaller than the sound horizon $c_s/H$. Hence, $c_sk\gg aH$ and \eqref{waveq} can be written as: $v''_k+c^2_sk^2v_k=0$, which is similar to \eqref{dispersion-r} if set $v_k(\tau)\propto\exp{(-i\omega\tau)}$.

%%%%%%%%%%%%%%%%%%%%%%%%%%%%

\end{document}